\PassOptionsToPackage{longnamesfirst}{natbib}

\documentclass[ecta,nameyear,draft]{econsocart}

\makeatletter
\let\numberlines@hook\relax
\let\set@numberlines@box\relax
\let\put@numberlines@box\relax
\makeatother

\RequirePackage[colorlinks,citecolor=blue,linkcolor=blue,urlcolor=blue,pagebackref]{hyperref}
\startlocaldefs

\theoremstyle{plain}

\newtheorem{theorem}{Theorem}
\newtheorem{proposition}{Proposition}

\newtheorem{lemma}{Lemma}

\theoremstyle{definition}
\newtheorem{definition}{Definition}

\newtheorem{remark}{Remark}

\newtheorem*{theorem*}{Blackwell's Theorem}

\usepackage[capitalise]{cleveref}
\crefname{figure}{Figure}{Figures}
\Crefname{figure}{Figure}{Figures}
\crefname{assumption}{Assumption}{Assumptions}
\crefname{footnote}{footnote}{footnotes}
\crefname{lemma}{Lemma}{Lemmas}

\usepackage{tikz}
\usepackage{tkz-euclide}
\usetikzlibrary{intersections}
\usepackage{pgfplots}
    \pgfplotsset{compat=1.15}
\usepackage{float}
\usepgfplotslibrary{fillbetween}
\definecolor{DodgerBlue}{RGB}{30,144,255}

\usepackage{standalone}
\usepackage{subcaption}
\usepackage{multirow}
\usepackage{nicematrix}
\usepackage{blkarray}
\usepackage{enumitem}

\usepackage{amssymb}

\usepackage[draft]{graphicx}

\newcommand{\td}{\widehat{\Delta}}
\newcommand{\bq}{\boldsymbol{q}}
\newcommand{\bp}{\boldsymbol{p}}
\newcommand{\bl}{\boldsymbol{\ell}}
\newcommand{\bmu}{\boldsymbol{\mu}}
\newcommand{\bdl}{\boldsymbol{\delta}}
\newcommand{\bnu}{\boldsymbol{\nu}}
\newcommand{\bb}{\boldsymbol{b}}

\newcommand{\bz}{\boldsymbol{z}}
\newcommand{\bu}{\boldsymbol{u}}
\newcommand{\bo}{\boldsymbol{0}}

\newcommand{\bff}{\boldsymbol{f}}
\newcommand{\blb}{\boldsymbol{\lambda}}
\newcommand{\cz}{\mathcal{Z}}

\newcommand{\cp}{\mathcal{P}}

\newcommand{\bmm}{\boldsymbol{m}}
\newcommand{\ux}{\underline{x}}
\newcommand{\ox}{\overline{x}}
\newcommand{\uy}{\underline{y}}
\newcommand{\oy}{\overline{y}}
\newcommand{\ua}{\underline{a}}
\newcommand{\oa}{\overline{a}}
\newcommand{\um}{\underline{m}}
\newcommand{\om}{\overline{m}}
\newcommand{\oX}{\overline{\mathcal{X}}}
\newcommand{\uu}{\underline{u}}
\newcommand{\ou}{\overline{u}}
\newcommand{\ha}{\hat{a}}
\newcommand{\htt}{\hat{t}}
\newcommand{\hs}{\hat{s}}
\newcommand{\hw}{\hat{w}}

\newcommand{\hS}{\widehat{S}}

\newcommand{\mr}{\mathbb{R}}
\newcommand{\me}{\mathbb{E}}
\newcommand{\cb}{\mathcal{B}}

\newcommand{\bdls}{\bdl^*}
\newcommand{\bd}{\boldsymbol{D}}
\newcommand{\cc}{\mathcal{C}}

\newcommand{\cX}{\mathcal{X}}
\newcommand{\cY}{\mathcal{Y}}
\newcommand{\bP}{\boldsymbol{P}}
\newcommand{\cA}{\mathcal{A}}

\usepackage{caption} 
\endlocaldefs

\makeatletter
\def\copyright@text{}
\def\@oddhead{\hfil\small\thepage}
\def\@evenhead{\small\thepage\hfil}
\makeatother

\begin{document}

\begin{frontmatter}

\title{Ranking Statistical Experiments via the Linear Convex Order and the Lorenz Zonoid: Economic Applications}

{\centering\normalsize
\href{https://arxiv.org/pdf/2502.06530}
{\textcolor{blue}{[Latest Version]}}\par}

\runtitle{Ranking Statistical Experiments via the Linear Convex Order: Economic Applications}

\begin{aug}
%
%
%
\author[add1]{\fnms{Kailin}~\snm{Chen}}
\address[add1]{%
\orgdiv{November 16, 2026}\\
}
\end{aug}

\begin{funding}
Chen: Toulouse School of Economics, \href{mailto:kailin.chen@tse-fr.eu}{kailin.chen@tse-fr.eu}. 

This paper was previously circulated under the title ``Experiments in the Linear Convex Order''. The author is indebted to Mehmet Ekmekci, Daniel Hauser, Johannes Hörner, Stephan Lauermann, Pauli Murto, and Juuso Välimäki for invaluable feedback and support. For helpful discussions, the author thanks Victor Augias, Sarah Auster, Yann Bramoulle, Francesc Dilmé, Théo Durandard, Alexis Ghersengorin, Zhengqing Gui, Nathan Hancart, Ryota Iijima, Ian Jewitt, Yonggyun Kim, Andreas Kleiner, Anton Kolotilin, Daniel Krähmer, Xiao Lin, Matti Liski, Qianjun Lyu, Thomas Mariotti, Benny Moldovanu, Mich\`ele Mueller-Itten, Axel Niemeyer, Lars Peter Østerdal, Justus Preusser, John Quah, Julia Salmi, Fedor Sandomirskiy, Anna Sanktjohanser, Ludvig Sinander, Alex Smolin, Peter Norman Sørensen, Mark Whitmeyer, Zizhe Xia, Kun Zhang, audiences at Oslo, Toulouse, the 3rd NET workshop, the 36th Stony Brook International Conference, the ESWC 2025, and the EEA 2025, and anonymous referees and audiences at the EC'25.
\end{funding}
\begin{abstract}

This paper introduces a novel ranking of statistical experiments, the linear-Blackwell (LB) order, which can equivalently be characterized by (i) the dispersion of the induced posterior and likelihood ratios in the sense of the linear convex order, (ii) the size of the Lorenz zonoid (the set of statewise expectation profiles), or (iii) the variability of the posterior mean. We apply the LB order to compare experiments in binary-action decision problems and in decision problems with quasi-concave payoffs. We also use it to compare experiments in moral hazard problems and in screening problems with ex post signals.
\end{abstract}


\end{frontmatter}

\section*{}

Statistical experiments capture information and uncertainty in mathematical models of economic problems. Formally, given a set of states $\Theta=\{\theta_0,\ldots,\theta_n\}$, an \emph{experiment} $F$ generates a random variable $X$ (which we call the signal) taking values in a space $\mathcal{X}$ (which we call the signal space). Conditional on the state being $\theta \in \Theta$, the signal $X$ is distributed according to $F(\cdot \mid \theta)$. We write $x$ for a generic realization of $X$. 


Blackwell’s celebrated theorem \citep[][]{blackwell1951comparison,blackwell1953equivalent} provides a ranking of experiments based on their informativeness. An experiment $F$ dominates an experiment $G$ in the \emph{Blackwell order} if $F$ induces posterior and likelihood ratios that are more dispersed than those of $G$ in the sense of the convex order.\footnote{This posterior-based characterization of the Blackwell order can be found in \citet{lehrer2008two} and \citet{denti2022experimental}.} Equivalently, $F$ dominates $G$ in the Blackwell order if $G$ is a \emph{garbling} of $F$---that is, if $G$ can be obtained by adding noise to $F$'s signal.

As is well known, the Blackwell order imposes excessively stringent requirements and often fails to apply in scenarios where intuition suggests that one experiment is more informative than the other. Furthermore, as emphasized by \citet{kim1995efficiency}, the Blackwell order is overly demanding in the principal–agent framework, since it is designed to compare experiments uniformly across all decision problems, whereas the agency model imposes additional structure on how information can be used. Finally, verifying the Blackwell order of two experiments is often computationally infeasible, especially when the experiments have infinite signal spaces.

This paper introduces a novel ranking of experiments, the \emph{linear-Blackwell (LB) order}, which has three equivalent characterizations: (\romannumeral 1) in terms of the dispersion of the induced posterior and likelihood ratios, in the sense of the linear convex order, (\romannumeral 2) in terms of the size of the Lorenz zonoid (the set of statewise expectation profiles), and (\romannumeral 3) in terms of the variability of the posterior mean. We take (i) as the definition of the LB order; (ii) and (iii) are equivalent reformulations.

Let us start by discussing the geometric characterization (ii). Fix an experiment $F$, and consider all Borel-measurable functions $h:\cX\to[0,1]$. Each such $h$ generates a statewise expectation profile $\bz_F(h) := \left( \mathbb{E}[h(X)\mid \theta_0], \ldots, \mathbb{E}[h(X)\mid \theta_n] \right)$, and these expectation profiles capture something about the informativeness of $F$. For example, in a moral hazard problem in which the state $\theta$ represents the agent's hidden action and the principal observes the signal $X$, the functions $h$ represent incentive schemes designed by the principal, which map the signal realizations to the agent’s utility. Note that $h$ shapes the agent’s incentives only through $\bz_F(h)$. When $F$ is informative, the principal can choose $h$ so that $\bz_F(h)$ has entries that vary substantially across states, thereby strengthening the agent’s incentives. By contrast, when $F$ is uninformative, every feasible $\bz_F(h)$ has nearly constant entries across states.

\begin{figure}
  \centering
   \begin{tabular}{@{}c@{\hspace{2cm}}c@{}}
    \subcaptionbox{$\cz(F)$}
      {\includegraphics[width=0.28\textwidth]{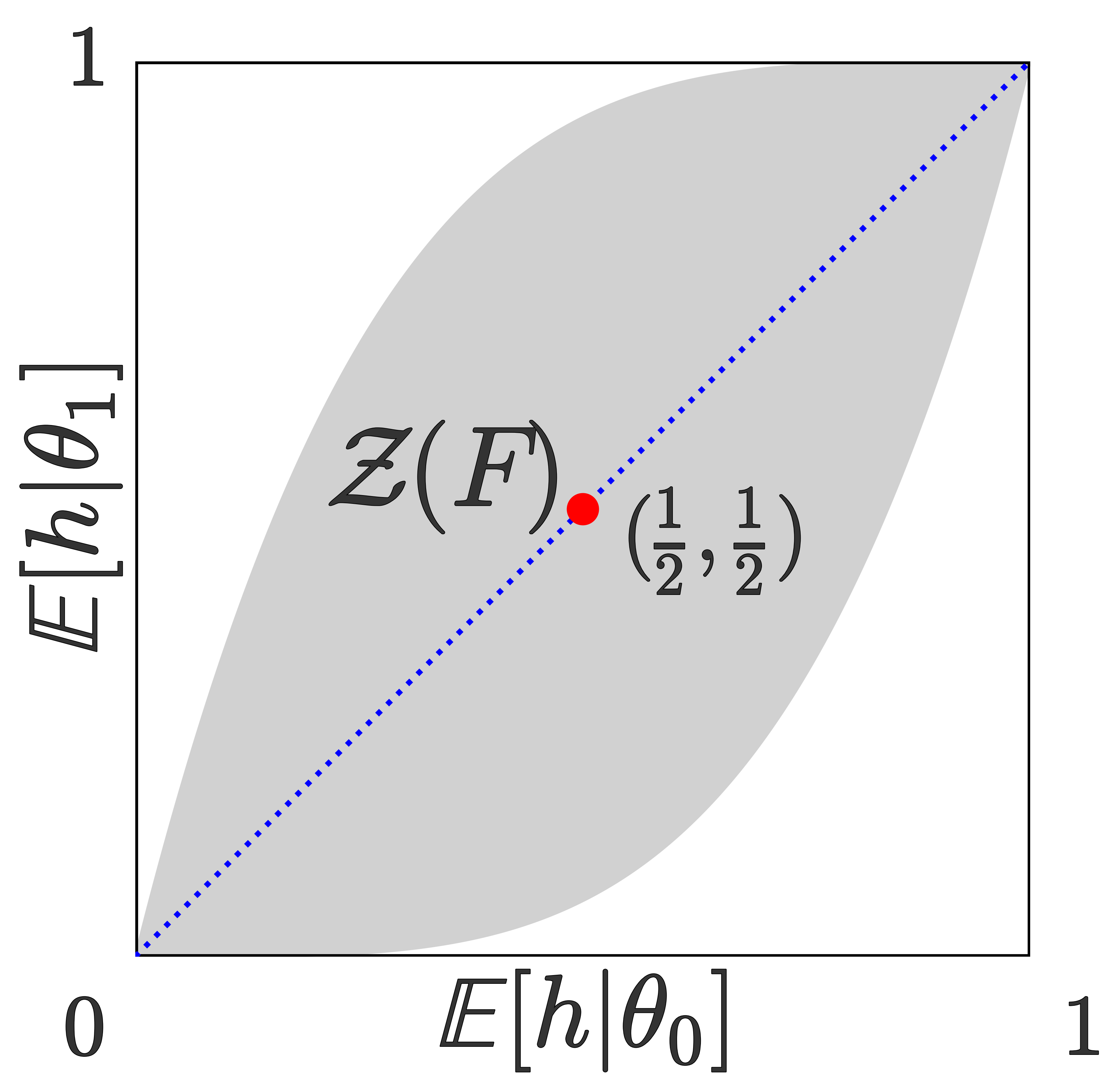}} &
    \subcaptionbox{$\cz(F)\supseteq \cz(G)$}
      {\includegraphics[width=0.28\textwidth]{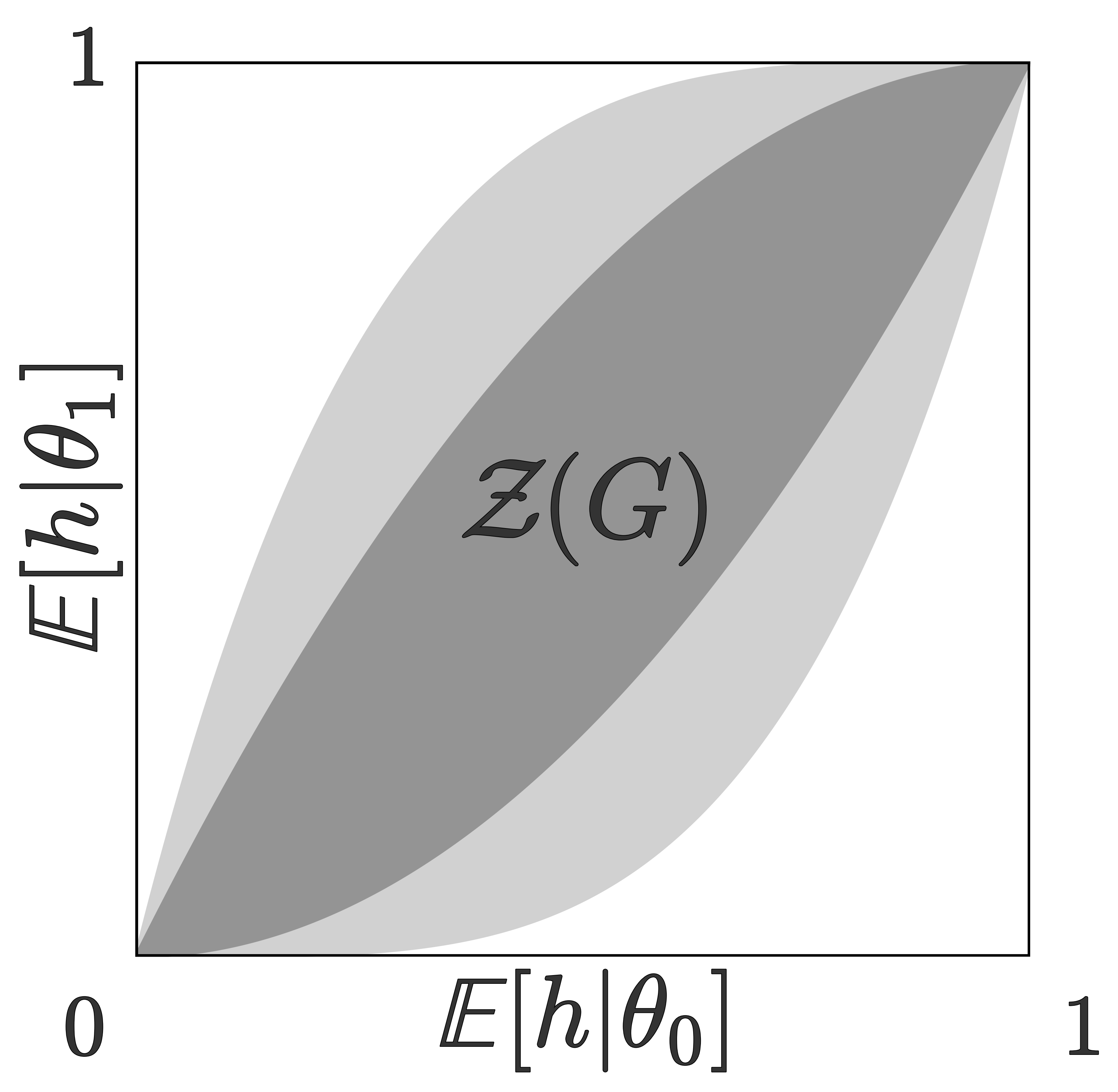}}
  \end{tabular}
  \caption{Left: the Lorenz zonoid $\cz(F)$ for an experiment $F$ with state space $\Theta=\{\theta_0,\theta_1\}$. Note that $\cz(F)$ is a subset of the unit square $[0,1]^{2}$ and contains the diagonal (the blue dotted line). Right: the Lorenz zonoids of $F$ and $G$ such that $F$ dominates $G$ in the LB order, which implies $\mathcal{Z}(F) \supseteq \mathcal{Z}(G)$.}
  \label{fig:1}
\end{figure}

Let $\cz(F):=\{\bz_F(h)\mid h\colon \cX\to[0,1]\}$ denote the set of all feasible statewise expectation profiles, termed the \emph{Lorenz zonoid} of $F$. The left panel of \cref{fig:1} shows an example Lorenz zonoid $\cz(F)$ for an experiment $F$ with state space $\Theta=\{\theta_0,\theta_1\}$. When $F$ is fully revealing, $\cz(F)$ is the entire unit cube. (In the example of \cref{fig:1}, this means the extreme profiles $(0,1)$ and $(1,0)$ are feasible.) When $F$ is completely uninformative (i.e., $F(\cdot\mid\theta)\equiv F(\cdot)$ for all $\theta$), $\cz(F)$ is simply the diagonal line segment $\{(t,\ldots,t) \mid t\in[0,1]\}$. Intuitively, a more informative experiment should have a larger Lorenz zonoid. This leads to our geometric characterization of the LB order: $F$ dominates $G$ in the LB order if and only if $\cz(F)$ contains $\cz(G)$, as shown in the right panel of \cref{fig:1}.

Note that the geometric characterization of the LB order requires \emph{expectation matching}: $F$ dominates $G$ if, for every $[0,1]$-valued function of $G$’s signal, there exists a $[0,1]$-valued function of $F$’s signal with the same conditional expectation in each state. In contrast, the garbling condition for the Blackwell order requires more stringent \emph{distribution matching}: $F$ dominates $G$ if, for every $[0,1]$-valued random variable derived from $G$’s signal, there exists a $[0,1]$-valued random variable derived from $F$’s signal with the same conditional distribution in each state.

Next we discuss the characterization (i) of the LB order, which is based on the dispersion of the posterior and likelihood ratios induced by an experiment. Fix an arbitrary prior on $\Theta$ with full support. By Bayes' rule, each signal realization $x\in\cX$ from an experiment $F$ induces a posterior vector $\bp_F(x)=(p_{F,1}(x),\ldots,p_{F,n}(x))$, where $p_{F,i}(x)$ denotes the probability of state $\theta_i$, and the probability of $\theta_0$ is determined residually. The experiment $F$ and the prior jointly generate the unconditional distribution of the random variable (signal) $X$ and thus a random posterior vector $\bp_F(X)$.

The characterization (i) states that $F$ dominates $G$ in the LB order if and only if the random vector $\bp_F(X)$ is more dispersed than the random vector\footnote{The experiment $G$ generates the random variable (signal) $Y$ that induces the random posterior vector $\bp_G(Y)$.} $\bp_G(Y)$ in the sense of the \emph{linear convex order}. The latter condition means that for each $\bb=(b_1,\ldots,b_n)\in\mathbb{R}^n$, the scalar $\boldsymbol{b}\cdot\bp_F(X)$ is a mean-preserving spread of $\boldsymbol{b}\cdot\bp_G(Y)$; that is,  $\mathbb{E}\{C[\sum_{i=1}^n b_i p_{F,i}(X)]\}\geq \mathbb{E}\{C[\sum_{i=1}^n b_i p_{G,i}(Y)]\}$ for every univariate convex function $C:\mathbb{R}\to\mathbb{R}$. In parallel, as every experiment can be represented as an $n$-dimensional random vector of likelihood ratios of $\theta_1,\ldots,\theta_n$ relative to $\theta_0$, we also show that $F$ dominates $G$ in the LB order if and only if the likelihood-ratio vector of $F$ dominates that of $G$ in the linear convex order. Finally, drawing on the geometric representation of the linear convex order provided by \citet{koshevoy1996lorenz,koshevoy1997zonoid}, we establish an equivalence between the inclusion of Lorenz zonoids described in the characterization (ii) and the linear convex order on posterior and likelihood-ratio vectors.

The characterization (i) enables a direct comparison between the LB order and the Blackwell order. Recall that $F$ dominates $G$ in the Blackwell order if and only if the random vector $\bp_F(X)$ is more dispersed than the random vector $\bp_G(Y)$ in the sense of the \emph{convex order}, that is, $\mathbb{E}\{C[\bp_F(X)]\}\geq \mathbb{E}\{C[\bp_G(Y)]\}$ for every convex function $C:\mathbb{R}^n\to\mathbb{R}$. It follows that when $n=1$ (so that $\Theta=\{\theta_0,\theta_1\}$), the Blackwell order and the LB order are equivalent. This is because the convex order and the linear convex order coincide in this case. Furthermore, when $n=1$ the Blackwell order is relatively easy to verify, because the set of univariate convex functions admits a simple characterization. However, when $n \geq 2$, the Blackwell order is far more complex, as noted by \citet{blackwell1954theory}: in general, it is computationally infeasible to check whether one experiment dominates another, and many pairs of experiments are not even comparable. This is because the set of multivariate convex functions has no tractable characterization (see, e.g., \citealp{johansen1972representation,johansen1974extremal}); thus, the convex order on multidimensional random vectors is difficult to verify. In contrast, the LB order is straightforward to work with for all $n$, since it relies on the linear convex order, which involves only univariate convex functions. In particular, the LB order of two experiments is fairly easy to check either analytically or through well-established numerical methods.

We now come to the characterization (iii) of the LB order, which captures an idea already widespread in the literature---namely, that an experiment is more informative if it induces greater variability in the posterior mean (see, e.g., \citealp{ganuza2010signal}, \citealp{gentzkow2016rothschild}, and \citealp{ravid2022learning}). Typically, however, the literature assumes that the state space is a fixed subset of $\mathbb{R}$. In contrast, our characterization applies for any assignment of statewise valuations: we show that $F$ dominates $G$ in the LB order if and only if, for every function $\phi:\Theta\to\mathbb{R}$, the posterior expectation $\mathbb{E}[\phi(\tilde{\theta})\mid X]$ under $F$ is a mean-preserving spread of that under $G$. Intuitively, varying $\phi$ should not alter the ranking of two experiments, since it does not alter the information they reveal. This characterization is equivalent to the characterization (i) because the posterior expectation is a linear combination of posterior beliefs, and it is equivalent to (ii) because the posterior expectation is dual to the statewise expectation profiles.

\subsection*{Economic Applications}

We present three economic applications of the LB order. First we consider the following decision problem: A decision-maker (DM) selects an action $a\in \cA$ and receives payoff $u(a,\theta)$ in state $\theta$. She is uncertain about the state and obtains information about it via a signal from an experiment. She then updates her belief using Bayes' rule and selects the action that maximizes her expected payoff. Blackwell's theorem \citep{blackwell1951comparison,blackwell1953equivalent} shows that an experiment $F$ dominates an experiment $G$ in the Blackwell order if and only if, in every decision problem, the DM's ex ante expected payoff is higher under $F$ than under $G$.

We show that $F$ dominates $G$ in the LB order if and only if, in every decision problem with $|\cA|=2$ (i.e., with a binary action space), the DM's ex ante expected payoff is higher under $F$ than under $G$. This result extends to all decision problems such that (\romannumeral 1) $\cA\subseteq \mathbb{R}$, and (\romannumeral 2) the DM’s expected payoff is quasi-concave in $a$ under each belief. The condition (\romannumeral 2) is equivalent to the standard assumption that the set of optimal actions is convex under each belief. \Citet{kolotilin2025persuasion} study a subset of these decision problems, in which the DM's expected payoff is strictly unimodal in $a$ under each belief.

Next we consider a moral hazard problem, that is, a hidden-action principal--agent model. We use the state-space formulation of \citet{grossman1983analysis}, in which the state variable $\theta$ denotes the agent’s action. More precisely, the agent privately selects an action $\theta\in\Theta$ and incurs a cost for doing so. Following \citet{hart1986theory}, we allow him to select a mixed action $\bdl\in\Delta(\Theta)$, which induces any signal distribution in the convex hull of the distributions conditional on each $\theta$. The principal observes a signal from an experiment, with the signal distribution determined by the agent's action. She commits to an incentive scheme mapping signal realizations to payments. Her aim is for the agent to select a certain target action. The agent's payoff is additively separable in his utility from the payment and the cost of the selected action. The principal incurs a certain disutility from the payment. Both parties may exhibit arbitrary risk attitudes.

We show that in this problem, the LB order captures both the breadth and the depth of the principal's control: dominance in the LB order corresponds to both a wider range of implementable target actions and a smaller minimal disutility to the principal from implementing any target action. Our results align with the seminal insight from \citet{holmstrom1979moral} and \citet{kim1995efficiency} that when the likelihood ratio is more dispersed, the principal is better able to deter deviations by the agent. Notably, we extend that insight to a more general setting without imposing the stringent conditions required for the validity of the first-order approach (FOA) \citep[see, e.g.,][]{rogerson1985first}.

Finally, we consider a screening problem, that is, a hidden-type principal–agent model. The agent holds private information about the state $\theta$, represented as a belief vector; we view this information as his type. The agent reports his type to the principal, who then selects an alternative $a \in \cA$. After the alternative is implemented, the principal observes an ex post signal correlated with the state (which is generated by an experiment) and then makes a transfer to the agent. Applications of this model include security design with contingent payments \citep[][]{hansen1985auctions,demarzo2005bidding,ekmekci2016just} and mechanism design with ex post auditing \citep[][]{townsend1979optimal,laffont1986using}. The principal commits to a mechanism consisting of an allocation rule and a transfer rule. The principal and the agent have conflicting interests, yet both possess additively separable utility functions: one component depends on the state and the alternative, and the other on the transfer.

To date, problems involving the informativeness of ex post signals have received limited attention. The existing literature (see, e.g., \citealp{cremer1988full}, \citealp{riordan1988optimal}, and \citealp{mcafee1992correlated}) typically assumes that the principal may impose arbitrarily large penalties on the agent. Under this assumption, the principal can generally implement any allocation rule and fully extract the agent's surplus. In such a setting, the informativeness of the ex post signal is irrelevant. In our model, by contrast, the principal cannot impose arbitrarily large penalties---that is, the agent has limited liability---so the informativeness of the ex post signal affects the principal's payoff. In particular, given experiments $F$ and $G$ generating ex post signals, we show that $F$ dominates $G$ in the LB order if and only if, for every pair of principal and agent utility functions, the principal's expected utility from her optimal mechanism is higher under $F$ than under $G$.

\subsection*{Related Literature}

This paper relates to the literature on ordinal comparisons of experiments (see, e.g., \citealp{blackwell1951comparison}, \citealp{blackwell1953equivalent}, \citealp{moscarini2002law}, \citealp{ganuza2010signal}, \citealp{cabrales2013entropy}, \citealp{mu2021blackwell}, \citealp{brooks2024comparisons}, \citealp{frick2024welfare}, and \citealp{whitmeyer2024blackwell}). The most closely related work is that of \citet{bertschinger2014blackwell}, who compare two experiments with finite signal spaces by examining the zonotopes generated by their likelihood vectors. Our geometric characterization of the LB order permits a similar comparison for a broader class of experiments: we interpret the order of two experiments in terms of their Lorenz zonoids (note that zonoids are a generalization of zonotopes). Furthermore, we establish that this interpretation is equivalent to interpretations in terms of the posterior and likelihood-ratio vectors, as well as the posterior means.

\citet{lehmann1988comparing} refines the Blackwell order by restricting to monotone decision problems; for such problems, he introduces a ranking of experiments satisfying the monotone likelihood ratio property (MLRP). The Lehmann order is further developed by \citet{persico2000information,jewitt2007information,quah2009comparative,athey2018value,di2021strategic,kim2023comparing}. However, MLRP imposes a highly restrictive requirement on the correlation between the signal and the state, which many natural distributions fail to satisfy (see \citealp{casella1987reconciling}, \citealp{rauh2005experimentation}, and \citealp{cheynel2020public}). Furthermore, MLRP lacks a solid microfoundation: as noted by \citet{mensch2021rational}, a rationally inattentive agent need not acquire MLRP-ordered signals in decision problems satisfying the single-crossing property. In the supplementary material for this paper, we introduce an extension of the Lehmann order called the \emph{monotone-posterior-expectation (MPE) order}. The MPE order coincides with the Lehmann order for experiments satisfying MLRP but also applies when MLRP does not hold. Furthermore, MLRP, the Lehmann order, and the MPE order are all defined with respect to an ordinal ranking of the state space. We show that $F$ dominates $G$ in the LB order if and only if $F$ dominates $G$ in the MPE order for every ordinal ranking of the state space. 

Our work also contributes to the literature on informativeness criteria in moral hazard problems (see, e.g., \citealp{holmstrom1979moral}, \citealp{gjesdal1982information}, \citealp{kim1995efficiency}, \citealp{jewitt1997information}, \citealp{jewitt2007information}, \citealp{dewatripont1999economics}, \citealp{fagart2007ranking}, \citealp{xie2017information}, and \citealp{frick2023monitoring}). We establish a new informativeness criterion based on the set of implementable actions and show that it is equivalent to the established criterion based on the principal's minimal disutility from implementing her target action. In addition, most of the existing literature assumes that the agent's action is a continuous, one-dimensional effort variable and that the FOA is valid. However, it is well known that the FOA is generally not valid. In particular, \citet{hart1986theory} emphasize that the FOA is valid primarily when the signal distribution conditional on each effort is a convex combination of two fixed distributions (the spanning condition), or equivalently, when the problem features a binary action space with randomization. \citet{jewitt2007information} studies informativeness criteria within the moral hazard framework of \citet{grossman1983analysis}, which allows for an arbitrary number of actions. We strengthen Jewitt's result \citep[Proposition 12 in][]{jewitt2007information} in two respects. First, we consider a more general framework allowing mixed actions, so any distribution in the convex hull of the set of distributions conditional on pure actions can be reached. Second, and more importantly, Jewitt assumes that strong duality holds under each experiment; however, this assumption fails in certain important situations (see \citealp{moroni2014existence}, and \citealp{chi2023dual}). In contrast, under our bounded-payment and convexity assumptions, we establish equality of primal and dual values even when strict feasibility fails. In a subsequent working paper, \citet{xia2025comparisons} proposes an additional geometric perspective on our result concerning the principal’s minimal disutility in the case of a risk-neutral principal and a risk-averse agent. \Citeauthor{xia2025comparisons} also establishes alternative rankings of experiments that are applicable when both parties are risk-neutral.

Finally, \citet{kim2023comparing} and \citet{asseyer2025information} study informativeness criteria for screening problems; however, they focus on ex ante signals (signals observed before the principal commits to a mechanism). In contrast, we study screening problems with ex post signals.

\section{Preliminaries}
\label{sec:2}

\subsection{Statistical Experiments}
\label{sec:2.1}

Let $\Theta=\{\theta_0,\ldots,\theta_n\}\subset \mathbb{R}$ be the set of states.\footnote{Appendix K of the supplementary material examines the case of a continuous state space.} A Bayesian decision-maker (DM) possesses a prior belief $\bq=(q_1,\ldots,q_n)\in\td_n$, where $\td_n:=\left\{\bq\in[0,1]^n\mid \sum_{i=1}^{n}q_i\leq 1\right\}$. She assigns probability $q_i$ to state $\theta_i$ for $i=1,\ldots,n$. To $\theta_0$ she assigns probability $q_0:=1-\sum_{i=1}^n q_i$; this allows $q_1,\ldots,q_n$ to vary freely. A prior $\bq$ has \emph{full support} if $\bq$ lies in $\operatorname{int} (\td_n)$, that is, the interior of $\td_n$.

A \textit{statistical experiment} generates a random variable and specifies its conditional distribution given each state $\theta\in\Theta$. More precisely, we denote by $F$ a statistical experiment which generates a random variable $X$ taking values in $\cX:=[\ux,\ox]$. Conditional on the state being $\theta \in \Theta$, the random variable $X$ is distributed according to $F(\cdot \mid \theta)$. We refer to $X$ as the signal and $\mathcal{X}$ as the signal space, and write $x$ for a generic realization of $X$. Similarly, we denote by $G$ an experiment which generates a random variable (signal) $Y$ taking values in $\cY:=[\uy,\oy]$, for which it specifies a family of conditional distributions $\left\{G(\cdot\mid \theta_0),\ldots,G(\cdot\mid \theta_n)\right\}$; here we let $y$ denote a generic realization of $Y$. For each $\theta\in\Theta$, both $F(\cdot\mid \theta)$ and $G(\cdot\mid \theta)$ are absolutely continuous with densities $f(\cdot\mid \theta)$ and $g(\cdot\mid \theta)$.\footnote{The case in which the signal spaces $\cX$ and $\cY$ are finite can be embedded into the continuous framework; see \citet{kim2023comparing}. All our results extend to the more general case in which $\cX$ and $\cY$ are Polish spaces.}


Given the prior $\bq$ and experiment $F$, let $F_{\bq}$ and $f_{\bq}$ respectively denote the unconditional distribution and density of $X$. For each signal realization $x\in\cX$, let $ \bp_F(x;\bq)=\left(p_{F,1}(x;\bq),\ldots,p_{F,n}(x;\bq)\right)$, where $p_{F,i}(x;\bq):=q_i\cdot f(x\mid \theta_i)/f_{\bq}(x)$, denote the DM's posterior belief. Furthermore, for each $x$ and $\phi:\Theta\to\mathbb{R}$, let $e_F(x;\phi,\bq):=\sum_{i=0}^{N}\mathbb{P}(\theta_i\mid X=x;\bq,F)\phi(\theta_i)$ denote the posterior expectation of $\phi$. The random variable $X$ distributed according to $F_{\bq}$ induces the random vector $\bp_{F} (X;\boldsymbol{q})$ and the random variable $e_F(X;\phi,\bq)$. Throughout the paper, we suppress $X$ from the notation for these and write them as $\bp_{F}(\bq)$ and $e_F(\phi,\bq)$.

If $f(x\mid\theta_0)>0$ for all $x \in \cX$, then for every $i$, the likelihood ratio $l_{F,i}(x):=f(x\mid\theta_i)/f(x\mid\theta_0)$ is well-defined everywhere. For each $x\in \cX$, let $\bl_F(x)=\left(l_{F,1}(x),\ldots,l_{F,n}(x)\right)$ denote the likelihood-ratio vector. Let the random variable $X$ be distributed according to $F(\cdot\mid \theta_0)$. It induces the random vector $\bl_F(X)$. We suppress $X$ from the notation for this vector and write it as $\bl_{F}$.


\subsection{Convex Order and Linear Convex Order}
\label{sec:2.2}

Consider two real-valued random vectors $\bmu=(\mu_1,\ldots,\mu_n)$ and $\bnu=(\nu_1,\ldots,\nu_n)$ such that $\mathbb{E}[\mu_i]=\mathbb{E}[\nu_i]$ for each $i$.

\begin{definition}
    \label{def:1}
    The random vector $\bmu$ dominates $\bnu$ in the \textit{convex order}, and we write $\bmu \succeq_{\textup{cx}} \bnu$, if $\mathbb{E}[C(\bmu)]\geq \mathbb{E}[C(\bnu)]$ for each convex function $C:\mathbb{R}^n\to \mathbb{R}$.
\end{definition}

When $n=1$, verifying the convex order \footnote{The convex order on univariate random variables is equivalent to majorization; see \citet{kleiner2021extreme} for further details.} is straightforward, because the set of univariate convex functions admits a simple characterization as a one-parameter family of extremal rays;\footnote{The restriction of any convex function $C:\mathbb{R}\to\mathbb{R}$ to a compact interval $[a,b]$ can be written in the form $C(x)=\alpha+\beta x+\int_a^b(x-t)_+\,\pi(dt)$ for some $\alpha,\beta\in\mathbb{R}$ and some finite nonnegative Borel measure $\pi$. In particular, after we remove the affine part, the extremal rays are generated by the hinges $(x-t)_+$, $t\in(a,b)$. On $\mathbb{R}$, the analogous representation, with the integral over $\mathbb{R}$, holds when both limiting slopes are finite and $C(x)-\beta x$ has a finite limit as $x\to-\infty$, where $\beta=\lim_{x\to-\infty}C'_+(x)$.} consequently, the random variable $\mu$ dominates $\nu$ in the convex order if and only if
\begin{equation}
    \label{eqn:1}
        \mathbb{E}\left[\left(\mu-t\right)_+\right]\geq \mathbb{E}\left[\left(\nu-t\right)_+\right]\quad\forall\, t\in\mathbb{R},
\end{equation}
where $\left(s\right)_+:=\max(s,0)$. 

When $n>1$, verifying the convex order is much harder, because the set of multivariate convex functions lacks a practical representation; conditions like \eqref{eqn:1} are no longer available. Hence one must check the inequalities in \cref{def:1} across an infinite-dimensional function space, or solve complex martingale optimal transport problems, which in general is computationally challenging. 

There is therefore a growing literature (see, e.g., \citealp{marshall1979inequalities}, \citealp{bhandari1988multivariate}, \citealp{joe1992multivariate}, and \citealp{scarsini1998multivariate}) examining an alternative order on random vectors:

\begin{definition}
    \label{def:2}
    The random vector $\bmu$ dominates $\bnu$ in the \textit{linear convex order}, and we write $\bmu \succeq_{\textup{lcx}} \bnu$, if $\bb\cdot\boldsymbol{\mu} \succeq_{\textup{cx}} \bb\cdot\boldsymbol{\nu}$ for all $\bb\in\mathbb{R}^n$.
    \end{definition}

More explicitly, $\bmu \succeq_{\textup{lcx}} \bnu$ if $\mathbb{E}[C(\sum_{i=1}^nb_i\mu_i)]\geq \mathbb{E}[C(\sum_{i=1}^nb_i\nu_i)]$ for every $\bb=(b_1,\ldots,b_n) \in \mathbb{R}^n$ and every univariate convex function $C:\mathbb{R}\to \mathbb{R}$. Since $C(\sum_{i=1}^nb_i\mu_i)$ is convex in $\bmu$, the convex order implies the linear convex order, but the converse need not hold.

Because the linear convex order involves only ``univariate'' comparisons, unlike the convex order, it admits a straightforward characterization for all $n$: the condition \eqref{eqn:1} implies that $\bmu \succeq_{\textup{lcx}} \bnu$ if and only if
\begin{equation}
    \label{eqn:2}
    \mathbb{E}\left[\left(\sum_{i=1}^nb_i\mu_i-t\right)_{+}\right]\geq \mathbb{E}\left[\left(\sum_{i=1}^nb_i\nu_i-t\right)_{+}\right]\quad\forall\, \bb\in\mathbb{R}^n,\ t\in \mathbb{R}.
\end{equation}

\section{Linear-Blackwell Order}
\label{sec:3}

\subsection{Definition}
\label{sec:3.1}

We introduce a new ranking of experiments based on the linear convex order:
\begin{definition}
\label{def:3}
    The experiment $F$ dominates the experiment $G$ in the \textit{linear-Blackwell (LB) order}, and we write $F\succeq_{\textup{LB}}G$, if $\bp_{F}(\bq)\succeq_{\textup{lcx}}\bp_{G}(\bq)$ for each $\bq\in\td_n$.
\end{definition}

Note that to establish that $F\succeq_{\textup{LB}}G$, it suffices to verify that $\bp_{F}(\bq)\succeq_{\textup{lcx}}\bp_{G}(\bq)$ for any one (arbitrary) $\bq\in\operatorname{int}(\td_n)$. Furthermore, \eqref{eqn:2} implies the following condition:
\begin{lemma}
    \label{lem:1}
    We have $F\succeq_{\textup{LB}}G$ if and only if, for every $\bb=(b_0,\ldots,b_n)\in\mr^{n+1},$
    \begin{equation}
    \label{eqn:3}
        \int_{\cX}\left(\sum_{i=0}^{n} b_i f(x\mid\theta_i)\right)_{+}\, dx
    \geq
    \int_{\cY}\left(\sum_{i=0}^{n} b_i g(y\mid\theta_i)\right)_{+}\, dy.
    \end{equation}
\end{lemma}
Verifying the LB order of two experiments entails checking \eqref{eqn:3} over a finite-dimensional space of parameters $\bb$. In the supplementary material, we provide computationally manageable numerical methods for doing so.

\subsection{Characterizations via Likelihood Ratio and Posterior Expectation}

\Cref{prop:1,prop:2} provide two additional characterizations of the LB order.
\begin{proposition}
    \label{prop:1}
    Assume that $f(x\mid \theta_0)>0$ for all $x\in \cX$. Then
    \begin{equation*}
        F\succeq_{\textup{LB}}G\iff \bl_F\succeq_{\textup{lcx}}\bl_G.
    \end{equation*}
\end{proposition}

By \cref{lem:1}, choosing $\theta_0$ as the baseline state entails no loss of generality. For any $r\in\{1,\ldots,n\}$, if we instead form the posterior vector by omitting $\theta_r$ and define the likelihood-ratio vector relative to $\theta_r$, then $F$ dominates $G$ in the LB order (according to \cref{def:3}) if and only if the corresponding posterior vector and likelihood-ratio vector associated with $F$ dominate those associated with $G$ in the linear convex order.

\begin{proposition}
    \label{prop:2} We have
    $F\succeq_{\textup{LB}}G$ if and only if $e_{F}(\phi,\bq)\succeq_{\textup{cx}}e_{G}(\phi,\bq)$ for every $\bq\in\td_n$ and $\phi:\Theta\to\mathbb{R}$.
\end{proposition}

As before, we observe that to establish that $F\succeq_{\textup{LB}}G$, it suffices to check that $e_{F}(\phi,\bq)\succeq_{\textup{cx}}e_{G}(\phi,\bq)$ for just one $\bq\in\operatorname{int}(\td_n)$. 

\cref{prop:2} captures the widespread idea that an experiment is more informative if it induces greater variability in posterior means (see, e.g., \citealp{ganuza2010signal}, \citealp{gentzkow2016rothschild}, and \citealp{ravid2022learning}). Note that the literature typically assumes that the state space is a fixed subset of $\mathbb{R}$ (i.e., $\phi$ is fixed), whereas the LB order reflects informativeness independent of the choice of $\phi$. Intuitively, varying $\phi$ should not alter the ranking of two experiments, since it does not alter the information they reveal.

\subsection{A Geometric Characterization}
\label{sec:3.2}

In this section we establish a geometric characterization of the LB order, drawing on the geometric interpretation of the linear convex order given by \citet{koshevoy1996lorenz}. Here the informativeness of an experiment is captured by the set of statewise expectation profiles associated with it. Given an experiment $F$, define $\mathcal{H}_F:=\{h:\cX\to[0,1] \mid h\text{ is Borel-measurable}\}$. For each $h\in\mathcal{H}_F$, let
\begin{equation*}
        \bz_F(h):=\left(\int_{\cX}h(x)dF(x\mid \theta_0),\ldots,\int_{\cX}h(x)dF(x\mid \theta_n)\right)
\end{equation*}
denote the vector of expectations of $h$ in each state. The \emph{Lorenz zonoid}\footnote{As introduced by \citet{koshevoy1996lorenz}, the \emph{lift Lorenz zonoid} of a random vector is a geometric representation of its variability. The lift Lorenz zonoid is a multivariate generalization of the classical Lorenz curve, which represents the variability of a univariate random variable. In our paper, the set \( \cz(F) \) coincides with the lift Lorenz zonoid of \( \bl_F \), the random vector of likelihood ratios associated with $F$. Hence, we refer to \( \cz(F) \) as the Lorenz zonoid of the experiment \( F \).
} of $F$ is the set 
\begin{equation*}
    \cz(F):=\{\bz_F(h)\mid h\in\mathcal{H}_F\},
\end{equation*}
that is, the set of expectation profiles obtained from all $h\in\mathcal {H}_F$. This set is a subset of the unit cube $[0,1]^{n+1}$; it is compact, convex, and centrally symmetric with respect to $(\frac{1}{2},\ldots,\frac{1}{2})$.\footnote{If $F$ features a finite signal space $\cX$, then $\cz(F)$ is a zonotope generated by the likelihood vectors of the signal. Specifically, assume $\cX=\{x_0,\ldots,x_m\}$. For $i=0,\ldots,m$, let $P_i=\left(\mathbb{P}(x_i\mid\theta_0),\ldots,\mathbb{P}(x_i\mid\theta_n)\right)$. Then $\cz(F)=\left\{\sum_{j=0}^m b_j P_j \mid b_j \in [0,1]\quad\forall\, j \right\}$, the Minkowski sum of the line segments $[\bo,P_0],\ldots,[\bo,P_m]$. Zonoids are a generalization of zonotopes; they may have infinitely many generators and can be characterized as Hausdorff limits of zonotopes.\label{ft:3}} Moreover, it contains the diagonal line segment $\{(t,\ldots,t)\mid t\in[0,1]\}$. 

\begin{figure}
  \centering
  \begin{tabular}{@{}c@{\hspace{2cm}}c@{}}
    \subcaptionbox{$\cz(F)$}
      {\includegraphics[width=0.28\textwidth]{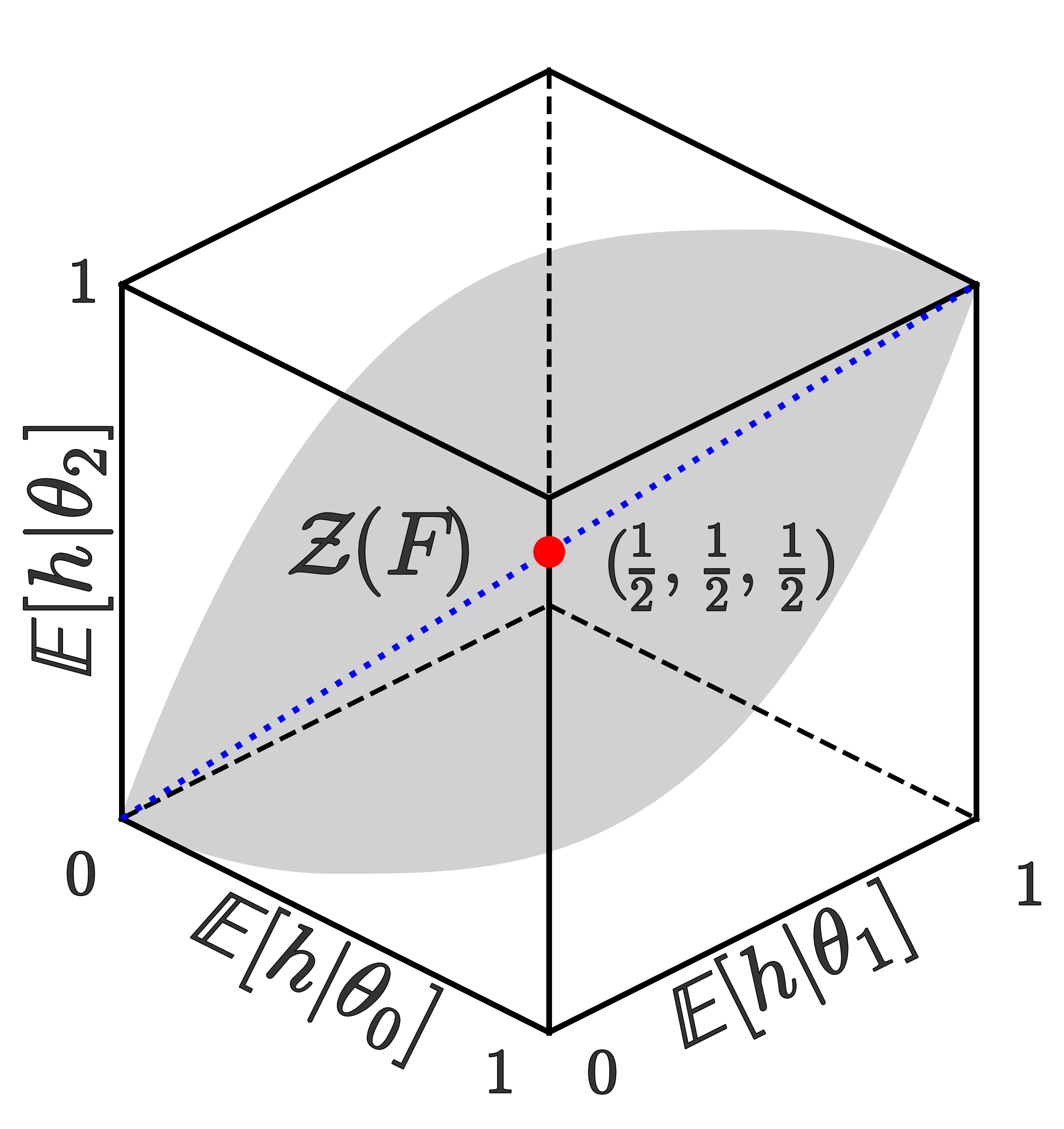}} &
    \subcaptionbox{$\cz(F)\supseteq \cz(G)$}
      {\includegraphics[width=0.28\textwidth]{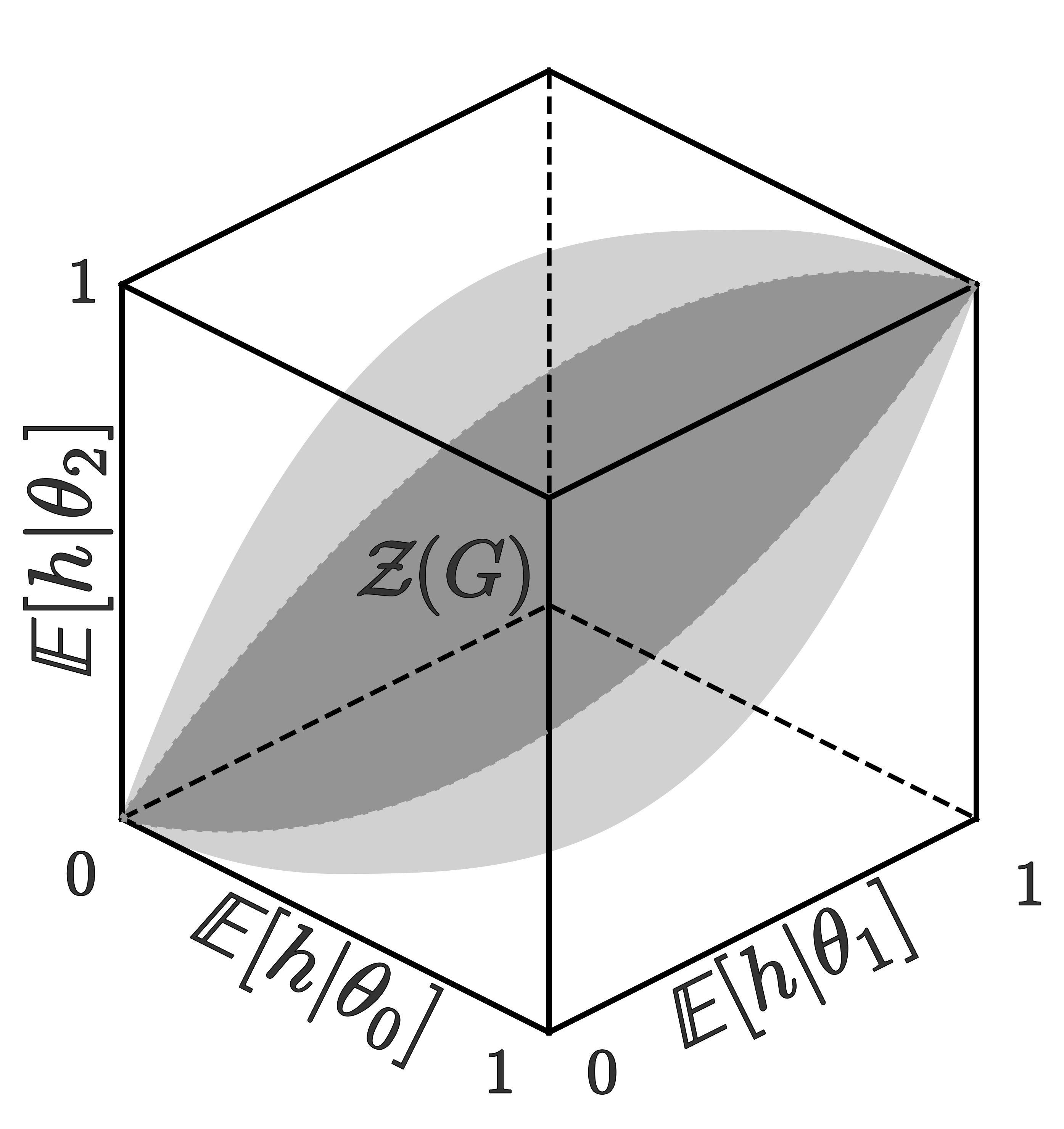}}
  \end{tabular}
  \caption{Left: the Lorenz zonoid $\cz(F)$ for an experiment $F$ with state space $\Theta=\{\theta_0,\theta_1,\theta_2\}$. Note that $\cz(F)$ is a subset of the unit cube $[0,1]^{3}$ and contains the diagonal (the blue dotted line). Right: the Lorenz zonoids of $F$ and $G$ with $F\succeq_{\textup{LB}}G$, which implies $\mathcal{Z}(F) \supseteq \mathcal{Z}(G)$.}
  \label{fig:2}
\end{figure}

The left panel of \cref{fig:2} shows an example Lorenz zonoid $\cz(F)$ for an experiment $F$ with state space $\{\theta_0,\theta_1,\theta_2\}$. When $F$ is completely uninformative (i.e., $F(\cdot\mid\theta)\equiv F(\cdot)$ for all $\theta$), $\cz(F)$ is simply the diagonal line segment. When $F$ is fully informative (i.e., it reveals the state), $\cz(F)$ is the entire unit cube. Intuitively, a more informative experiment should have a larger Lorenz zonoid. This intuition is illustrated in the right panel of \cref{fig:2} and made precise in \cref{prop:3}.

\begin{proposition}
    \label{prop:3}
    We have $F\succeq_{\textup{LB}}G$ if and only if $\cz(F)\supseteq \cz(G)$.
\end{proposition}

\begin{proof}
Since $\cz(F)$ and $\cz(G)$ are compact and convex,
$\cz(F)\supseteq \cz(G)$ if and only if, for every $\bb=(b_0,\ldots,b_n)\in\mr^{n+1}$,
their support functions satisfy
\begin{equation*}
\max_{\bz_F(h)\in \cz(F)} \bb\cdot \bz_F(h)
\geq
\max_{\bz_G(h)\in \cz(G)} \bb\cdot \bz_G(h).
\end{equation*}
By the definition of $\bz_F(h)$ and $\bz_G(h)$, this condition is equivalent to
\begin{equation}
\label{eqn:3.5}
\max_{h:\cX\to[0,1]} \sum_{i=0}^{n} b_i \int_{\cX} h(x) f(x\mid \theta_i)\,dx
\geq
\max_{h:\cY\to[0,1]} \sum_{i=0}^{n} b_i \int_{\cY} h(y) g(y\mid \theta_i)\,dy.
\end{equation}
The left-hand side of \eqref{eqn:3.5} can be computed as
\begin{equation*}
\begin{split}
\max_{h:\cX\to[0,1]} \sum_{i=0}^{n} b_i \int_{\cX} h(x) f(x\mid \theta_i)\,dx
&= \max_{h:\cX\to[0,1]} \int_{\cX} h(x)\Bigl[\sum_{i=0}^{n} b_i f(x\mid \theta_i)\Bigr]\,dx \\
&= \int_{\cX} \Bigl(\sum_{i=0}^{n} b_i f(x\mid \theta_i)\Bigr)_{+}\,dx,
\end{split}
\end{equation*}
where the last equality follows from choosing $h(x)=1$ whenever
$\sum_{i=0}^{n} b_i f(x\mid \theta_i)\ge 0$ and $h(x)=0$ otherwise.
An analogous expression holds for the right-hand side of \eqref{eqn:3.5}.
Hence \eqref{eqn:3.5} is equivalent to \eqref{eqn:3}.
\end{proof}

\cref{prop:3} plays an important role in applications. For example, in the binary-action decision problem of \cref{sec:4}, each $h\in\mathcal{H}_F$ corresponds to a strategy mapping signal realizations to probabilities of selecting a given action; in the principal--agent models of \cref{sec:5,sec:6}, each $h\in\mathcal{H}_F$ corresponds to an incentive scheme mapping signal realizations to the agent's utility. Thus, in the binary-action decision problem, $\mathcal{Z}(F)$ captures the set of statewise expected probability profiles generated by all strategies, and in the principal--agent models it captures the set of statewise expected utility profiles induced by all incentive schemes. In other words, $\mathcal{Z}(F)$ represents the extent to which $F$ enables the DM in the binary-action decision problem to correlate actions with states, or enables the principal in the principal--agent problem to provide incentives. Increasing informativeness in the sense of the LB order expands the Lorenz zonoid, strengthening the abilities of the DM or the principal.

\begin{remark}
In the definitions above, we may replace $[0,1]$ by any compact interval in $\mathbb{R}$, and $\mathcal{H}_F$ by the set of Borel-measurable functions from $\cX$ to that interval; all of our arguments can immediately be extended to this setting via normalization. However, if we replace $[0,1]$ by an unbounded interval (i.e., $\mr$, $(-\infty,b]$, or $[b,+\infty)$ for some $b \in \mr$), which allows $\mathcal{H}_F$ to include functions that can take very high or low values at signal realizations that are rare but induce extreme posterior beliefs, then the set of expectation profiles will be driven solely by those signal realizations rather than by the experiment as a whole. Consequently, a larger set of expectation profiles will no longer imply more dispersed posterior and likelihood ratios under the linear convex order. \Citet{azrieli2025elicitability}, \citet{kim2024weighted,kim2025asymptotic}, and \citet{xia2025comparisons} analyze cases which allow for unbounded intervals.
\end{remark}

\subsection{Comparison with Blackwell Order}
\label{sec:3.4}

The experiment $F$ dominates $G$ in the \textit{Blackwell order}, and we write $F\succeq_{\textup{B}}G$, if $\bp_{F}(\bq)\succeq_{\textup{cx}}\bp_{G}(\bq)$ for all $ \bq\in\td_n$. Equivalently, $F\succeq_{\textup{B}}G$ if and only if $G$ is a \emph{garbling} of $F$, that is, there exists a stochastic kernel $\kappa(y\mid x)$ satisfying
\begin{equation*}
    \int_{\cY}\kappa(y\mid x)\,dy=1 \quad \forall\, x\in \cX \quad \text{and} \quad g(y|\theta)=\int_{\cX}\kappa(y\mid x)f(x|\theta)\,dx\quad \forall\, y\in \cY,\ \theta\in\Theta.
\end{equation*}

The Blackwell order is less tractable than the LB order, since, as discussed earlier, the convex order is much harder to verify than the linear convex order. Checking whether one experiment dominates another in the Blackwell order requires checking the associated inequalities for all multivariate convex functions—which comprise an infinite-dimensional space—or solving complex martingale optimal transport problems, which may be computationally infeasible. In contrast, the LB order admits the characterization \eqref{eqn:3}, which can be verified analytically or using well-established numerical methods.

Furthermore, the Blackwell order requires \emph{distribution matching}: $G$ is a garbling of $F$ if and only if, for every randomization rule $H:\cY\to\Delta([0,1])$ that induces a random variable $\omega$ from $G$’s signal, there exists $H':\cX\to\Delta([0,1])$ inducing a random variable $\omega'$ from $F$’s signal such that $\omega\mid\theta \overset{d}{=} \omega'\mid\theta$ for all $\theta$---i.e., such that $\omega$ and $\omega'$ follow the same distribution in each state $\theta$. In contrast, by \cref{prop:3}, the LB order requires only \emph{expectation matching}; that is, it requires $\me[h(Y)\mid\theta]=\me[h'(X)\mid\theta]$ for all $\theta$, where $h(y):=\me_{H(y)}(\omega)$ and $h'(x):=\me_{H'(x)}(\omega')$.

When the state space is binary, the Blackwell order coincides with the LB order. When there are more than two states, the Blackwell order implies the LB order, but the converse need not hold. In the supplementary material, we present three further results relating the LB and Blackwell orders. First, we show that the equivalence of the two orders in the binary-state case extends as follows: $F\succeq_{\textup{LB}}G$ if and only if $F\succeq_{\textup{B}}G$ for every \emph{weighted dichotomy} on the state space. A weighted dichotomy is a partition of the state space into two subsets, together with an assignment of weights within each, according to which experiments on the original state space can be transformed into experiments over a binary state space. Second, we show that for experiments with finite signal space, the Blackwell order coincides with the LB order whenever the dominating experiment is irredundant, that is, its likelihood vectors are linearly independent. Finally, we show that two experiments are LB-equivalent (mutually dominating) if and only if they are Blackwell-equivalent.

\begin{table}
    \centering
    \setlength{\abovedisplayskip}{0pt}
    \setlength{\belowdisplayskip}{0pt}
    \begin{minipage}{0.45\textwidth}
        \centering
        \[
        \begin{blockarray}{c *{6}{c}}
            & x_0 & x_1 & \cdots & x_n & x_{n+1} \\
            \begin{block}{c[ *{6}{c} ]}
                \theta_0 & 1-\epsilon & 0 & \cdots & 0 & \epsilon \\
                \theta_1 & 0 & 1-\epsilon & \cdots & 0 & \epsilon \\
                \vdots   & \vdots & \vdots & \ddots & \vdots & \vdots \\
                \theta_n & 0 & 0 & \cdots & 1-\epsilon & \epsilon \\
            \end{block}
        \end{blockarray}
        \]
        \subcaption{$F(\epsilon)$}
    \end{minipage}
    \hfill
    \begin{minipage}{0.45\textwidth}
        \centering
        \[
        \begin{blockarray}{c *{5}{c}}
            & y_0 & y_1 & \cdots & y_n \\
            \begin{block}{c[ *{5}{c} ]}
                \theta_0 & 0 & 1/n & \cdots & 1/n \\
                \theta_1 & 1/n & 0 & \cdots & 1/n \\
                \vdots   & \vdots & \vdots & \ddots & \vdots \\
                \theta_n & 1/n & 1/n & \cdots & 0 \\
            \end{block}
        \end{blockarray}
        \]
        \subcaption{$\widehat{G}$}
    \end{minipage}
        \caption{}
        \label{tab:1}
\end{table}

Because the LB order is defined in a less restrictive way, it measures the informativeness of experiments in a broader range of scenarios than the Blackwell order encompasses. As an example, consider the experiments $F(\epsilon)$ and $\widehat{G}$ shown in \cref{tab:1}, for $n>1$.\footnote{This example builds on that in \citet{bertschinger2014blackwell}. The author thanks Xiao Lin for the suggestion.} The experiment $F(\epsilon)$ generates a state-revealing signal with probability $1-\epsilon$ and an uninformative signal with probability $\epsilon$. In experiment $\widehat{G}$, each signal realization rules out one state and provides no additional information. Intuitively, for small $\epsilon$, $F(\epsilon)$ should be more informative than $\widehat{G}$, since $F(\epsilon)$ becomes perfectly informative as $\epsilon\to 0$. However, $F(\epsilon)$ does not dominate $\widehat{G}$ in the Blackwell order for any $\epsilon>0$, because $\hat{G}$ cannot be a garbling of $F(\epsilon)$.\footnote{If it were, we would have $\mathbb{P}(y_i \mid x_{n+1}) = 0$ for each $i$, which contradicts the fact that $\sum_{i=0}^{n}\mathbb{P}(y_i \mid x_{n+1}) = 1$.} In contrast, $F(\epsilon)\succeq_{\textup{LB}}\widehat{G}$ for all $\epsilon\leq\frac{n-1}{n}$, since it can be shown that $\cz(\widehat{G})\subseteq\cz(F(\tfrac{n-1}{n}))$.

\section{Decisions Under Uncertainty}
\label{sec:4}

\subsection{Decision Problems}
\label{sec:4.1}

We now come to our first economic application of the LB order.

A \textit{decision problem} $\{\cA,u\}$ consists of (\romannumeral 1) a set of actions $\cA$, and (\romannumeral 2) a payoff function $u:\cA\times\Theta\to\mathbb{R}$ (where $\Theta \subset \mathbb{R}$ continues to denote a finite state space). We let $a$ denote a generic action in $\cA$. We assume $\cA$ is a compact subset of $\mathbb{R}^k$, and $u(a,\theta)$ is continuous in $a$ for each $\theta\in\Theta$. The DM selects an action $a\in \cA$ and receives payoff $u(a,\theta)$ when the state is $\theta$. She is uncertain about the state and possesses a prior $\bq$. Upon observing a signal from an experiment, she updates her belief via Bayes’ rule and selects the action that maximizes her expected payoff.

Given a belief $\bp$, we let $u(a,\bp):=\sum_{i=0}^{n}p_iu(a,\theta_i)$ denote the expected payoff of selecting action $a$. We let $\cA^*(\bp):=\operatorname*{arg\,max}_{a\in \cA}u(a,\bp)$ denote the set of optimal actions under $\bp$. Finally, we define the \textit{value function}—the maximal payoff the DM can receive—as $V(\bp):=\max_{a\in \cA}u(a,\bp)$. An experiment $F$ and a prior $\bq$ induce the unconditional distribution of the random variable $X$ and the random vector $\bp_{F}(\bq)$.  The DM's ex ante expected payoff is $\overline{V}_{F}(\bq):=\mathbb{E}[V(\bp_{F}(\bq))]$. 

Under the experiment $F$, the DM's strategy is a mapping $\sigma_{F}:\cX\to\Delta(\cA)$. With a slight abuse of notation, we let $\sigma_{F}(x,a)$ denote the probability (or the density) of the DM's selecting action $a$ upon observing signal realization $x\in\cX$. The expected payoff from the strategy $\sigma_F$ in state $\theta$ is $u(\sigma_F,\theta):=\int_{\cX}\int_{\cA}\sigma_{F}(x,a)u(a,\theta)f(x\mid\theta)\,da \,dx$.

The Blackwell order, as defined in \cref{sec:3.4}, captures the relative preferability of two experiments in terms of both the DM's ex ante expected payoff and her expected payoff from any given strategy (see \citealp{blackwell1951comparison,blackwell1953equivalent}).
\begin{theorem*}
    \label{bthm}
    The following statements are equivalent:
    \begin{enumerate}[label=\textup{(\roman*)}]
        \item $F\succeq_{\textup{B}}G$;
        \item $\overline{V}_{F}(\bq)\geq \overline{V}_{G}(\bq)$ for every decision problem $\{\cA,u\}$ and every $\bq\in\td_n$;
        \item for every decision problem $\{\cA,u\}$ and every strategy $\sigma_G$ under $G$, there exists a strategy $\sigma_F$ under $F$ with $u(\sigma_F,\theta)\geq u(\sigma_G,\theta)$ for each $\theta\in\Theta$.
    \end{enumerate}
\end{theorem*}

\subsection{Binary-Action Decision Problems}

For binary-action decision problems, where $\cA = \{a_0, a_1\}$, the LB order captures the relative preferability of two experiments in terms of both the DM's ex ante expected payoff and her expected payoff from any given strategy. \cref{lem:2} extends Proposition 9 in \citet{bertschinger2014blackwell} to general experiments.
\begin{lemma}
    \label{lem:2}
    The following statements are equivalent:
    \begin{enumerate}[label=\textup{(\roman*)}]
        \item $F\succeq_{\textup{LB}}G$;
        \item $\overline{V}_{F}(\bq)\geq \overline{V}_{G}(\bq)$ for every binary-action decision problem $\{\cA,u\}$ and every $\bq\in\td_n$;
        \item for every binary-action decision problem $\{\cA,u\}$ and every strategy $\sigma_G$ under $G$, there exists a strategy $\sigma_F$ under $F$ with $u(\sigma_F,\theta)\geq u(\sigma_G,\theta)$ for each $\theta\in\Theta$.
    \end{enumerate}
\end{lemma}

\begin{proof}
We prove \((\textup{i}) \Rightarrow (\textup{iii})\) and \((\textup{ii}) \Rightarrow (\textup{i})\) using the characterizations of the LB order established in \cref{sec:3}. The implication \((\textup{iii}) \Rightarrow (\textup{ii})\) is immediate.

\noindent
\textbf{Step 1:} \((\textup{i}) \Rightarrow (\textup{iii})\).  Each strategy \(\sigma_G\) induces a function \(h_{\sigma_G} : \cY \to [0,1]\) defined by $h_{\sigma_G}(y) = \sigma_G(y, a_0)$. If $F\succeq_{\textup{LB}}G$, then by \cref{prop:3} there exists $h\in\mathcal{H}_F$ such that 
\begin{equation}
\label{eqn:4}
    \bz_F(h) = \bz_G(h_{\sigma_G}).
\end{equation}
Construct a strategy $\sigma_F$ with $\sigma_F(x, a_0)=h(x)$. By \eqref{eqn:4}, $\sigma_F$ and $\sigma_G$ yield the same probability of selecting $a_0$ in each state.  Hence, $u(\sigma_F, \theta) = u(\sigma_G, \theta)$ for each $\theta$.

\noindent
\textbf{Step 2:} \((\textup{ii}) \Rightarrow (\textup{i})\).  
Assume for the sake of contradiction that \(F \nsucceq_{\textup{LB}} G\).  Then, by \eqref{eqn:3},
there exists \(\bb = (b_0, \ldots, b_n)\) such that 
\begin{equation}
\label{eqn:5}
    \int_{\cX}\left(\sum_{i=0}^{n} b_i f(x\mid\theta_i)\right)_{+} dx
    \;<\;
    \int_{\cY}\left(\sum_{i=0}^{n} b_i g(y\mid\theta_i)\right)_{+} dy.
\end{equation}
Take the uniform prior \(\bq = \left( \frac{1}{n+1}, \ldots, \frac{1}{n+1} \right)\), and consider $u$ such that $u(a_0, \theta_i) = 0$ and $u(a_1, \theta_i) = b_i$ for each $i$. Then it is optimal for the DM to select \(a_1\) upon observing \(x\) if $\sum_{i=0}^{n} u(a_1, \theta_i) f(x\mid\theta_i) > 0$. Thus, $\overline{V}_{F}(\bq) = \frac{1}{n+1} \int_{\cX} \left( \sum_{i=0}^{n} b_i f(x\mid\theta_i) \right)_{+}\, dx$. However, \eqref{eqn:5} implies \(\overline{V}_{F}(\bq) < \overline{V}_{G}(\bq)\), contradicting \((\textup{ii})\).  
\end{proof}

\subsection{Quasi-Concavity}

We now consider decision problems with $\cA\subset \mathbb{R}$, so that $\cA$ is totally ordered. For simplicity, we assume either (\romannumeral 1) $|\cA|<\infty$, or (\romannumeral 2) $\cA$ is a compact interval in $\mathbb{R}$ and $u(a,\theta)$ is differentiable as a function of $a$ for all $\theta$.

\begin{definition}
\label{def:4}
    A decision problem $\{\cA,u\}$ is \textit{quasi-concave (QCC)} if $u(a,\bp)$ is quasi-concave as a function of $a$ for each $\bp\in\td_n$.
\end{definition}

Note that $\cA\subset \mathbb{R}$ and the quasi-concavity of $u(a,\bp)$ in $a$ imply that $u(a,\bp)$ is unimodal in $a$. The left panel of \cref{fig:3} depicts the belief simplex for a QCC problem with $\cA=\{a_0,a_1,a_2\}$. A key feature of such a problem is that the indifference line between $a_0$ and $a_1$ does not intersect that between $a_1$ and $a_2$ at any $\bp\in\operatorname{int}(\td_2)$. The right panel of \cref{fig:3} depicts the belief simplex for a non-QCC problem. Here, the indifference line between $a_0$ and $a_1$ intersects that between $a_1$ and $a_2$ at $\bp^*\in\operatorname{int}(\td_2)$. For each $\bp'$ on the boundary between the dark grey area and the white area, it follows that $u(a_0,\bp')=u(a_2,\bp')>u(a_1,\bp')$, violating quasi-concavity.

\begin{figure}
    \centering
    \begin{tikzpicture}[scale=1]

    \begin{scope}[shift={(0,0)}] 
        \draw[->] (0,0) -- (4.5,0) node[right] {$p_2$}; 
        \draw[->] (0,0) -- (0,4.5) node[above] {$p_1$}; 

        \node[below left] at (0,0) {$0$}; 
        \node[left] at (0,4) {$1$};       
        \node[below] at (4,0) {$1$};      

        \fill[darkgray, opacity=0.7] (0,0) -- (0,2) -- (1.5,0) -- cycle; 
        \fill[lightgray, opacity=0.5] (0,2) -- (1.5,0) -- (2.5,0) -- (1.8,2.2) -- (0,4) -- cycle; 

        \draw (0,4) -- (4,0);       
        \draw (0,2) -- (1.5,0);     
        \draw (1.8,2.2) -- (2.5,0); 
    \end{scope}

    \begin{scope}[shift={(10,0)}] 
        \draw[->] (0,0) -- (4.5,0) node[right] {$p_2$}; 
        \draw[->] (0,0) -- (0,4.5) node[above] {$p_1$}; 

        \node[below left] at (0,0) {$0$};
        \node[left] at (0,4) {$1$};
        \node[below] at (4,0) {$1$};
        \node[right] at (1.5,1.2) {$\bp^*$};

        \filldraw[black] (1.5,1.2) circle (1pt);

        \fill[darkgray, opacity=0.7] (0,0) -- (0,2) -- (1.5,1.2) -- (2.6,0) -- cycle; 
        \fill[lightgray, opacity=0.5] (0,2) -- (0,4) -- (1,3) -- (1.5,1.2) -- cycle; 

        \draw (0,4) -- (4,0);
        \draw (0,2) -- (1.5,1.2);
        \draw (1,3) -- (1.5,1.2);
        \draw (1.5,1.2) -- (2.6,0);

        \filldraw[black] (2.05,0.6) circle (1pt) node[right,yshift=3pt] {$\bp'$};
    \end{scope}

\end{tikzpicture}
    \caption{QCC and non-QCC problems. Let $\Theta=\{\theta_0,\theta_1,\theta_2\}$. A decision problem with $\cA=\{a_0,a_1,a_2\}$ is characterized by a partition of the belief simplex $\td_2$ into three regions according to the DM's preferred actions. The left panel shows the belief simplex for a QCC problem, while the right panel shows that of a non-QCC problem. The dark grey, light grey, and white areas contain the beliefs under which the DM prefers $a_0$, $a_1$, and $a_2$, respectively.}
    \label{fig:3}
\end{figure}

The property of quasi-concavity plays a crucial role in many problems in economics---for example, in simplifying optimization problems to first-order conditions, modeling preferences for diversification, and justifying the median voter theorem (see \citealp{black1948rationale}, \citealp{jewitt1988justifying}, and \citealp{schmeidler1989subjective}). Quasi-concavity ensures that the set of optimal actions is well-behaved, which allows for consistent and tractable analysis. Specifically, if a decision problem is QCC, then the set of optimal actions $\cA^*(\bp)$ is convex relative to $\cA$ for each $\bp$.\footnote{That is, if $a,a'\in \cA^*(\bp)$ with $a<a'$, then $a''\in \cA^*(\bp) $ for each $a''\in(a,a')\cap \cA$.} The converse holds under mild conditions \citep[see][]{chen2025quasiconcavity}. Finally, a simple sufficient condition for QCC is that $u(\cdot,\theta)$ be concave on $A$ for every $\theta \in \Theta$, since nonnegative weighted sums preserve concavity.

\Citet{kolotilin2025persuasion} examine a special class of QCC problems in which $u(a,\bp)$ is strictly unimodal in $a$ for each $\bp$.\footnote{Appendix A of \citet{kolotilin2025persuasion} provides sufficient conditions for strict unimodality, building on \citet{quah2012aggregating} and \citet{choi2017ordinal}. These conditions can be adapted to yield sufficient conditions for a decision problem to be QCC.} This property ensures that the optimal action is always unique and satisfies the first-order condition (when $\cA$ is an interval). The assumption that the optimal action is unique for each $\bp$ appears frequently in the literature (see, e.g., \citealp{persico2000information}, \citealp{gentzkow2016competition}, and \citealp{farboodi2025good}), and it implies quasi-concavity, since a singleton set is convex.

Building on \cref{lem:2}, we show that the LB order provides an effective ranking of experiments in all QCC problems.

\begin{theorem}
    \label{thm:1}
    The following statements are equivalent:
    \begin{enumerate}[label=\textup{(\roman*)}]
        \item $F\succeq_{\textup{LB}}G$;
        \item $\overline{V}_{F}(\bq)\geq \overline{V}_{G}(\bq)$ for every QCC problem and every $\bq\in\td_n$;
        \item for every QCC problem and every strategy $\sigma_G$ under $G$, there exists a strategy $\sigma_F$ under $F$ with $u(\sigma_F,\theta)\geq u(\sigma_G,\theta)$ for each $\theta\in\Theta$.
    \end{enumerate}
\end{theorem}

We prove \((\textup{i}) \Rightarrow (\textup{ii})\) by decomposing an arbitrary QCC problem into binary-action subproblems and showing that its value function equals the sum of the subproblems' value functions. For illustration, consider the QCC problem $\{\cA,u\}$ shown in the left panel of \cref{fig:3}. Define the subproblems $(\cA',u')$ and $(\cA'',u'')$ as follows:
\begin{align*}
    \cA'&=\{a_0,a_1\}, \quad u'(a,\theta)=u(a,\theta)\ \ \forall\,\theta\in \Theta,\ a\in \cA', \\
    \cA''&=\{a_1,a_2\}, \quad u''(a_1,\theta)=0\ \ \text{and}\ \
    u''(a_2,\theta)=u(a_2,\theta)-u(a_1,\theta)\ \ \forall\,\theta\in \Theta.
\end{align*}
Let $V'$ and $V''$ denote the value functions of $(\cA',u')$ and $(\cA'',u'')$, respectively. It can be shown that $V(\bp)=V'(\bp)+V''(\bp)$ for each $\bp\in\td_2$. For example, for each $\bp$ in the dark grey area, since $u(a_0,\bp)>u(a_1,\bp)>u(a_2,\bp)$, we have $V'(\bp)=u'(a_0,\bp)=u(a_0,\bp)$, $V''(\bp)=u''(a_1,\bp)=0$, and $V(\bp)=u(a_0,\bp)$. 

The implication \((\textup{ii}) \Rightarrow (\textup{iii})\) follows from a separating hyperplane argument, while \((\textup{iii}) \Rightarrow (\textup{i})\) holds because all binary-action problems are QCC.

\cref{thm:1} extends to all decision problems in which payoffs are aggregated from a collection of binary-action subproblems.\footnote{Equivalently, up to an affine term, the value function belongs to the closed convex cone generated by the hinge functions \((\alpha\cdot \bp-t)_+\). That is, $V(\bp)=\ell(\bp)+\int (\alpha\cdot \bp-t)_+\,\mu(d\alpha,dt)$, for some affine function \(\ell\) and finite nonnegative measure \(\mu\). Each hinge is the non-affine part of a binary-action value function.}\footnote{In particular, \cref{thm:1} also extends to multidimensional action spaces when the problem is additively separable into QCC coordinate subproblems. Specifically, let $A=\prod_{j=1}^m A_j$, with $A_j\subset \mathbb{R}$, and suppose
$u(a,\theta)=\sum_{j=1}^m u_j(a_j,\theta)$ for each $a=(a_1,\ldots,a_m)$, where each coordinate problem $\{A_j,u_j\}$ is QCC. For a more detailed discussion, see Section~4.3 of the \href{https://arxiv.org/pdf/2502.06530v6}{earlier version of this paper}, which further examines the indifference set $\{\bp\in\widehat{\Delta}^n: |A^*(\bp)|>1\}$.} \Citet{de2023robust} show that when the state space is binary, all decision problems can be decomposed into binary-action subproblems. However, this result does not hold when there are more than two states.

We also note that, by the equivalence between (i) and (iii) of \cref{thm:1}, a DM with \textit{maxmin expected utility} preferences \citep{gilboa1989maxmin} prefers $F$ to $G$ in every QCC problem if and only if $F\succeq_{\textup{LB}}G$. This result extends to a broader family of ambiguity preferences characterized by \citet{cerreia2011uncertainty}; the proof follows from arguments similar to those of \citet{li2020information}.

\section{Moral Hazard}
\label{sec:5}

\subsection{Model Setting}
\label{sec:5.1}

\subsubsection*{Model with Hidden Actions}

In this section we examine a moral hazard problem, i.e., a hidden-action principal–agent model. Here, the state space is the agent's action space: the agent (he) privately selects an action $\theta\in\Theta=\{\theta_0,\ldots,\theta_n\}$. Moreover, he may randomize, choosing a mixed action given by $\bdl=(\delta_1,\ldots,\delta_n)\in \td_n$, where $\delta_i$ is the probability of selecting $\theta_i$ for $i=1,\ldots,n$, and $\delta_0:=1-\sum_{i=1}^n \delta_i$ is the probability of selecting $\theta_0$. The agent incurs a cost of $c(\bdl)$ to select $\bdl$. \footnote{Appendix~L of the supplementary material studies the case
in which the agent is restricted to pure actions. In that setting,
a weaker informativeness order is necessary and sufficient for
comparing the principal's minimum disutility across all target
actions and environments.}

After the agent has chosen his action, a publicly observable signal $X$ taking values in $\cX=[\ux,\ox]$ is generated by an experiment $F$. The signal $X$ is distributed as $F(\cdot\mid\bdl)=\sum_{i=0}^n \delta_i F(\cdot\mid\theta_i)$. The principal (she) aims for the agent to select a target action $\bdl^*\in\td_n$. To achieve this, she chooses an incentive scheme $\hs:\cX\to\Delta(\mathbb{R})$ that maps signal realizations to lotteries over payments. If the agent selects $\bdl$ and receives a payment $s$, his utility is $u(s)-c(\bdl)$, and the principal incurs disutility $v(s)$.

Our model is based on the state-space formulation introduced by \citet{wilson1967structure}, \citet{spence1971insurance}, and \citet{ross1973economic}.\footnote{Section~5.1 of the \href{https://arxiv.org/pdf/2502.06530v1}{earlier version of this paper} studies the state-space formulation, in which the agent chooses some $\theta_i\in\Theta$ and cannot randomize. That formulation is nested in the present one by taking $c(\bdl)=\sum_{i=0}^n \delta_i c_i$, in which $c_i$ is the cost of choosing $\theta_i$.} We extend that formulation in the spirit of \citet{hart1986theory}, who propose a framework in which the agent directly selects a signal distribution from a convex set.\footnote{See also \citet{georgiadis2020optimal}, \citet{georgiadis2024flexible}, and \citet{sinander2024optimism}.} In our model, we permit mixed actions, which can induce any signal distribution in the convex hull of distributions conditional on each $\theta\in\Theta$.\footnote{Our model can also be interpreted as a situation in which the principal requires the agent to allocate time across multiple tasks, as in \citet{holmstrom1991multitask}.} Finally, following \citet{grossman1983analysis}, we assume the agent's utility is additively separable in the payment and the cost of the selected action, and we restrict our attention to the problem of implementing a target action.

\subsubsection*{Assumptions}

We assume there exists $\bmm=(\um,\om)\in\mathbb{R}^2$ with $\um\leq \om$  such that the principal must select $\hs:\cX\to\Delta([\um,\om])$. That is, the principal cannot impose arbitrarily large penalties or rewards. The lower bound $\um$ reflects the agent’s limited liability, while the upper bound $\om$ reflects the principal’s financial constraint.\footnote{See \citet{jewitt2008moral} for further justifications for this assumption.}

Furthermore, we assume $c(\cdot)$ is convex. We may do so without loss of generality: given arbitrary $c(\cdot)$, for each $\bdl$, the agent can select a convex combination of actions to induce $F(\cdot\mid\bdl)$ and thus incur cost $\operatorname{con}(c)(\bdl)$, where $\operatorname{con}(c)(\cdot)$ is the convex envelope of $c(\cdot)$. \footnote{We can also drop the convexity assumption and let the agent directly incur the primitive cost $c(\bdl)$ when choosing $\bdl$. Any action $\bdl$ with 
$c(\bdl)>\operatorname{con}(c)(\bdl)$ is not implementable under any experiment: under any incentive scheme, the agent has a strictly profitable deviation to some $\bdl'$. Hence it suffices to restrict attention to actions satisfying $c(\bdl)=\operatorname{con}(c)(\bdl)$; on this restricted domain, the analysis and results in this section remain valid.}

Finally, we assume both $u(\cdot)$ and $v(\cdot)$ are continuous on $[\um,\om]$. We allow the agent and the principal to exhibit arbitrary risk attitudes.

\subsubsection*{Simplification}

Let $\Gamma:=\{(u(s),v(s))\mid s\in[\um,\om]\}$ be the set of achievable utility–disutility profiles. Let $\operatorname{co}(\Gamma)$ denote its convex hull. Define $\uu:=\min_{s\in[\um,\om]}u(s)$, $\ou:=\max_{s\in[\um,\om]}u(s)$, and $\bu:=(\uu,\ou)$. For each $\tilde{u}\in[\uu,\ou]$, define 
\begin{equation*}
    \gamma(\tilde{u}):=\min\{\tilde{v}\mid (\tilde{u},\tilde{v})\in\operatorname{co}(\Gamma)\};
\end{equation*}
this is the minimal expected disutility for the principal from a lottery over payments that yields expected utility $\tilde{u}$ to the agent.\footnote{By Carathéodory’s theorem, for each $\tilde{u}\in[\uu,\ou]$, there exists a probability distribution $H_{\tilde{u}}$ supported on at most two points in $[\um,\om]$ such that $\mathbb{E}_{H_{\tilde{u}}}[u(S)]=\tilde{u}$ and $\mathbb{E}_{H_{\tilde{u}}}[v(S)]=\gamma(\tilde{u})$.} That is, the principal effectively incurs disutility $\gamma(\tilde{u})$ when delivering utility $\tilde{u}$ to the agent via the payment. The function $\gamma$ is convex and continuous, as it is the lower boundary of $\operatorname{co}(\Gamma)$. \cref{fig:4} illustrates the relationship between $\Gamma$ and $\gamma$.

\begin{figure}
    \centering
    \centering

\pgfmathsetmacro{\xa}{2}
\pgfmathsetmacro{\xb}{4}

\pgfmathsetmacro{\aL}{0.18}
\pgfmathsetmacro{\bL}{0.02}
\pgfmathsetmacro{\A}{1.9}
\pgfmathsetmacro{\kk}{1.35}
\pgfmathsetmacro{\B}{0.10}
\pgfmathsetmacro{\s}{1.05}
\pgfmathsetmacro{\c}{0.22}

\pgfmathsetmacro{\fAtwo}{1 + \aL*(\xa)^2 + \bL*(\xa)}
\pgfmathsetmacro{\fBfour}{\fAtwo + \A*(1 - exp(-\kk*(\xb - \xa))) + \B*(\xb - \xa)}
\pgfmathsetmacro{\slope}{(\fBfour - \fAtwo)/(\xb - \xa)}

\pgfmathsetmacro{\aTouch}{1.2}
\pgfmathsetmacro{\bTouch}{5.2}
\pgfmathsetmacro{\mTouch}{(\aTouch + \bTouch)/2}
\pgfmathsetmacro{\alphaU}{0.025}
\pgfmathsetmacro{\betaU}{3.5}
\pgfmathsetmacro{\muU}{0.45}
\pgfmathsetmacro{\gammaU}{0.05}

\pgfmathsetmacro{\xLab}{3.2}
\pgfmathsetmacro{\lowLab}{\fAtwo + \A*(1 - exp(-\kk*(\xLab - \xa))) + \B*(\xLab - \xa)}
\pgfmathsetmacro{\deltaLab}{\alphaU*(\xLab - \aTouch)^2*(\xLab - \bTouch)^2*(1 + \betaU*exp(-\muU*(\xLab - \mTouch)^2))/(1 + \gammaU*(\xLab - \mTouch)^2)}
\pgfmathsetmacro{\upLab}{\lowLab + \deltaLab}
\pgfmathsetmacro{\yLabelLeft}{\lowLab + 0.40*\deltaLab} 
\pgfmathsetmacro{\gammaLab}{\fAtwo + \slope*(\xLab - \xa)}  
\pgfmathsetmacro{\yGammaText}{\gammaLab - 0.8}

\begin{tikzpicture}
\begin{axis}[
    width=12cm, height=8cm,
    axis lines=left,
    xmin=0, xmax=6,
    ymin=0.8, ymax=7.0,
    xtick=\empty, ytick=\empty,
    clip=false
]

\addplot[name path=lowA, domain=0.3:2, samples=300, very thin] {1 + \aL*x^2 + \bL*x};
\addplot[name path=lowB, domain=2:4, samples=300, very thin] {\fAtwo + \A*(1 - exp(-\kk*(x - \xa))) + \B*(x - \xa)};
\addplot[name path=lowC, domain=4:5.1, samples=300, very thin] {\fBfour + \s*(x - \xb) + \c*(x - \xb)^2};

\addplot[name path=upA, domain=0.3:2, samples=300, very thin]
  {1 + \aL*x^2 + \bL*x
   + \alphaU*(x - \aTouch)^2*(x - \bTouch)^2*(1 + \betaU*exp(-\muU*(x - \mTouch)^2))/(1 + \gammaU*(x - \mTouch)^2)};
\addplot[name path=upB, domain=2:4, samples=300, very thin]
  {\fAtwo + \A*(1 - exp(-\kk*(x - \xa))) + \B*(x - \xa)
   + \alphaU*(x - \aTouch)^2*(x - \bTouch)^2*(1 + \betaU*exp(-\muU*(x - \mTouch)^2))/(1 + \gammaU*(x - \mTouch)^2)};
\addplot[name path=upC, domain=4:5.1, samples=300, very thin]
  {\fBfour + \s*(x - \xb) + \c*(x - \xb)^2
   + \alphaU*(x - \aTouch)^2*(x - \bTouch)^2*(1 + \betaU*exp(-\muU*(x - \mTouch)^2))/(1 + \gammaU*(x - \mTouch)^2)};

\addplot[gray!18] fill between[of=lowA and upA];
\addplot[gray!18] fill between[of=lowB and upB];
\addplot[gray!18] fill between[of=lowC and upC];

\addplot[color=blue, semithick, domain=0.3:2, samples=300]
  {1 + \aL*x^2 + \bL*x}; 
\addplot[color=blue, semithick, domain=2:4, samples=2]
  {\fAtwo + \slope*(x - \xa)}; 
\addplot[color=blue, semithick, domain=4:5.1, samples=300]
  {\fBfour + \s*(x - \xb) + \c*(x - \xb)^2}; 

\addplot[black, very thin, dashed, domain=2:4, samples=300]
  {\fAtwo + \A*(1 - exp(-\kk*(x - \xa))) + \B*(x - \xa)};

\node[font=\Large] at (axis cs:\xLab,\yLabelLeft) {$\Gamma$};
\node[text=blue, font=\Large] at (axis cs:\xLab,\yGammaText) {$\gamma$};

\addplot[black, dashed, very thin] coordinates {(0.3,0.8) (0.3,7.0)};
\addplot[black, dashed, very thin] coordinates {(5.1,0.8) (5.1,7.0)};

\node[font=\Large] at (axis cs:0.3,0.3) {$\underline{u}$};
\node[font=\Large] at (axis cs:5.1,0.3) {$\overline{u}$};

\node[font=\Large] at (axis description cs:1.05,-0.04) {$\tilde{u}$};
\node[font=\Large] at (axis description cs:-0.03,0.95) {$\tilde{v}$};

\end{axis}
\end{tikzpicture}\hspace{10mm}
    \caption{The set of achievable utility–disutility profiles $\Gamma$ and the effective disutility function $\gamma$. The curve $\gamma$ is the lower boundary of $\operatorname{co}(\Gamma)$.}
    \label{fig:4}
\end{figure}

Using $\gamma$, we can simplify our original model as follows: instead of an incentive scheme $\hs$ as defined earlier, we let the principal select a function $\hw:\cX\to[\uu,\ou]$ that maps signal realizations directly to the agent's utility from the payment. If the agent selects action $\bdl$ and receives utility $w$ from the payment, his net payoff is $w-c(\bdl)$, while the principal incurs disutility $\gamma(w)$. Let $\mathcal{W}^{\bu}_{F}$ denote the set of Borel-measurable functions $\hw:\cX\to[\uu,\ou]$.

Under this simpler framework, we define a \emph{moral hazard environment} as a triple $\mathcal{M}:=\{\bu, c,\gamma\}$ consisting of (\romannumeral 1) a vector $\bu=(\uu,\ou)\in\mathbb{R}^2$ with $\uu\leq\ou$, which specifies the bounds on the agent's utility from payment; (\romannumeral 2) a convex function $c:\td_n\to\mathbb{R}$, which specifies the agent's cost for the selected action; and (\romannumeral 3) a continuous convex function $\gamma:[\uu,\ou]\to\mathbb{R}$, which specifies the principal's disutility from delivering a given utility level to the agent through the payment. We let $\mathfrak{M}$ denote the set of all moral hazard environments.

\subsubsection*{Principal's Problem}

The principal aims to implement $\bdl^*$ by selecting $\hw$ that satisfies both individual rationality (IR) and incentive compatibility (IC), while minimizing her expected disutility. More precisely, given a target action $\bdl^*\in\td_n$, an environment $\mathcal{M}\in\mathfrak{M}$, and an experiment $F$, the principal's optimization problem is
\begin{align}
    &I_F(\bdl^*,\mathcal{M}):=\inf_{\hw\in\mathcal{W}^{\bu}_{F}}\int_{\cX}\gamma(\hw(x))\,dF(x\mid\bdl^*) 
    \quad\text{subject to} \label{eqn:6} \\
    &\text{(IR)} \quad \int_{\cX}\hw(x)\,dF(x\mid\bdl^*) - c(\bdl^*) \geq 0,
    \label{eqn:7} \\
    &\text{(IC)} \quad \bdl^* \in \operatorname*{arg\,max}_{\bdl\in\td_n}
        \left[\int_{\cX}\hw(x)\,dF(x\mid\bdl) - c(\bdl)\right].
     \label{eqn:8}
\end{align}
The action $\bdl^*$ is \emph{implementable under $F$} if the principal's problem is feasible, meaning there exists $\hw\in\mathcal{W}^{\bu}_{F}$ satisfying \eqref{eqn:7} and \eqref{eqn:8}. We set $I_F(\bdl^*,\mathcal{M})=+\infty$ if $\bdl^*$ is not implementable under $F$.

\subsection{Informativeness Criteria}

To compare experiments in moral hazard problems, we introduce two informativeness criteria:
\begin{definition}
    \label{def:5}
    The experiment $F$ is \textit{more flexible} than $G$ if for each $\bdl^*\in\td_n$ and $\mathcal{M}\in\mathfrak{M}$, whenever $\bdl^*$ is implementable under $G$, it is also implementable under $F$.    
\end{definition}
\begin{definition}
\label{def:6}
    The experiment $F$ is \textit{more effective} than $G$ if $I_F(\bdl^*,\mathcal{M})\leq I_G(\bdl^*,\mathcal{M})$ for each $\bdl^*\in\td_n$ and $\mathcal{M}\in\mathfrak{M}$.
\end{definition}
The first criterion, flexibility, measures the set of implementable actions—the principal's \emph{breadth of control}. The second, effectiveness, measures the principal’s minimal disutility for implementing the target action—her \emph{depth of control}.

\begin{theorem}
\label{thm:2}
    The following statements are equivalent:
    \begin{enumerate}[label=\textup{(\roman*)}]
        \item $F\succeq_{\textup{LB}} G$;
        \item $F$ is more flexible than $G$;
        \item $F$ is more effective than $G$.
    \end{enumerate}
\end{theorem}

The intuition behind \cref{thm:2} is as follows. Dominance in the LB order implies greater flexibility because a larger Lorenz zonoid means a larger set of feasible statewise expected utility profiles. Specifically, if $F\succeq_{\textup{LB}} G$ and $\hw$ implements $\bdl^*$ under $G$, then by \cref{prop:3} there exists $\hw'$ under $F$ such that $\mathbb{E}[\hw(Y)\mid\theta]=\mathbb{E}[\hw'(X)\mid\theta]$ for each $\theta$ (expectation matching). Since \eqref{eqn:7} and \eqref{eqn:8} are determined by these conditional expectations, $\hw'$ also satisfies the IR and IC constraints, and hence implements $\bdl^*$. Importantly, however, expectation matching does not guarantee that the principal incurs lower disutility, since the principal and the agent may have different risk attitudes, as reflected in the curvature of $\gamma$.

Dominance in the LB order also implies that the principal incurs lower disutility when implementing any target action, because when the likelihood ratio is more dispersed, the principal is better able to deter deviations by the agent. Specifically, to prevent deviation from $\bdl^*$ to some $\bdl'$, the principal should reward the agent when the signal realization $x\in\cX$ yields a low likelihood ratio $\frac{f(x\mid\bdl')}{f(x\mid\bdl^*)}$, and punish the agent when $\frac{f(x\mid\bdl')}{f(x\mid\bdl^*)}$ is high. A more dispersed likelihood ratio means a wider expected utility gap between $\bdl^*$ and $\bdl'$, which reinforces the agent’s incentive compatibility. \Citet{holmstrom1979moral} and \citet{kim1995efficiency} apply this reasoning to moral hazard problems in which the agent's action is a one-dimensional effort variable. They focus on experiments satisfying MLRP and assume the validity of the FOA; hence, in their setting, the principal only needs to guard against local downward deviations, in which the agent marginally reduces his effort. In our setting, by contrast, the agent may deviate in any direction, that is, toward any $\theta$. By combining all possible deviations, we show that greater dispersion of the likelihood-ratio vector in the sense of the linear convex order (\cref{prop:1}) strengthens incentive compatibility and thereby lowers the principal’s disutility.

Finally, the equivalence between greater flexibility (increased breadth of control) and greater effectiveness (increased depth of control) follows from the equivalence between a larger Lorenz zonoid and a more dispersed likelihood ratio.

\subsection{Robustness}
\label{sec:5.3}

\subsubsection*{Bounds on Penalties and Rewards}

We now relax the assumption that the principal must select $\hs:\cX\to\Delta([\um,\om])$ (i.e., we drop the fixed bounds $\um,\om$ on the payments the principal may offer).

An incentive scheme $\hs:\cX \to \Delta(\mathbb{R})$ induces a random variable $\hS$—the payment—with conditional distribution $\hs(x)$ given the signal realization $x\in\cX$. Recall that the distribution of $X$ is determined by the agent’s action. The scheme $\hs$ is \emph{uniformly bounded} if there exists $\bmm(\hs)=(\um(\hs),\om(\hs))\in\mathbb{R}^2$ such that $\mathbb{P}\left(\hS\in[\um(\hs),\om(\hs)]\mid \theta\right)=1$ for each $\theta\in\Theta$. For a deterministic scheme $\hs:\cX \to \mathbb{R}$, this is equivalent to requiring $\hs\in\bigcap_{\theta\in\Theta}L^{\infty}(F(\cdot|\theta))$, meaning $\hs$ is essentially bounded at all $\theta$.

\cref{thm:2} remains valid when the principal is required to select a uniformly bounded scheme $\hs$, with scheme-specific bounds $\bmm(\hs)=(\um(\hs),\om(\hs))$ in place of common bounds $\bmm=(\um,\om)$. If $F\succeq_{\textup{LB}} G$ and a uniformly bounded $\hs$ under $G$ implements $\bdl^*$, then by \cref{thm:2}, there exists $\hs':\cX\to\Delta([\um(\hs),\om(\hs)])$ under $F$ implementing $\bdl^*$ at lower disutility for the principal. Hence the principal's minimal disutility from implementing $\bdl^*$ is lower under $F$ than under $G$. Conversely, if $F\nsucceq_{\textup{LB}} G$, we can restrict our attention to bounded utility functions such that the function $\hw$ induced by each $\hs$ maps from the signal space to a compact set, and show that $F$ is neither more flexible nor more effective than $G$.

\subsubsection*{Additive Separability}

Following \citet{grossman1983analysis}, we have assumed that the agent’s utility is additively separable in the payment and the cost of the selected action, and our focus is to minimize the principal's disutility from implementing her target action, which implicitly requires her utility to be additively separable in the disutility from the payment and the payoff to her from the selected action. If additive separability fails, then the LB order (and even the Blackwell order) may fail to rank experiments in accordance with the principal's utility; see \citet{gjesdal1982information} for a counterexample.

\section{Screening with Ex Post Signals}
\label{sec:6}

\subsection{Model Setting}
\label{sec:6.1}

\subsubsection*{Model with Hidden Types and Ex Post Signals}

    We now study a screening problem, i.e., a hidden-type principal--agent model, in which an experiment is used to generate an ex post signal. As before, $\Theta=\{\theta_0,\ldots,\theta_n\}$ denotes the state space, and $F$ denotes an experiment with signal space $\cX$. The model also includes a set of \emph{alternatives} $\cA=\{a_0,\ldots,a_k\}$.\footnote{Appendix~L of the supplementary material studies the case
in which the agent observes the state. In that setting, a weaker
informativeness order is necessary and sufficient for comparing
implementability, the principal's minimum disutility, and her
maximal expected payoff across all screening environments.} The principal and the agent share a common prior $\bq$ about the realized state $\theta$. The game proceeds as follows:
\begin{enumerate}
    \item The agent receives private information about $\theta$. Specifically, he draws a belief $\boldsymbol{P}$, which is a random vector taking values in a finite\footnote{The finiteness of the type set $\mathcal P$ is imposed only for expositional convenience. The analysis extends to general standard Borel type spaces $\mathcal P\subseteq\Delta(\Theta)$ by applying the finite-type result to arbitrary finite restrictions of $\mathcal P$ and passing to the limit.} set $\mathcal{P}$ according to $\pi\in\Delta(\mathcal{P})$ with $E_{\pi}(\boldsymbol{P})=\bq$. We refer to $\bP$ as the agent’s private type and write $\bp$ for a generic realization of $\bP$.
    \item The agent sends a message to the principal.
    \item The principal selects an alternative $a\in \cA$.
    \item An ex post signal correlated with $\theta$ is publicly observed: with probability $\psi(a)$, a signal $X$ taking values in $\cX\in[\ux,\ox]$ is generated by the experiment $F$. With probability $1-\psi(a)$, an uninformative signal $\perp$ is realized. Let $\oX:=\cX\cup\{\perp\}$.
    \item The principal makes a transfer $t\in\mathbb{R}$ to the agent.
\end{enumerate}

In state $\theta$, if the principal selects $a$ and makes transfer $t$, she receives utility $v_1(a,\theta)-v_2(t)$, while the agent receives utility $u_1(a,\theta)+u_2(t)$. 

The principal possesses full commitment power. By the revelation principle, she commits to a mechanism $\{\ha,\htt\}$ consisting of an \emph{allocation rule} $\ha:\mathcal{P}\to \Delta(\cA)$ and a \emph{transfer rule} $\htt:\mathcal{P}\times \cA\times \oX \to \Delta(\mathbb{R})$. The allocation rule maps reported types to lotteries over alternatives, while the transfer rule maps triples consisting of a reported type, a selected alternative, and an ex post signal to lotteries over transfers.

\subsubsection*{Assumptions}

Following the simplification step of \cref{sec:5.1}, we assume that $v_2(t)$ is continuous and convex in $t$, and $u_2(t)\equiv t$. Consequently, we can restrict our attention to deterministic transfer rules $\htt:\mathcal{P}\times \cA\times \oX \to \mathbb{R}$.

We assume that the agent has limited liability and the principal is financially constrained, so that the principal cannot impose arbitrarily large penalties or rewards.\footnote{By the arguments in \cref{sec:5.3}, our main result (\cref{thm:3}) extends to the case where $\htt$ is essentially bounded.} Specifically, we assume there exists $\bmm=(\um,\om)\in\mathbb{R}^2$ with $\um<\om$ such that\footnote{This assumption can be relaxed to allow $\bmm$ to depend on $\bp$ and $a$.} $\htt$ must satisfy
\begin{equation}
\label{eqn:9}
    \htt(\bp,a,x)\in[\um,\om] \quad \forall\, \bp\in\mathcal{P},\ a\in \cA,\ x\in \oX.
\end{equation}
Let $\mathcal{T}_F^{\bmm}$ denote the set of Borel-measurable functions $\htt$ satisfying \eqref{eqn:9}. 

Let $\mathcal{E}:=\{\cA,\bq,\mathcal{P},\pi,\psi,v_1,v_2,u_1,\bmm\}$ denote a screening environment as specified above, omitting the experiment $F$ that generates the ex post signals. Let $\mathfrak{E}$ denote the collection of all such environments.

\subsubsection*{Principal's Problem}

The principal aims to maximize her expected utility by selecting $\ha$ and $\htt$ that satisfy both individual rationality (IR) and incentive compatibility (IC). For each $\bp\in\cp$, the allocation rule $\ha$ generates a random variable $\hat{A}(\bp)$ that is distributed according to $\ha(\bp)$. Given a screening environment $\mathcal{E}$ and an experiment $F$ generating the ex post signals, the principal's optimization problem is
\begin{equation}
    \label{eqn:10}
W_F(\mathcal{E})=\sup_{\ha\in(\Delta(\cA))^{\mathcal{P}},\,\htt\in\mathcal{T}_F^{\bmm}}\mathbb{E}\left\{\mathbb{E}\left\{v_1(\hat{A}(\bP),\tilde{\theta})-\mathbb{E}[v_2(\htt(\bP,\hat{A}(\bP),x))\mid \tilde{\theta},\hat{A}(\bP)]\middle | \bP\right\}\right\},
\end{equation}
subject to
\begin{align}
    &\text{(IR)} \quad \mathbb{E}\left\{u_1(\hat{A}(\bp),\tilde{\theta})+\mathbb{E}[\htt(\bp,\hat{A}(\bp),x)\mid \tilde{\theta},\hat{A}(\bp)]\right\}\geq 0\quad \forall\,\bp\in\mathcal{P}, \label{eqn:11} \\[1ex]
    &\text{(IC)} \quad \bp\in\operatorname*{arg\,max}_{\bp'\in\mathcal{P}}\left\{\mathbb{E}\left\{u_1(\hat{A}(\bp'),\tilde{\theta})+\mathbb{E}[\htt(\bp',\hat{A}(\bp'),x)\mid \tilde{\theta},\hat{A}(\bp')]\right\}\right\}\quad \forall\,\bp\in\mathcal{P}.  \label{eqn:12} 
\end{align}
The agent's realized type $\bp$ determines the distribution of $\tilde{\theta}$ in \eqref{eqn:11} and \eqref{eqn:12},  while his reported type $\bp'$ induces the allocation $\hat{A}(\bp')$ and governs the transfer rule. The realized state and the allocation jointly determine the distribution of $x$, thereby pinning down the transfer.

\subsubsection*{Discussion}

The model described above relates to the literature on security design with contingent payments (see, e.g., \citealp{hansen1985auctions}, \citealp{demarzo2005bidding}, and \citealp{ekmekci2016just}). It captures a scenario in which an investor (principal) decides whether to acquire an asset—such as equity or debt—from an entrepreneur (agent) of unknown quality (state) to finance his project. The investor’s return is contingent on the project’s future performance (an ex post signal), which is known (i.e., $\psi>0$) only if the project is funded. The experiment $F$ reveals the information on which the returns are based. It reflects the investor’s accounting capacity and the degree of financial transparency mandated by policy.

The model also relates to the literature on mechanism design with ex post auditing (see, e.g., \citealp{townsend1979optimal}, \citealp{laffont1986using}, \citealp{mookherjee1989optimal}, and \citealp{mylovanov2017optimal}). It describes the process of procurement with auditing, in which an organization (e.g., a government agency) decides whether to allocate a project to a contractor, then conducts an audit of the project’s quality upon completion. Here, the experiment $F$ reflects the organization’s auditing ability (as determined, for example, by the level of technology employed).

To date, problems involving the informativeness of ex post signals have received limited attention. The existing literature (see, e.g., \citealp{cremer1988full}, \citealp{riordan1988optimal}, and \citealp{mcafee1992correlated}) typically allows the principal to impose arbitrarily large penalties on the agent. With unlimited penalties, as long as the ex post signal satisfies the full-rank condition \citep[see][]{riordan1988optimal}, the principal can implement any allocation rule and can thus fully extract the agent's surplus. That is, she can achieve the expected utility from solving the optimization problem \eqref{eqn:10} subject solely to the IR constraint \eqref{eqn:11}. In such a setting, the informativeness of the ex post signal is irrelevant. By contrast, in our setting, the principal cannot impose arbitrarily large penalties. Consequently, the informativeness of the ex post signal is crucial for incentive provision and directly affects the principal's payoff.

\subsection{Comparison of Experiments}

The LB order enables the comparison of experiments in terms of the principal's expected utility under her optimal mechanism:

\begin{theorem}
\label{thm:3}
    We have $F\succeq_{\textup{LB}}G$ if and only if $W_F(\mathcal{E})\geq W_G(\mathcal{E})$ for every screening environment $\mathcal{E}\in\mathfrak{E}$.
\end{theorem}

The main step in establishing \cref{thm:3} is the following: when $F\succeq_{\textup{LB}}G$, for each allocation rule $\ha$, if there exists a payment rule $\htt:\mathcal{P}\times \cA\times \overline{\cY} \to [\um,\om]$ under $G$ implementing $\ha$ (i.e., such that $(\ha,\htt)$ satisfies \eqref{eqn:11} and \eqref{eqn:12}), then there exists a payment rule $\htt':\mathcal{P}\times \cA\times \oX \to [\um,\om]$ under $F$ implementing $\ha$ at lower expected disutility for the principal. The proof parallels that of \cref{thm:2} and relies on the same ideas: (\romannumeral 1) expanding the Lorenz zonoid enlarges the set of implementable allocation rules, since \eqref{eqn:11} and \eqref{eqn:12} are determined by the conditional expectations of $\htt$ given each $\theta$; and (\romannumeral 2) when the likelihood ratio is more dispersed, in the sense of the linear convex order, the principal can more effectively deter misreporting and can thus induce truthtelling with lower disutility.

\Cref{thm:3} establishes the LB order as a useful tool for the investment and procurement applications described in the previous section. For example, it shows that if the disclosure or accounting system used by an investor is more informative in the LB sense, then the investor has a higher ex ante expected payoff from each investment opportunity. Moreover, we can extend \cref{thm:3} to a setting in which the investor faces ambiguity about the distribution of entrepreneurial quality; the theorem then shows that greater informativeness in the LB sense leads to a higher payoff guarantee for the investor. Likewise, in procurement with ex post auditing, \cref{thm:3} provides a means of ranking auditing methods: the use of an LB-dominant method expands the set of implementable contracts and reduces expected enforcement costs.

\section{Concluding Remarks}

The LB order has numerous further applications. For example, combining the ideas of \cref{sec:5,sec:6}, we can use it to study moral hazard problems under adverse selection (i.e., principal–agent models with hidden actions and hidden types). Here, the state space $\Theta$ would be the product of the agent's type space and his action space. We can show that in such a setting, increased informativeness in the sense of the LB order enables the principal to implement every mapping from the agent's types to his actions at lower disutility.  We can also apply the LB order to repeated moral hazard problems, in which the principal can incentivize the agent through his continuation payoff. Finally, the LB order can be applied to a broader class of problems with additively separable utility functions, the key feature shared by \cref{sec:5,sec:6}.

%


\begin{appendix}

\section{Proofs of Results in Section 2}
\label{ap:a}

\subsection{Proof of Lemma 1}

For each $\bq\in\td_n$, we have $\bp_F(\bq)\succeq_{\textup{lcx}}\bp_G(\bq)$ if and only if, for every $\bb'=(b'_0,\ldots,b'_n)\in\mathbb{R}^{n+1}$,
\begin{equation}\label{eqn:13}
    \int_{\cX}\!\Big(\sum_{i=1}^{n} b'_i\, p_{F,i}(x;\bq)+b'_0\Big)_{+}\, dF_{\bq}(x)
    \;\geq\;
    \int_{\cY}\!\Big(\sum_{i=1}^{n} b'_i\, p_{G,i}(y;\bq)+b'_0\Big)_{+}\, dG_{\bq}(y),
\end{equation}
where, for $H\in\{F,G\}$, the marginal distribution is $H_{\bq}(\cdot)=\sum_{i=0}^{n}q_i H(\cdot\mid\theta_i)$ with density $h_{\bq}(\cdot)=\sum_{i=0}^{n}q_i h(\cdot\mid\theta_i)$, and 
$p_{H,i}(\cdot;\bq)=\frac{q_i h(\cdot\mid\theta_i)}{h_{\bq}(\cdot)}$. Equivalently, \eqref{eqn:13} can be written as
\begin{equation}\label{eqn:14}
\begin{split}
    &\int_{\cX}\!\Big[b'_0 q_0\, f(x\mid\theta_0)+\sum_{i=1}^{n}(b'_0+b'_i) q_i\, f(x\mid\theta_i)\Big]_{+}\, dx\\[2pt]
    \;\geq\;&\int_{\cY}\!\Big[b'_0 q_0\, g(y\mid\theta_0)+\sum_{i=1}^{n}(b'_0+b'_i) q_i\, g(y\mid\theta_i)\Big]_{+}\, dy
    \qquad \forall\,\bb'\in\mr^{n+1}.
\end{split}
\end{equation}
By \eqref{eqn:3}, the condition \eqref{eqn:14} holds for each $\bq\in\td_n$, since we can set 
$b_0=b'_0 q_0$ and $b_i=(b'_0+b'_i) q_i$. Conversely, if \eqref{eqn:14} holds for some 
$\bq\in\operatorname{int}(\td_n)$ (so $q_i>0$ for all $i\in\{0,\ldots,n\}$), then \eqref{eqn:3} follows from choosing 
$b'_0=b_0/q_0$ and $b'_i=b_i/q_i-b_0/q_0$.

\subsection{Proof of Proposition 1}

By \eqref{eqn:1}, we have $\bl_F\succeq_{\textup{lcx}}\bl_G$ if and only if, for every 
$\bb=(b_0,\ldots,b_n)\in\mathbb{R}^{n+1}$,
\begin{equation*}
    \int_{\cX}\left(\sum_{i=1}^{n}b_i \, l_{F,i}(x)+b_0\right)_{+} \, dF(x\mid\theta_0)
    \;\geq\; 
    \int_{\cY}\left(\sum_{i=1}^{n}b_i \, l_{G,i}(y)+b_0\right)_{+} \, dG(y\mid\theta_0),
\end{equation*}
which is exactly \eqref{eqn:3}.

\subsection{Proof of Proposition 2}

For each $\bq\in\td_n$,
\begin{equation*}
    e_{F}(\phi,\bq)\succeq_{\textup{cx}}e_{G}(\phi,\bq)\quad \forall\, \phi:\Theta\to\mathbb{R}
\end{equation*}
if and only if
\begin{equation}\label{eqn:15}
\begin{split}
    &\int_{\cX}\!\left\{\phi(\theta_0)\!\left[1-\sum_{i=1}^{n}p_{F,i}(x;\bq)\right]+\sum_{i=1}^{n} \phi(\theta_i)\,p_{F,i}(x;\bq)-t\right\}_{+}\, dF_{\bq}(x)\\ 
    \geq\;&\int_{\cY}\!\left\{\phi(\theta_0)\!\left[1-\sum_{i=1}^{n} p_{G,i}(y;\bq)\right]+\sum_{i=1}^{n} \phi(\theta_i)\,p_{G,i}(y;\bq)-t\right\}_{+}\, dG_{\bq}(y) \quad \forall\, \phi:\Theta\to\mathbb{R},~ t\in\mr.
\end{split}
\end{equation}
It is immediate that \eqref{eqn:13} is equivalent to \eqref{eqn:15}.

\section{Proof of Theorem 1}
\label{ap:b}

We prove $(\textup{i}) \Rightarrow (\textup{ii})$ and $(\textup{ii}) \Rightarrow (\textup{iii})$. The implication $(\textup{iii}) \Rightarrow (\textup{i})$ is immediate from \cref{lem:2}, since all binary‑action problems are QCC.

\subsection{Implication $(\textup{i}) \Rightarrow (\textup{ii})$}
\label{ap:b1}

First, fix a finite‑action QCC problem $\{\cA,u\}$ with $\cA=\{a_0,\ldots,a_k\}$, $k\ge 1$, and $a_0<\cdots<a_k$. Recall that (i) $u(a,\bp)$ is the expected payoff of $a$ at belief $\bp$; (ii) $\cA^*(\bp)$ is the set of optimal actions; (iii) $V(\bp)=\max_{a\in \cA} u(a,\bp)$ is the value function; and (iv) $\overline V_F(\bq)=\mathbb{E}[V(p_F(\bq))]$ is the ex ante value under experiment $F$ and prior $\bq$.

For each $i\in\{0,\ldots,k-1\}$, define a binary-action subproblem $\{\cA_i,u_i\}$ with $\cA_i=\{a_i,a_{i+1}\}$ and
\begin{equation*}
  u_0(a_0,\theta)=u(a_0,\theta),\qquad
  u_0(a_1,\theta)=u(a_1,\theta)\quad \forall\,\theta\in\Theta,
\end{equation*}
\begin{equation*}
  u_i(a_i,\theta)=0,\qquad
  u_i(a_{i+1},\theta)=u(a_{i+1},\theta)-u(a_i,\theta)\quad
  \forall\, i\in\{1,\ldots,k-1\},~\theta\in\Theta.
\end{equation*}
Let $u_i(a,\bp)$ be the induced expected payoff, $\cA_i^*(\bp)$ the set of optimal actions, $V_i(\bp)$ the value function, and $\overline V_{F,i}(\bq)$ the corresponding ex ante value.

We claim that $V(\bp)=\sum_{i=0}^{k-1} V_i(\bp)$ for all $\bp\in\td_n$. To prove this, fix $\bp$ and suppose $a_i\in \cA^*(\bp)$ with $i\geq 1$. The quasi-concavity of $u(\cdot,\bp)$ implies that the sequence $\{u(a_j,\bp)\}_{j=0}^k$ is unimodal:
\begin{equation*}
  u(a_0,\bp)\le\cdots\le u(a_i,\bp)\quad\text{and}\quad
  u(a_i,\bp)\ge u(a_{i+1},\bp)\ge\cdots\ge u(a_k,\bp).
\end{equation*}
Thus $a_{j+1}\in \cA_j^*(\bp)$ for each $j<i$, while $a_j\in \cA_j^*(\bp)$ for each $j\ge i$. Consequently,
\begin{equation*}
  V_0(\bp)=u(a_1,\bp),\qquad
  V_j(\bp)=u(a_{j+1},\bp)-u(a_j,\bp)~\text{for } j=1,\ldots,i-1,\qquad
  V_j(\bp)=0~\text{for } j\ge i.
\end{equation*}
Summing gives
\begin{equation*}
  \sum_{j=0}^{k-1} V_j(\bp)
  =u(a_1,\bp)+\sum_{j=1}^{i-1}\big(u(a_{j+1},\bp)-u(a_j,\bp)\big)
  =u(a_i,\bp)=V(\bp).
\end{equation*}
If $i=0$, then $V_0(\bp)=u(a_0,\bp)$ and $V_j(\bp)=0$ for $j\ge1$.

By linearity of expectation,
\begin{equation*}
  \overline V_F(\bq)=\sum_{i=0}^{k-1}\overline V_{F,i}(\bq)\qquad \forall\,\bq\in\td_n.
\end{equation*}
If $F\succeq_{\textup{LB}} G$, then by \cref{lem:2} we have $\overline V_{F,i}(\bq)\ge \overline V_{G,i}(\bq)$ for every $i$ and every $\bq$. Summing over $i$ yields $\overline V_F(\bq)\ge \overline V_G(\bq)$ for each $\bq\in\td_n$, which proves $(\textup{i})\Rightarrow(\textup{ii})$.

The case where $\cA$ is a compact interval $[\ua,\oa]$ and $u(a,\theta)$ is differentiable in $a$ is analogous; we simply define the binary-action subproblems using $\partial u(a,\theta)/\partial a$ in place of the finite differences $u(a_{i+1},\theta)-u(a_i,\theta)$. Let $a^*(\bp):=\min \cA^*(\bp)$. We use the decomposition
\begin{equation*}
    V(\bp)=u(a^*(\bp),\bp)=u(\ua,\bp)+\int_{\ua}^{a^*(\bp)}\frac{\partial u(a,\bp)}{\partial a}da=u(\ua,\bp)+\int_{\ua}^{\oa}\left(\frac{\partial u(a,\bp)}{\partial a}\right)_{+}da.
\end{equation*}
The last equality holds because $u(a,\bp)$ is quasi-concave.

\subsection{Implication $(\textup{ii}) \Rightarrow (\textup{iii})$}
\label{ap:b2}

Fix a QCC problem $\{\cA,u\}$ and any strategy $\sigma_F$ under $F$. Define the statewise payoff vector $\boldsymbol{u}_F(\sigma_F)\;:=\;\big(u(\sigma_F,\theta_0),\ldots,u(\sigma_F,\theta_n)\big)\in\mathbb{R}^{n+1}$ and the set of all implementable vectors $\mathcal{U}_F\;:=\;\{\,\boldsymbol{u}_F(\sigma_F)\;|\;\sigma_F:\cX\!\to\!\Delta(\cA)\text{ is measurable}\}$. Note that $\mathcal{U}_F$ is nonempty and convex. It is also closed, as $\cA$ is compact and $u(\cdot,\theta)$ is continuous, implying the continuity of the map $\sigma_F \mapsto u_F(\sigma_F)$ under $L^1$-convergence. For a prior $\boldsymbol{q}=(q_0,\ldots,q_n)$, the ex ante value is the support function
\begin{equation*}
  \overline V_F(\boldsymbol{q})\;=\;\sup_{\boldsymbol{u}\in\mathcal{U}_F}\;\boldsymbol{q}\cdot \boldsymbol{u}.
\end{equation*}
The assumption $(\textup{ii})$ states that $\overline V_F(\boldsymbol{q})\ge \overline V_G(\boldsymbol{q})$ for every $\boldsymbol{q}$.

Suppose, for the sake of contradiction, that there exists a strategy $\sigma_G$ under $G$ such that no strategy $\sigma_F$ under $F$ satisfies $\boldsymbol{u}_F(\sigma_F)\ge \boldsymbol{u}_G(\sigma_G)$ componentwise. Then
\begin{equation*}
  \boldsymbol{u}_G(\sigma_G)\notin \mathcal{U}_F-\mathbb{R}^{n+1}_+\;=\;\big\{\,\boldsymbol{t}\in\mathbb{R}^{n+1}\;\big|\;\exists\,\boldsymbol{u}\in\mathcal{U}_F\text{ with }\boldsymbol{t}\le \boldsymbol{u}\text{ componentwise}\,\big\}.
\end{equation*}
The set $\mathcal{U}_F-\mathbb{R}^{n+1}_+$ is closed, convex, and downward closed. Hence, by the separating hyperplane theorem, there exist $\boldsymbol{\alpha}=(\alpha_0,\ldots,\alpha_n)\in\mathbb{R}^{n+1}_+\setminus\{\bo\}$ and $c\in\mathbb{R}$ such that
\begin{equation*}
  \boldsymbol{\alpha}\cdot\boldsymbol{u}\;\le\;c\quad\forall\,\boldsymbol{u}\in \mathcal{U}_F-\mathbb{R}^{n+1}_+,
  \qquad\text{and}\qquad
  \boldsymbol{\alpha}\cdot\boldsymbol{u}_G(\sigma_G)\;>\;c.
\end{equation*}
Since $\mathcal{U}_F\subset \mathcal{U}_F-\mathbb{R}^{n+1}_+$,
\begin{equation*}
  \sup_{\boldsymbol{u}\in\mathcal{U}_F}\boldsymbol{\alpha}\cdot\boldsymbol{u}\;\le\;c\;<\;\boldsymbol{\alpha}\cdot\boldsymbol{u}_G(\sigma_G)\;\le\;\sup_{\boldsymbol{u}\in\mathcal{U}_G}\boldsymbol{\alpha}\cdot\boldsymbol{u}.
\end{equation*}
Setting $\boldsymbol{q}:=\boldsymbol{\alpha}/\sum_{i=0}^n\alpha_i$ yields a prior $\boldsymbol{q}$ with
\begin{equation*}
  \sup_{\boldsymbol{u}\in\mathcal{U}_F} \boldsymbol{q}\cdot \boldsymbol{u}
  \;<\;\boldsymbol{q}\cdot \boldsymbol{u}_G(\sigma_G)
  \;\le\;\sup_{\boldsymbol{u}\in\mathcal{U}_G} \boldsymbol{q}\cdot \boldsymbol{u},
\end{equation*}
that is, $\overline V_F(\boldsymbol{q})<\overline V_G(\boldsymbol{q})$, contradicting $(\textup{ii})$. This completes the proof.

\section{Proof of Theorem 2}
\label{ap:c}

For simplicity, we assume that (\romannumeral 1 ) the cost function $c(\cdot)$ is differentiable, and (\romannumeral 2)~$f(x|\theta)>0$ and $g(y|\theta)>0$ for all $x\in \cX$, $y\in \cY$, and $\theta\in\Theta$, which ensures that the likelihood ratios are well defined. We address the general case without these assumptions in the supplementary material.

The proof proceeds in three steps. We first compute an explicit representation of the IC constraint \eqref{eqn:8}, then prove $(\textup{i}) \Leftrightarrow(\textup{ii})$, and finally prove $(\textup{i}) \Leftrightarrow(\textup{iii})$. 

\subsection{Representation of the IC Constraint}
\label{ap:c1}

The agent’s expected payoff from selecting $\bdl$ is
\begin{equation*}
J(\bdl)\;:=\;\int_\cX \hw(x)\, dF(x\mid \bdl)\;-\;c(\bdl),
\end{equation*}
which is concave in $\bdl$ because the first term is linear in $\bdl$ and $c$ is convex. Recall that $F(\cdot\mid \bdl)=\sum_{i=0}^n \delta_i F(\cdot\mid \theta_i)$ with $\delta_0=1-\sum_{i=1}^n\delta_i$. 

Thus, for each $\bdl^\ast\in \operatorname{int}(\td_n)$,
\begin{equation*}
\bdl^\ast \in \operatorname*{arg\,max}_{\bdl\in \td_n} J(\bdl)
\quad\iff\quad
\frac{\partial J}{\partial \delta_i}(\bdl^\ast)=0\quad \forall\, i\in\{1,\ldots,n\}.
\end{equation*}
Computing the gradient of the linear term using $\delta_0=1-\sum_{j=1}^n\delta_j$ gives
\begin{equation*}
\frac{\partial}{\partial \delta_i}\!\left[\int_\cX \hw(x)\,dF(x\mid \bdl)\right]_{\bdl=\bdl^\ast}
= \int_\cX \hw(x)\,\big[f(x\mid \theta_i)-f(x\mid \theta_0)\big]\,dx,
\end{equation*}
so we can replace the IC constraint \eqref{eqn:8} by the following $n$ equalities (one for each $i\in\{1,\ldots,n\}$):
\begin{equation*}
\int_\cX \hw(x)\,\big[f(x\mid \theta_i)-f(x\mid \theta_0)\big]\,dx\;-\;c_i(\bdl^\ast)=0,
\tag{$IC_i$}\label{eqn:IC_i}
\end{equation*}
where $c_i(\bdl^*):=\frac{\partial}{\partial \delta_i}c(\bdl^*)$.

When $\bdl^\ast$ lies on the boundary of $\td_n$, the IC constraint can be represented in the same way, except with the equalities replaced by weak inequalities in the directions that would move $\bdl^\ast$ outside $\td_n$. For example, if the principal aims to implement $\boldsymbol{0}=(0,\ldots,0)$ (the pure action $\theta_0$), the IC condition can be replaced by the following inequalities (one for each $i\in\{1,\ldots,n\}$):
\begin{equation*}
\int_\cX \hw(x)\,\big[f(x\mid \theta_i)-f(x\mid \theta_0)\big]\,dx\;-\;c_i(\boldsymbol{0})\;\le 0.
\tag{$IC^0_i$}\label{eqn:IC^0_i}
\end{equation*}

\subsection{Equivalence of $(\textup{i})$ and $(\textup{ii})$}
\label{ap:c2}

\subsubsection*{Step 1: $(\textup{i}) \Rightarrow (\textup{ii})$}

For simplicity, assume $\bdl^\ast\in\operatorname{int}(\td_n)$. Suppose there exists $\hw\in\mathcal{W}^{\bu}_{G}$ implementing $\bdl^\ast$ under $G$. Such $\hw$ satisfies the IR constraint \eqref{eqn:7} and the IC constraints \eqref{eqn:IC_i} for all $i$.
If $F\succeq_{\textup{LB}}G$, then by \cref{prop:3} there exists $\hat{h}'\in \mathcal{H}_F$ with $\int_\cX \hat{h}'(x)\,f(x\mid\theta_i)\,dx=\int_{\cY} \hat{h}(y)\,g(y\mid\theta_i)\,dy$ for each $i$, where $\hat{h}(y):=\big(\hw(y)-\underline u\big)/\big(\overline u-\underline u\big)\in[0,1]$. 
Define $\hw'(x):=\underline u+(\overline u-\underline u)\hat{h}'(x)$, so that $\hw'\in\mathcal{W}^{\bu}_{F}$ and
\begin{equation*}
\int_\cX \hw'(x)\,f(x\mid\theta_i)\,dx=\int_{\cY} \hw(y)\,g(y\mid\theta_i)\,dy\quad\forall\, i\in\{0,\ldots,n\}.
\end{equation*}
By linearity in $\bdl^\ast$, this equality also holds under the mixed action $\bdl^\ast$, i.e., $\int_\cX \hw' \, dF(\cdot\mid\bdl^\ast)=\int_{\cY} \hw \, dG(\cdot\mid\bdl^\ast)$, so IR carries over from $G$ to $F$. 
Moreover, subtracting the $i$- and $0$-components yields
\begin{equation*}
\int_\cX \hw'(x)\big[f(x\mid\theta_i)-f(x\mid\theta_0)\big]\,dx
=\int_{\cY} \hw(y)\big[g(y\mid\theta_i)-g(y\mid\theta_0)\big]\,dy
= c_i(\bdl^\ast),
\end{equation*}
so each \eqref{eqn:IC_i} also carries over. Hence $\hw'$ satisfies IR and all IC constraints under $F$, and therefore implements $\bdl^\ast$.

\subsubsection*{Step 2: $(\textup{ii}) \Rightarrow (\textup{i})$}

Assume for the sake of contradiction that $F \nsucceq_{\textup{LB}} G$. By \cref{prop:3}, $\mathcal{Z}(F)$ does not contain $\mathcal{Z}(G)$. Since $\mathcal{Z}(F)$ is compact and convex, the separating hyperplane theorem yields $\bz=(z_0,\ldots,z_n)\in \mathcal{Z}(G)$ and $\bb=(b_0,\ldots,b_n)\in\mathbb{R}^{n+1}$ such that
\begin{equation}
  \bb\cdot\bz \;>\; \bb\cdot\bz' \quad \forall\, \bz'\in \mathcal{Z}(F).
  \label{eqn:16}
\end{equation}
Since $\mathcal{Z}(F)$ and $\mathcal{Z}(G)$ are centrally symmetric about $(\tfrac12,\ldots,\tfrac12)$, if $\bb$ separates $\bz$ then $-\bb$ separates the reflected point. Thus, we can assume $\sum_{i=0}^n b_i \ge 0$. We now consider two cases:

\medskip
\noindent\emph{Case 1: $\sum_{i=0}^n b_i>0$.} Rescale $\bb$ so that $\sum_{i=0}^n b_i=1$. Fix any $\bdl^*=(\delta_1,\ldots,\delta_n)\in \operatorname{int}(\td_n)$ and set $\delta_0:=1-\sum_{i=1}^n \delta_i$. Then, using $b_0=1-\sum_{i=1}^n b_i$ and $\delta_0=1-\sum_{i=1}^n \delta_i$, we have
\begin{equation}
  \bb\cdot\bz
  = \sum_{i=0}^n b_i z_i
  = \sum_{i=0}^n \delta_i z_i + \sum_{i=0}^n (b_i-\delta_i)z_i
  = \sum_{i=0}^n \delta_i z_i + \sum_{i=1}^n (b_i-\delta_i)(z_i-z_0).
  \label{eqn:17}
\end{equation}
Choose $c(\cdot)$ so that $c(\bdl^*)=\sum_{i=0}^n \delta_i z_i$ and $c_i(\bdl^*)=z_i-z_0$ for each $i\in\{1,\ldots,n\}$. Let $\bu=(0,1)$. Since $\bz\in \mathcal{Z}(G)$, there exists $\hw\in \mathcal{W}^{\bu}_G$ with $z_i=\int_{\cY} \hw(y)\,dG(y\mid \theta_i)$ for each $i$, which satisfies the IR constraint \eqref{eqn:7} with equality and satisfies all \eqref{eqn:IC_i} at $\bdl^*$, and hence implements $\bdl^*$ under $G$.

We claim there is no $\hw'\in \mathcal{W}^{\bu}_F$ implementing $\bdl^*$ under $F$. If there were, we could define $z'_i:=\int_\cX \hw'(x)\,dF(x\mid \theta_i)$. Then the IR constraint \eqref{eqn:7} would imply $\sum_{i=0}^n \delta_i z'_i \ge c(\bdl^*)=\sum_{i=0}^n \delta_i z_i$, and each \eqref{eqn:IC_i} would yield $z'_i-z'_0=c_i(\bdl^*)=z_i-z_0$. Plugging these into \eqref{eqn:17} would give
\begin{equation*}
  \bb\cdot\bz' = \sum_{i=0}^n \delta_i z'_i + \sum_{i=1}^n (b_i-\delta_i)(z_i-z_0)
  \ge \sum_{i=0}^n \delta_i z_i \;+\; \sum_{i=1}^n (b_i-\delta_i)(z_i-z_0)
  = \bb\cdot\bz,
\end{equation*}
contradicting \eqref{eqn:16}.

\medskip
\noindent\emph{Case 2: $\sum_{i=0}^n b_i=0$.} Then $b_0=-\sum_{i=1}^n b_i$ and
\begin{equation}
  \bb\cdot\bz = \sum_{i=0}^n b_i z_i
  = \sum_{i=1}^n b_i z_i + b_0 z_0
  = \sum_{i=1}^n b_i (z_i - z_0).
  \label{eqn:18}
\end{equation}
Fix any $\bdl^*\in \operatorname{int}(\td_n)$ and choose $c(\cdot)$ and $\bu=(0,1)$ as above. Then there exists $\hw\in \mathcal{W}^{\bu}_G$ with $z_i=\int_{\cY} \hw(y)\,dG(y\mid \theta_i)$ for each $i$, satisfying the IR constraint \eqref{eqn:7} and all \eqref{eqn:IC_i}, and hence implementing $\bdl^*$ under $G$. If there were $\hw'\in \mathcal{W}^{\bu}_F$ implementing $\bdl^*$ under $F$, we could define $z'_i:=\int_\cX \hw'(x)\,dF(x\mid \theta_i)$. The IC equalities would imply $z'_i-z'_0=z_i-z_0$ for each $i$, and then \eqref{eqn:18} would give $\bb\cdot\bz'=\bb\cdot\bz$, contradicting \eqref{eqn:16}.

Therefore, $F$ is not more flexible than $G$.

\subsection{Equivalence of $(\textup{i})$ and $(\textup{iii})$}
\label{ap:c3}

The implication $(\textup{iii}) \Rightarrow (\textup{i})$ holds because $(\textup{iii}) \Rightarrow (\textup{ii})$ is immediate by feasibility, and we have already proved $(\textup{ii}) \Rightarrow (\textup{i})$. We now prove $(\textup{i}) \Rightarrow (\textup{iii})$. It suffices to treat the action $\boldsymbol{0}=(0,\ldots,0)$ (the pure action $\theta_0$); the arguments for other actions are analogous.

\subsubsection*{Case 1: Strict Feasibility Under $G$}
\label{ap:c31}

We absorb the range constraint into the objective by defining
\begin{equation*}
  \gamma_{\bu}(w):=
  \begin{cases}
    \gamma(w), & w\in[\underline u,\overline u],\\[2pt]
    +\infty & \text{otherwise}.
  \end{cases}
\end{equation*}
At the target action $\bo$, the principal’s problem under $G$ is
\begin{align*}
  \inf_{\hat w\in\mathcal W_G^{\bu}}
  &\ \int_{\cY} \gamma_{\bu}\!\big(\hat w(y)\big)\,g(y\mid\theta_0)\,dy \\
  \text{s.t.}\quad
  &\ \int_{\cY} \hat w(y)\,g(y\mid\theta_0)\,dy\ -\ c(\bo)\ \ge 0,\tag{$IR$}\label{eqn:IR}\\
  &\ \int_{\cY} \hat w(y)\,\big[g(y\mid \theta_i)-g(y\mid \theta_0)\big]\,dy\ -\ c_i(\bo)\ \le 0,\quad i=1,\ldots,n.\tag{$IC^0_i$}\label{eqn:IC'^0_i}
\end{align*}

Let $\lambda\geq 0$ be the Lagrangian multiplier for \eqref{eqn:IR} and $\mu_i \geq 0$ that for \eqref{eqn:IC'^0_i}. The Lagrangian is
\begin{align*}
  L_G(\hat w,\lambda,\boldsymbol\mu)
  &= \int_{\cY} \gamma_{\bu}\!\big(\hat w(y)\big)\,g(y\mid\theta_0)\,dy
     + \lambda\!\left\{ c(\bo) - \int_{\cY} \hat w(y)\,g(y\mid\theta_0)\,dy \right\} \\
  &\quad + \sum_{i=1}^n \mu_i \left\{ \int_{\cY} \hat w(y)\,\big[g(y\mid\theta_i)-g(y\mid\theta_0)\big]\,dy - c_i(\bo) \right\}.
\end{align*}
Recall that $\ell_{G,i}(y) =g(y\mid\theta_i)/g(y\mid\theta_0)$. Hence,
\begin{equation}
\label{eqn:19}
L_G(\hat w,\lambda,\boldsymbol\mu)
  = \lambda\,c(\bo) - \sum_{i=1}^n \mu_i c_i(\bo)
    + \int_{\cY} \Big\{ \gamma_{\bu}\!\big(\hat w(y)\big)
     - \hat w(y)\big[\lambda+\textstyle\sum_{i=1}^n \mu_i\big(1-\ell_{G,i}(y)\big)\big] \Big\}\,dG(y\mid\theta_0).
\end{equation}

Assume \emph{strict feasibility} under $G$: there exists $\hat w\in\mathcal W_G^{\bu}$ making \eqref{eqn:IR} and all \eqref{eqn:IC'^0_i} strictly slack. Strong duality holds;\footnote{For $H\in\{F,G\}$, let $\nu_H:=(n+1)^{-1}\sum_{i=0}^nH(\cdot\mid\theta_i)$. Modulo $\nu_H$-a.e.\ equality, the bounded scheme set $\mathcal W_H^{\bu}$ is compact and convex in the weak-$*$ topology of $L^\infty(\nu_H)$, by Banach--Alaoglu. The integral objective is convex and weak-$*$ lower semicontinuous, since it is a supremum of continuous affine integral functionals, and all statewise moment constraints are weak-$*$ continuous and affine. Sion's minimax theorem therefore gives equality of primal and dual values for any finite collection of these equalities and inequalities, with unrestricted multipliers for equality constraints. The same argument applies to finite products of bounded scheme sets. Attainment of the dual supremum is not required.}
 hence
\begin{equation*}
  I_G(\bo,\mathcal M)=\inf_{\hat w\in\mathcal W_G^{\bu}}\ \sup_{\lambda\geq 0,\ \boldsymbol\bmu\geq 0}\,L_G(\hat w,\lambda,\boldsymbol\mu)
  \;=\;
  \sup_{\lambda\geq 0,\ \boldsymbol\bmu\geq 0}\ \inf_{\hat w\in\mathcal W_G^{\bu}}\,L_G(\hat w,\lambda,\boldsymbol\mu).
\end{equation*}
Now take the pointwise infimum in $\hat w$. Let
\begin{equation}
\label{eqn:20}
  \rho(t):=\sup_{w\in\mathbb{R}}[t w-\gamma_{\bu}(w)]
  \;=\;\sup_{w\in[\underline u,\overline u]}[t w-\gamma(w)]
\end{equation}
be the convex conjugate of $\gamma_{\bu}$. Recall that $\gamma_{\bu}$ is proper, lower semicontinuous, and convex. Inserting \eqref{eqn:20} into \eqref{eqn:19} yields
\begin{align*}
  \inf_{\hat w\in\mathcal W_G^{\bu}} L_G(\hat w,\lambda,\boldsymbol\mu)
  = \lambda\,c(\bo) - \sum_{i=1}^n \mu_i c_i(\bo)
      - \int_{\cY} \rho\!\Big(\lambda+\sum_{i=1}^n \mu_i\big(1-\ell_{G,i}(y)\big)\Big)\,dG(y\mid\theta_0).
\end{align*}
Denote the right-hand side of this equation by $K_G(\lambda,\boldsymbol\mu)$. It follows that
\begin{equation*}
  I_G(\bo,\mathcal M)
  \;=\;
  \sup_{\lambda\geq 0,\ \boldsymbol\bmu\geq 0}\,K_G(\lambda,\boldsymbol\mu).
\end{equation*}

Suppose $F\succeq_{\textup{LB}}G$. By the expectation-matching construction established in Step~1 of the proof of $(\textup{i})\Rightarrow(\textup{ii})$, any strictly feasible statewise expectation profile achieved under $G$ can be replicated under $F$ using some $\hat w'\in\mathcal W_F^{\bu}$; hence strict feasibility also holds under $F$, and $I_F(\bo,\mathcal M)=\sup_{\lambda\geq 0,\ \boldsymbol\bmu\geq 0}\,K_F(\lambda,\boldsymbol\mu)$ with
\begin{align*}
  K_F(\lambda,\boldsymbol\mu)
  &:= \lambda\,c(\bo) - \sum_{i=1}^n \mu_i c_i(\bo)
      - \int_\cX \rho\!\Big(\lambda+\sum_{i=1}^n \mu_i\big(1-\ell_{F,i}(x)\big)\Big)\,dF(x\mid\theta_0).
\end{align*}
Since $\rho$ is convex and $\bl_F\succeq_{\textup{lcx}}\bl_G$ by \cref{prop:1}, it follows that $K_F(\lambda,\boldsymbol\mu)\le K_G(\lambda,\boldsymbol\mu)$ for every $(\lambda,\boldsymbol\mu)$, which implies $I_F(\bo,\mathcal M)\ \le\ I_G(\bo,\mathcal M)$.

\begin{remark}[Finite signal spaces]
If $F$ and $G$ have finite signal spaces, we can skip the next step, because feasibility without strict slackness then guarantees strong duality, since all constraints are affine.
\end{remark}

\subsubsection*{Case 2: Without Strict Feasibility}
\label{ap:c32}

Index all the constraints—\eqref{eqn:IR} and all \eqref{eqn:IC'^0_i}—by $\mathcal I:=\{0,1,\ldots,n\}$, where $0$ denotes \eqref{eqn:IR} and $i\in\{1,\ldots,n\}$ denotes \eqref{eqn:IC'^0_i}. For any $\mathcal S\subseteq\mathcal I$, let $\mathcal M(\mathcal S)$ be the variant of the principal's problem in which the constraints with indices in $\mathcal S$ are imposed as equalities, while those with indices in $\mathcal I\setminus\mathcal S$ remain as in \eqref{eqn:IR}–\eqref{eqn:IC'^0_i}. Define $I_G(\bo,\mathcal M(\mathcal S))$ as the infimum of the objective subject to the constraints in $\mathcal M(\mathcal S)$ under $G$. Since $\emptyset\subseteq\mathcal I$ and $\mathcal M(\emptyset)$ is the original principal's problem,
\begin{equation}
\label{eqn:21}
  I_G(\bo,\mathcal M)\;=\;\min_{\mathcal S\subseteq\mathcal I} I_G(\bo,\mathcal M(\mathcal S)).
\end{equation}

Note that $\mathcal M(\mathcal S)$ is strictly feasible under $G$ if there exists $\hat w\in\mathcal W_G^{\bu}$ that satisfies all of the constraints of $\mathcal M(\mathcal S)$ and makes every inequality constraint strictly slack. Now take $\mathcal S'\subseteq\mathcal I$ such that $\mathcal M(\mathcal S')$ is feasible but not strictly feasible under $G$. For any $\hat w$ satisfying all constraints of $\mathcal M(\mathcal S')$, let
\begin{equation*}
  \widehat{\mathcal S}(\hat w):=\{\,i\in\mathcal I:\ \text{constraint $i$ binds at }\hat w\,\}
  \quad\text{and}\quad
  \mathcal S^*:=\mathcal S'\cup \widehat{\mathcal S}(\hat w).
\end{equation*}
Then $\hat w$ satisfies all constraints in $\mathcal M(\mathcal S^*)$, and every inequality constraint is strictly slack at $\hat w$; hence $\mathcal M(\mathcal S^*)$ is strictly feasible under $G$. Therefore, for each feasible point $\hat{w}$ of $M(S')$ (where $M(S')$ is feasible but not strictly feasible under $G$), $\hat{w}$ is also feasible for some variant $M(S^*)$ that is strictly feasible under $G$. Consequently, the feasible set of $M(S')$ is contained in the union of the feasible sets of strictly feasible variants $M(S)$, and hence taking infima implies
\begin{equation*}
  I_G(\bo,\mathcal M(\mathcal S'))\;\ge\;
  \min_{\substack{\mathcal S\subseteq\mathcal I:\\ \mathcal M(\mathcal S)\ \text{strictly feasible under }G}}
  I_G(\bo,\mathcal M(\mathcal S)).
\end{equation*}
Hence, for the minimization problem in \eqref{eqn:21}, we can restrict attention to strictly feasible problems. Thus, 
\begin{equation}
\label{eqn:22}
  I_G(\bo,\mathcal M)\;=\;
  \min_{\substack{\mathcal S\subseteq\mathcal I:\\ \mathcal M(\mathcal S)\ \text{strictly feasible under }G}}
  I_G(\bo,\mathcal M(\mathcal S)).
\end{equation}

For any $\mathcal S$ with $\mathcal M(\mathcal S)$ strictly feasible under $G$, the expectation-matching argument from Step~1 of the proof of $(\textup{i})\Rightarrow(\textup{ii})$ makes $\mathcal M(\mathcal S)$ strictly feasible under $F$ as well. By the compactness argument in Case~1, strong duality holds under both $G$ and $F$, and by the dual comparison established in Case~1, it follows that $I_F(\bo,\mathcal M(\mathcal S)) \le I_G(\bo,\mathcal M(\mathcal S))$. Therefore,
\begin{equation*}
  \min_{\substack{\mathcal S\subseteq\mathcal I:\\ \mathcal M(\mathcal S)\ \text{strictly feasible under }G}}
  I_F(\bo,\mathcal M(\mathcal S))
  \;\le\;
  \min_{\substack{\mathcal S\subseteq\mathcal I:\\ \mathcal M(\mathcal S)\ \text{strictly feasible under }G}}
  I_G(\bo,\mathcal M(\mathcal S)).
\end{equation*}
Moreover,
\begin{equation*}
  I_F(\bo,\mathcal M)
  \;=\;\min_{\mathcal S\subseteq\mathcal I} I_F(\bo,\mathcal M(\mathcal S))
  \;\le\;\min_{\substack{\mathcal S\subseteq\mathcal I:\\ \mathcal M(\mathcal S)\ \text{strictly feasible under }G}}
            I_F(\bo,\mathcal M(\mathcal S)).
\end{equation*}
Hence, by \eqref{eqn:22}, it follows that $I_F(\bo,\mathcal M)\ \le\ I_G(\bo,\mathcal M)$.

\end{appendix}

\bibliographystyle{ecta-fullname} 
\bibliography{main}  




\end{document}


\begin{frontmatter}

\title{Supplement to ``Ranking Statistical Experiments via the Linear Convex Order and the Lorenz Zonoid: Economic Applications''}
\runtitle{Supplementary Material}

\begin{aug}
%
%
%
\author[add1]{\fnms{Kailin}~\snm{Chen}\ead[label=e1]{kailin.chen@aalto.fi}}
\address[add1]{%
\orgdiv{Department of Economics},
\orgname{Aalto University}}
\end{aug}


\end{frontmatter}

\section*{Introduction}

This supplementary file contains \cref{ap:e,ap:f,ap:g,ap:h,ap:i,ap:j,ap:d}; \cref{M-ap:a,M-ap:b,M-ap:c} appear in the main text. \cref{ap:e} introduces the \emph{monotone–posterior-expectation (MPE) order}. We show that the MPE order coincides with the Lehmann order under MLRP and remains applicable when MLRP fails. We also clarify the relationships among the Lehmann, MPE, LB, and Blackwell orders. \cref{ap:f} presents numerical methods for verifying the LB order. \cref{ap:g} discusses the LB order with respect to the product and mixture of experiments. \cref{ap:h} establishes additional relations between the LB and Blackwell orders, complementing Section \hyperref[M-sec:3.4]{2.4} of the main text. \cref{ap:i} revisits the assumption in \cref{M-ap:c} on the differentiability of the cost function. We show that \cref{M-thm:2} holds without differentiability. \cref{ap:j} relaxes the assumption in \cref{M-ap:c} on the positivity of signal densities. We show that \cref{M-thm:2} holds even when densities may vanish. \cref{ap:d} presents the proof of \cref{M-thm:3}. \cref{ap:k} extends the analysis and results to the case of a continuous state space. \cref{ap:l} introduces the \emph{weak-linear-Blackwell (wLB) order}
and provides geometric and likelihood-ratio characterizations.
We show that this weaker order is necessary and sufficient for
uniform comparisons of the principal's minimum disutility in
moral hazard with pure actions, and of implementability and the
principal's maximal expected payoff in screening when the agent
observes the state.


\begin{appendix}
\setcounter{section}{3}
\setcounter{definition}{6}
\setcounter{proposition}{3}
\setcounter{lemma}{2}
\setcounter{equation}{23}
\setcounter{remark}{2}
\setcounter{figure}{4}
\setcounter{theorem}{3}
\setcounter{footnote}{26}

\section{Monotone-Posterior-Expectation Order}
\label{ap:e}

\subsection{Definition and Characterization}

Fix an ordinal ranking $\theta_0\prec\theta_1\prec\cdots\prec\theta_n$ of $\Theta$.
\begin{definition}
    \label{def:7} Experiment $F$ dominates $G$ in the \textit{monotone-posterior-expectation (MPE) order}, i.e., $F\succeq_{\textup{MPE}}G$, if $e_{F}(\phi,\bq)\succeq_{\textup{cx}}e_{G}(\phi,\bq)$ for each $\bq\in\td_n$ and each $\phi:\Theta\to\mathbb{R}$ that is (weakly) increasing in $\theta$.
\end{definition}

The MPE order admits two key properties. First, it must hold \emph{for all priors}, delivering a prior‑free, robust comparison of experiments. In contrast, the literature \citep[see, e.g.,][]{ganuza2010signal,gentzkow2016rothschild,ravid2022learning} typically \emph{assumes} a fixed prior. Second, the MPE order depends only on the \emph{ordinal ranking} of states, rendering it invariant to monotone relabelings of the state space, an appealing property when only the ordinal ranking of states is economically meaningful; see \citet{brandt2014information} for a related discussion under MLRP.

\citet{chi2014value} examines the MPE order (termed “generates more dispersion prediction”) for experiments satisfying MLRP, while we consider the general case where MLRP need not hold.

A vector $\bb=(b_0,\ldots,b_n)\in\mathbb{R}^{n+1}$ is \emph{quasi-monotone} if there exists $k\in\{1,\ldots,n\}$ such that $b_i\le 0$ for all $i<k$ and $b_i\ge 0$ for all $i\ge k$. Analogous to \cref{M-lem:1}, we have:

\begin{lemma}
    \label{lem:3}
    $F\succeq_{\textup{MPE}}G \iff $
    \begin{equation}
        \label{eqn:29}
        \int_{\cX}\!\!\left(\sum_{i=0}^{n} b_i f(x\mid\theta_i)\right)_{+}\!dx
        \;\ge\;
        \int_{\cY}\!\!\left(\sum_{i=0}^{n} b_i g(y\mid\theta_i)\right)_{+}\!dy
        \quad\forall\ \text{quasi-monotone }\bb\in\mathbb{R}^{n+1}.
    \end{equation}
\end{lemma}

\begin{proof}
Let $p_{F,0}(x;\bq)=1-\sum_{i=1}^{n}p_{F,i}(x;\bq)$.  
The relation $e_{F}(\phi,\bq)\succeq_{\textup{cx}}e_{G}(\phi,\bq)$ holds if and only if
\begin{equation}
    \label{eqn:30}
    \int_{\cX}\!\left\{\sum_{i=0}^{n}\phi(\theta_i)p_{F,i}(x;\bq)-t\right\}_{+}f_{\bq}(x)dx
    \ge
    \int_{\cY}\!\left\{\sum_{i=0}^{n}\phi(\theta_i)p_{G,i}(y;\bq)-t\right\}_{+}g_{\bq}(y)dy
    \quad\forall t\in\mathbb{R}.
\end{equation}
Using $h_{\bq}(\cdot)=\sum_{i=0}^{n}q_i h(\cdot\mid\theta_i)$ for $h\in\{f,g\}$ and
$p_{H,i}(\cdot)=q_i h(\cdot\mid\theta_i)/h_{\bq}(\cdot)$ for $H\in\{F,G\}$, we can rewrite \eqref{eqn:30} as
\begin{equation}
    \label{eqn:31}
    \int_{\cX}\!\left\{\sum_{i=0}^{n}[\phi(\theta_i)-t]q_i f(x\mid\theta_i)\right\}_{+}\!dx
    \ge
    \int_{\cY}\!\left\{\sum_{i=0}^{n}[\phi(\theta_i)-t]q_i g(y\mid\theta_i)\right\}_{+}\!dy
    \quad\forall t\in\mathbb{R}.
\end{equation}
Condition \eqref{eqn:29} implies \eqref{eqn:31} for every $\bq\in\td_n$ and every $\phi$ increasing in $\theta$, since the vector $\big([\phi(\theta_i)-t]q_i\big)_{i=0}^n$ is quasi-monotone. Conversely, for each quasi-monotone $\bb$, one can choose any strictly increasing $\phi$ and find $t$ and $\bq$ such that $b_i=[\phi(\theta_i)-t]q_i$ for all $i$.
\end{proof}

\begin{remark}
    The proof of \cref{lem:3} shows that it suffices to verify $e_{F}(\phi,\bq)\succeq_{\textup{cx}}e_{G}(\phi,\bq)$ for one arbitrary strictly increasing $\phi$ to establish $F\succeq_{\textup{MPE}}G$.
\end{remark}

\subsection{Relations to Other Orders}

\subsubsection*{Relation to LB order}

The MPE order is defined with respect to the ordinal ranking $\theta_0\prec\cdots\prec\theta_n$ and restricts attention to increasing $\phi$. In contrast, by \cref{M-prop:2}, $F\succeq_{\textup{LB}}G$ if and only if $e_{F}(\phi,\bq)\succeq_{\textup{cx}}e_{G}(\phi,\bq)$ for every $\phi:\Theta\to\mathbb{R}$.

\begin{proposition}
\label{prop:5}
$F\succeq_{\textup{LB}}G$ if and only if $F\succeq_{\textup{MPE}}G$ for the state space relabeled by each bijection $\beta:\Theta\to\Theta$.
\end{proposition}

\subsubsection*{Relation to Lehmann Order}

Fix the ordinal ranking $\theta_0\prec\theta_1\prec\cdots\prec\theta_n$. An experiment $F$ with signal space $\cX=[\ux,\ox]\subset\mr$ satisfies the \emph{Monotone Likelihood Ratio Property (MLRP)} if $f(x\mid\theta')f(x'\mid\theta)\leq f(x'\mid\theta')f(x\mid\theta)$ for each $\theta\prec\theta'$ and $x<x'$.

\citet{lehmann1988comparing} introduces a ranking for experiments satisfying MLRP. For each $y\in \cY$ and $\theta\in\Theta$, define $\chi(y,\theta)$ by $F(\chi(y,\theta)\mid\theta)=G(y\mid\theta)$.
\begin{definition}
    Experiment $F$ dominates $G$ in the \textit{Lehmann order}, i.e., $F\succeq_{\textup{L}}G$, if $\chi(y,\theta)$ is (weakly) increasing in $\theta$ for each $y\in \cY$.
\end{definition}

By the equivalence theorem in \citet{chi2014value},
\begin{proposition}
\label{prop:6}
    Suppose $F$ and $G$ satisfy MLRP. Then $F\succeq_{\textup{MPE}}G\iff F\succeq_{\textup{L}}G$.
\end{proposition}
\cref{fig:5} illustrates the relations among the Lehmann order, the MPE order, the LB order, and the Blackwell order.

\begin{figure}[H]
    \centering
\begin{tikzpicture}[node distance=4.2cm, thick, >=Stealth] 

\node[draw, circle, minimum size=2.4cm, text width=2cm, align=center] (lehmann) {Lehmann\\order}; 
\node[draw, circle, minimum size=2.4cm] (pm) [right of=lehmann] {MPE order}; 
\node[draw, circle, minimum size=2.4cm] (lb) [right of=pm] {LB order}; 
\node[draw, circle, minimum size=2.4cm, text width=2cm, align=center] (blackwell) [right of=lb] {Blackwell\\order}; 

\draw[->] (lehmann) -- node[above, font=\small] {no MLRP} (pm); 
\draw[->] (pm) -- node[above, font=\small] {all ordinal} 
              node[below, font=\small] {rankings} (lb); 
\draw[->] (lb) -- node[above, font=\small] {$lcx$ to $cx$} 
               (blackwell); 

\end{tikzpicture}
\caption{Relations among the Lehmann, MPE, LB, and Blackwell orders. MPE extends Lehmann beyond MLRP. LB holds when MPE holds for every ordinal ranking of $\Theta$ (i.e., for every permutation of $\Theta$). Blackwell is stricter than LB since it uses the convex (rather than linear convex) order.}
\label{fig:5}
\end{figure}

\subsection{Decision under Uncertainty}

Consider the decision problem $\{\cA,u\}$ in the main text with $\cA\subseteq\mathbb{R}$. Assume either (i) $\cA=\{a_0,\ldots,a_k\}$, or (ii) $\cA=[\underline a,\overline a]$ and $u(a,\theta)$ is differentiable in $a$ for all $\theta$. In case (i), define $\Delta u(a_i,\theta):=u(a_i,\theta)-u(a_{i-1},\theta)$ for each $i\ge 1$ and let $\Delta u(a_0,\theta)\equiv 0$. In case (ii), define $\Delta u(a,\theta):=\partial u(a,\theta)/\partial a$ for each $a\in \cA$. The problem is \emph{locally-single-crossing (LSC)} if, for each $a\in \cA$, the vector $\big(\Delta u(a,\theta_0),\ldots,\Delta u(a,\theta_n)\big)$ is quasi-monotone. Note that LSC is implied by the single-crossing property of \citet{milgrom1994monotone} and the interval dominance order property of \citet{quah2009comparative}, while the converse need not hold.

\begin{lemma}
    \label{lem:4}
    The following are equivalent:
    \begin{enumerate}[label=\textup{(\roman*)}]
        \item $F\succeq_{\textup{MPE}}G$.
        \item $\overline{V}_{F}(\bq)\geq \overline{V}_{G}(\bq)$ for every binary-action LSC problem and every $\bq\in\td_n$.
        \item For every binary-action LSC problem and every strategy $\sigma_G$ under $G$, there exists a strategy $\sigma_F$ under $F$ with $u(\sigma_F,\theta)\geq u(\sigma_G,\theta)$ for each $\theta\in\Theta$.
    \end{enumerate}
\end{lemma}

\begin{proof}
For each $\bp=(p_1,\ldots,p_n)\in\td_n$ with $p_0=1-\sum_{i=1}^n p_i$, a binary-action problem $\{a_0,a_1\}$ yields
\[
    V(\bp)=\sum_{i=0}^{n}p_i u(a_0,\theta_i)+\Big(\sum_{i=0}^{n}p_i\,\Delta u(a_1,\theta_i)\Big)_{+}.
\]
Hence,
\[
    \overline{V}_{F}(\bq)=\mathbb{E}\!\big[V(\bp_F(X;\bq))\big]
    =\sum_{i=0}^{n}q_i u(a_0,\theta_i)+\int_{\cX}\!\Big(\sum_{i=0}^{n}q_i\,\Delta u(a_1,\theta_i) f(x\mid\theta_i)\Big)_{+}dx,
\]
and analogously for $G$. Under LSC, the vector $\big(q_i\,\Delta u(a_1,\theta_i)\big)_{i}$ is quasi-monotone. Therefore, by Lemma~\ref{lem:3}, we establish (i)$\Leftrightarrow$(ii). Finally, (ii)$\Leftrightarrow$(iii) follows from the separating-hyperplane argument in \cref{M-ap:b2}
\end{proof}

Using the binary-action decomposition in \cref{M-ap:b1}, it follows that
\begin{theorem}
    \label{thm:4}
        The following are equivalent:
    \begin{enumerate}[label=\textup{(\roman*)}]
        \item $F\succeq_{\textup{MPE}}G$.
        \item $\overline{V}_{F}(\bq)\geq \overline{V}_{G}(\bq)$ for every decision problem that is QCC and LSC and every $\bq\in\td_n$.
        \item For every decision problem that is QCC and LSC and every strategy $\sigma_G$ under $G$, there exists a strategy $\sigma_F$ under $F$ with $u(\sigma_F,\theta)\geq u(\sigma_G,\theta)$ for each $\theta\in\Theta$.
    \end{enumerate}
\end{theorem}

Finally, note that single-crossing and LSC are defined with respect to the ordinal ranking $\theta_0\prec\cdots\prec\theta_n$. \citet{chen2025quasiconcavity} shows that if a decision problem is QCC, then there exists a bijection $\beta:\Theta\to\Theta$ such that the problem is LSC on the state space relabeled by $\beta$. That is, QCC implies LSC after relabeling states.

\section{Numerical Methods for Verifying LB order}
\label{ap:f}

For each $\bb=(b_0,\ldots,b_n)\in\mr^{n+1}$, define
\[
    \xi(\bb;F,G)
    :=\int_{\cX}\!\!\left(\sum_{i=0}^{n} b_i\, f(x\mid\theta_i)\right)_{+}\!dx
      -\int_{\cY}\!\!\left(\sum_{i=0}^{n} b_i\, g(y\mid\theta_i)\right)_{+}\!dy.
\]
By \cref{M-lem:1}, $F\succeq_{\textup{LB}}G$ if and only if $\inf_{\bb\in\mr^{n+1}}\xi(\bb;F,G)\ge 0$. 
Note that for every $\bb\in\mr^{n+1}$: (i) $\xi(\lambda\bb;F,G)=\lambda\,\xi(\bb;F,G)$ for each $\lambda>0$; and (ii) $\xi(\bb;F,G)=\xi(-\bb;F,G)$. 
Therefore, it suffices to check $\bb$ in a subset of $\mr^{n+1}$. 
Let $\cb:=\{\bb\in\mr^{n+1}\mid \|\bb\|_2=1,\ b_0\ge 0\}$ denote the (closed) hemisphere of the unit sphere under the Euclidean norm. 
Then $F\succeq_{\textup{LB}}G$ if and only if
\begin{equation}
\label{eqn:32}
    \min_{\bb\in \cb}\ \xi(\bb;F,G)\ \ge\ 0.
\end{equation}
For convenience, one may equivalently work with a different normalized set, 
such as a (closed) half of the surface of the unit cube.

If $\xi(\bb;F,G)$ admits an analytic form, we can verify \eqref{eqn:32} either analytically or through numerical optimization on smooth manifolds; see \citet{boumal2023introduction} for a textbook reference. 
If $\xi(\bb;F,G)$ does not admit an analytic form, we can still approximate its value for any $\bb$ using numerical integration methods, including classical quadrature \citep[][]{davis2007methods}, Bayesian quadrature \citep[][]{hennig2022probabilistic}, sparse-grid methods \citep[][]{bungartz2004sparse}, and Monte Carlo sampling \citep[][]{hung2024review}. 
We then verify \eqref{eqn:32} using numerical optimization methods for black-box objectives, including Bayesian optimization \citep[][]{shahriari2015taking}, evolutionary algorithms \citep[][]{jin2005evolutionary}, and the Dividing RECTangles (DIRECT) algorithm \citep[][]{jones2021direct}.

Moreover, as noted in \cref{M-ft:3} of the main text, any zonoid can be approximated arbitrarily well by zonotopes \citep[see, e.g.,][]{bourgain1989approximation,matouvsek1996improved,siegel2025optimal}. Consequently, verification of the LB order reduces to checking zonotope inclusion; see \citet{kulmburg2024approximability} for a survey of available methods and recent developments.

\section{Product and Mixture of Experiments}
\label{ap:g}

Consider two experiments $F_1$ and $F_2$. Let $F_1\otimes F_2$ denote the experiment where the DM receives two conditionally independent signals from $F_1$ and $F_2$, respectively. For each $t\in[0,1]$, let $tF_1+(1-t)F_2$ denote the experiment where the DM receives a signal from $F_1$ with probability $t$ and from $F_2$ with probability $1-t$.

\begin{proposition}\label{prop:7}
For any experiments $F_1, F_2, G_1, G_2$ with $F_1\succeq_{\textup{LB}}G_1$ and $F_2\succeq_{\textup{LB}}G_2$, we have
\begin{equation}\label{eqn:33}
F_1\otimes F_2 \succeq_{\textup{LB}} G_1\otimes G_2,
\end{equation}
and
\begin{equation}\label{eqn:34}
tF_1+(1-t)F_2 \succeq_{\textup{LB}} tG_1+(1-t)G_2 \quad \forall\, t\in[0,1].
\end{equation}
\end{proposition}

\begin{proof}
The implication \eqref{eqn:34} follows from Theorem 5.6 in \citet{koshevoy1998lift}. We now prove \eqref{eqn:33}. Let $\cX_1,\cX_2,{\cY}_1,{\cY}_2$ be the signal spaces, let $x_1,x_2,y_1,y_2$ denote generic signal realizations, and let $f_1,f_2,g_1,g_2$ be the densities of $F_1,F_2,G_1,G_2$, respectively. By \cref{M-lem:1}, $F_1\succeq_{\textup{LB}}G_1$ implies that, for every $\bb=(b_0,\ldots,b_n)\in\mathbb{R}^{n+1}$ and for each fixed $x_2\in \cX_2$,
\[
\int_{\cX_1}\Big(\sum_{i=0}^n b_i\, f_2(x_2\mid\theta_i)\, f_1(x_1\mid\theta_i)\Big)_+\,dx_1
\;\ge\;
\int_{{\cY}_1}\Big(\sum_{i=0}^n b_i\, f_2(x_2\mid\theta_i)\, g_1(y_1\mid\theta_i)\Big)_+\,dy_1.
\]
Integrating both sides over $x_2\in \cX_2$ yields
\begin{equation}\label{eq:step1}
\int_{\cX_2}\!\!\int_{\cX_1}\Big(\sum_{i=0}^n b_i\, f_2(x_2\mid\theta_i)\, f_1(x_1\mid\theta_i)\Big)_+\,dx_1dx_2
\;\ge\;
\int_{\cX_2}\!\!\int_{{\cY}_1}\Big(\sum_{i=0}^n b_i\, f_2(x_2\mid\theta_i)\, g_1(y_1\mid\theta_i)\Big)_+\,dy_1dx_2.
\end{equation}

Similarly, since $F_2\succeq_{\textup{LB}}G_2$, \cref{M-lem:1} implies that for each fixed $y_1\in {\cY}_1$ and every $\bb\in\mathbb{R}^{n+1}$,
\[
\int_{\cX_2}\Big(\sum_{i=0}^n b_i\, g_1(y_1\mid\theta_i)\, f_2(x_2\mid\theta_i)\Big)_+\,dx_2
\;\ge\;
\int_{\cY_2}\Big(\sum_{i=0}^n b_i\, g_1(y_1\mid\theta_i)\, g_2(y_2\mid\theta_i)\Big)_+\,dy_2.
\]
Integrating both sides over $y_1\in \cY_1$ and swapping the order of integration gives
\begin{equation}\label{eq:step2}
\int_{\cY_1}\!\!\int_{\cX_2}\Big(\sum_{i=0}^n b_i\, g_1(y_1\mid\theta_i)\, f_2(x_2\mid\theta_i)\Big)_+\,dx_2dy_1
\;\ge\;
\int_{\cY_1}\!\!\int_{\cY_2}\Big(\sum_{i=0}^n b_i\, g_1(y_1\mid\theta_i)\, g_2(y_2\mid\theta_i)\Big)_+\,dy_2dy_1.
\end{equation}
 Combining \eqref{eq:step1} and \eqref{eq:step2} yields, for every $\bb\in\mathbb{R}^{n+1}$,
\[
\int_{\cX_1}\!\!\int_{\cX_2}\Big(\sum_{i=0}^n b_i\, f_1(x_1\mid\theta_i)\, f_2(x_2\mid\theta_i)\Big)_+\,dx_2dx_1
\;\ge\;
\int_{\cY_1}\!\!\int_{\cY_2}\Big(\sum_{i=0}^n b_i\, g_1(y_1\mid\theta_i)\, g_2(y_2\mid\theta_i)\Big)_+\,dy_2dy_1.
\]
By \cref{M-lem:1} and the conditional independence in the product experiment, this is precisely $F_1\otimes F_2 \succeq_{\textup{LB}} G_1\otimes G_2$.
\end{proof}

\section{Relations between LB and Blackwell}
\label{ap:h}

\subsection{Extension from Binary-State Case}

Consider a \emph{weighted dichotomy} $\Omega=\{\Omega_0,\Omega_1,W_0(\cdot),W_1(\cdot),\omega_0,\omega_1\}$ constructed on the state space $\Theta$. The sets $\Omega_0$ and $\Omega_1$ form a partition of $\Theta$ with $\Omega_0 \cup \Omega_1=\Theta$ and $\Omega_0 \cap \Omega_1=\emptyset$. The weighting functions $W_0$ and $W_1$ assign nonnegative weights to each state in $\Omega_0$ and $\Omega_1$, respectively, such that $\sum_{\theta\in \Omega_0}W_0(\theta)=1$ and $\sum_{\theta\in \Omega_1}W_1(\theta)=1$. The partition and the weighting functions generate a dichotomy—equivalently, a binary state space $\{\omega_0,\omega_1\}$—in which an experiment $F$ on $\Theta$ maps to an experiment $F_{\Omega}$ on $\{\omega_0,\omega_1\}$ defined by $F_{\Omega}(x\mid\omega_0)=\sum_{\theta\in \Omega_0}W_0(\theta)F(x\mid\theta)$ and $F_{\Omega}(x\mid\omega_1)=\sum_{\theta\in \Omega_1}W_1(\theta)F(x\mid\theta)$ for each $x\in \cX$. Let $f_{\Omega}(\cdot\mid\omega_0)$ and $f_{\Omega}(\cdot\mid\omega_1)$ be the corresponding densities.

\begin{proposition}\label{prop:8}
    $F\succeq_{\textup{LB}}G \iff F_{\Omega}\succeq_{\textup{B}}G_{\Omega}\quad \forall\,\Omega$.
\end{proposition}

\begin{proof}
$F_{\Omega}\succeq_{\textup{B}}G_{\Omega}$ if and only if $F_{\Omega}\succeq_{\textup{LB}}G_{\Omega}$ since the state space is binary under $\Omega$. By \cref{M-lem:1}, $F_{\Omega}\succeq_{\textup{LB}}G_{\Omega}$ if and only if for each $\bb=(b_0,b_1)\in\mathbb{R}^2$,
\[
\int_\cX \big(b_0 f_{\Omega}(x\mid \omega_0)+b_1 f_{\Omega}(x\mid \omega_1)\big)_{+}\,dx
\;\geq\;
\int_{\cY} \big(b_0 g_{\Omega}(y\mid \omega_0)+b_1 g_{\Omega}(y\mid \omega_1)\big)_{+}\,dy.
\]
Equivalently,
\begin{equation}
\label{eqn:37}
\begin{split}
\int_\cX \Big(b_0\!\!\sum_{\theta\in \Omega_0}\!W_0(\theta)f(x\mid \theta)+b_1\!\!\sum_{\theta\in \Omega_1}\!W_1(\theta)f(x\mid \theta)\Big)_{+}\,dx \\
\geq \int_{\cY} \Big(b_0\!\!\sum_{\theta\in \Omega_0}\!W_0(\theta)g(y\mid \theta)+b_1\!\!\sum_{\theta\in \Omega_1}\!W_1(\theta)g(y\mid \theta)\Big)_{+}\,dy.
\end{split}
\end{equation}
It is immediate that \eqref{M-eqn:3} implies \eqref{eqn:37}, establishing “$\Rightarrow$.” For “$\Leftarrow$,” assume for the sake of contradiction that $F\nsucceq_{\textup{LB}}G$. In step 2 of the proof of \cref{prop:4} (\cref{ap:d1}), we construct an $\Omega$ such that $F_{\Omega}\nsucceq_{\textup{LB}}G_{\Omega}$, completing the proof.
\end{proof}

\cref{prop:8} enables generalizations of the established results on the Blackwell order in the binary case to the multi-state setting within the LB framework. Additionally, for the MPE order defined in \cref{ap:e}, we have $F\succeq_{\textup{MPE}}G$ iff $F_{\Omega}\succeq_{\textup{B}}G_{\Omega}$ for every weighted dichotomy $\Omega$ where $\Omega_0$ and $\Omega_1$ form a monotone partition of $\Theta$.\footnote{Given the ranking $\theta_0\prec\theta_1\prec\cdots\prec\theta_n$, there exists $\kappa$ such that $\Omega_0=\{\theta_i:i\le\kappa\}$ and $\Omega_1=\{\theta_i:i>\kappa\}$.}

\subsection{Equivalence under Irredundancy}

Consider an experiment $F$ with a finite signal set $\cX=\{x_0,\ldots,x_k\}$. Then $F$ is characterized by the likelihood matrix $P=(f_{ij})_{(n+1)\times(k+1)}$ with entries $f_{ij}:=\mathbb{P}(x_j\mid \theta_i)$. For each $j\in\{0,\ldots,k\}$, let $\bff_j:=(f_{0j},\ldots,f_{nj})$ denote the likelihood vector of $x_j$. Row-stochasticity implies $\sum_{j=0}^{k}\bff_j=\mathbf{1}_{n+1}$. We say that $F$ is \emph{irredundant} if $\{\bff_0,\ldots,\bff_k\}$ is linearly independent, that is, no $\bff_j$ lies in the linear span of the others. Equivalently, the posterior vectors associated with $x_0,\ldots,x_k$ are affinely independent under any full-support prior.

\begin{proposition}
\label{prop:9}
Consider $F$ and $G$ with finite signal spaces. If $F$ is irredundant, then
\[
F\succeq_{\textup{LB}}G \iff F\succeq_{\textup{B}}G.
\]
\end{proposition}

\cref{prop:9} follows from the proof of Theorem 2 in \citet{wu2023geometric}. Specifically, if $F$ is irredundant and $F \nsucceq_{\textup{B}} G$, Wu constructs a binary-action decision problem in which the DM attains a strictly higher ex-ante payoff under $G$ than under $F$. By \cref{M-lem:2}, this yields $F \nsucceq_{\textup{LB}} G$.

\subsection{Coincidence of Equivalences}

Define \(F\sim_{\textup{LB}}G\) if \(F\succeq_{\textup{LB}}G\) and \(G\succeq_{\textup{LB}}F\).
Analogously, \(F\sim_{\textup{B}}G\) if \(F\succeq_{\textup{B}}G\) and \(G\succeq_{\textup{B}}F\).

\begin{proposition}\label{prop:10}
\(F\sim_{\textup{LB}}G \iff F\sim_{\textup{B}}G\).
\end{proposition}
\begin{proof}
    By \cref{M-prop:3}, we have \(\cz(F)=\cz(G)\). By Theorem 2.1 of \citet{koshevoy1997zonoid}, equality of lift-zonoids implies that \(F\) and \(G\) yield the same posterior distributions for every prior and the same distribution of likelihood-ratio vectors. Hence \(F\sim_{\textup{B}}G\).  The converse is immediate, since \(\succeq_{\textup{B}}\) implies \(\succeq_{\textup{LB}}\).
\end{proof}

\section{General Cost Functions}
\label{ap:i}

In \cref{M-ap:c}, we assume that the cost function $c(\cdot)$ is differentiable. In this section, we show that \cref{M-thm:2} holds for any convex cost function. As discussed in \cref{M-sec:5.1}, the convexity assumption is without loss of generality.

\subsection{Continuity}

For a convex cost function $c(\cdot)$, it is continuous on $\operatorname{int}\td_n$. For each $\bdl\in\partial\td_n$, define $\overline{c}(\bdl):=\liminf_{\tilde{\bdl}\to \bdl,\,\tilde{\bdl}\in\operatorname{int}\td_n} c(\tilde{\bdl})$. By convexity, $c(\bdl)\ge \overline{c}(\bdl)$. Suppose there exists $\bdl^*\in\partial\td_n$ at which $c(\cdot)$ is not continuous, that is, $c(\bdl^*)>\overline{c}(\bdl^*)$. We claim that $\bdl^*$ is not implementable under either experiment. Recall that
\begin{equation}
\label{eqn:38}
J(\bdl^*)=\int_\cX \hw(x)\, dF(x\mid \bdl^*)\;-\;c(\bdl^*).
\end{equation}
The first term is continuous in $\bdl^*$ (it is linear in $\bdl$), so for each scheme $\hw\in\mathcal{W}^{\bu}_{F}$ and each $\varepsilon>0$, there exists $\tilde{\bdl}\in\operatorname{int}\td_n$ with $\tilde{\bdl}\approx\bdl^*$ such that $\bigl|\int_\cX \hw(x)\, dF(x\mid \bdl^*)-\int_\cX \hw(x)\, dF(x\mid \tilde{\bdl})\bigr|<\varepsilon$, and by $c(\bdl^*)>\overline{c}(\bdl^*)$ we may choose $\tilde{\bdl}$ so that $c(\tilde{\bdl})<c(\bdl^*)$. Therefore, for sufficiently small $\varepsilon$, $J(\bdl^*)<J(\tilde{\bdl})$, so $\bdl^*$ cannot satisfy IC for any scheme. Thus $\bdl^*$ is not implementable under $F$ or $G$, and $I_F(\bdl^*,\mathcal{M})=I_G(\bdl^*,\mathcal{M})=+\infty$. Consequently, \cref{M-thm:2} holds trivially for $\bdl^*$. We can therefore restrict attention to cost functions $c(\cdot)$ that are continuous on $\td_n$.

\subsection{Differentiability}

We now assume that $c(\cdot)$ is continuous on $\td_n$. In \cref{M-ap:c}, we rely on the differentiability of $c$ to represent the IC constraint \eqref{M-eqn:8} by $n$ constraints that are linear in $\hw$ (that is, \eqref{M-eqn:IC_i} or \eqref{M-eqn:IC^0_i} defined in \cref{M-ap:c1} for $i\in\{1,\ldots,n\}$). We now employ an approach based on subgradients.

\subsubsection*{IC on Subgradients}
Note that 
\begin{equation*}
\begin{split}
    \int_\cX \hw(x)\, dF(x\mid \bdl)&=\sum_{i=1}^n\int_\cX \hw(x)\,\delta_i f(x\mid\theta_i)\,dx+\int_\cX \hw(x)\,\Bigl(1-\sum_{i=1}^{n}\delta_i\Bigr)f(x\mid\theta_0)\,dx\\
    &=\int_\cX \hw(x)f(x\mid\theta_0)\,dx+\sum_{i=1}^n\delta_i\int_{\cX} \hw(x)\,[f(x\mid\theta_i)-f(x\mid\theta_0)]\,dx\\
    &=D_0(\hw)+\bd(\hw)\cdot\bdl,
\end{split}
\end{equation*}
where $D_0(\hw):=\int_{\cX} \hw(x)f(x\mid\theta_0)\,dx$ and $\bd(\hw):=(D_1(\hw),\ldots,D_n(\hw))$ with $D_i(\hw):=\int_{\cX}\hw(x)\,[f(x\mid\theta_i)-f(x\mid\theta_0)]\,dx$ for each $i\in\{1,\ldots,n\}$.

Using \eqref{eqn:38}, we have 
\begin{equation*}
    J(\bdl)=D_0(\hw)+\bd(\hw)\cdot\bdl-c(\bdl).
\end{equation*}
Rewrite the IC constraint \eqref{M-eqn:8} as
\begin{equation*}
    J(\bdls)\ge J(\bdl')\quad \forall \bdl'\in\td_n.
\end{equation*}
Let $K(\bdl):=-J(\bdl)+\iota_{\td_n}(\bdl)$, where $\iota_{\td_n}(\bdl)=0$ if $\bdl\in\td_n$ and $\iota_{\td_n}(\bdl)=+\infty$ otherwise. Then the IC constraint becomes
\begin{equation}
\label{eqn:39}
    K(\bdls)\le K(\bdl')\quad \forall \bdl'\in \mr^n.
\end{equation}
Note that $K$ is convex. Define its subdifferential by
\begin{equation*}
    \partial K(\bdl):=\{\bb\in\mr^n\mid K(\bdl')\ge K(\bdl)+\bb\cdot(\bdl'-\bdl)\ \forall \bdl'\}.
\end{equation*}
The IC constraint \eqref{eqn:39} is equivalent to
\begin{equation}
\label{eqn:40}
    \bo\in\partial K(\bdls).
\end{equation}
Using the Moreau–Rockafellar sum rule,
\begin{equation*}
    \partial K(\bdl)=\partial c(\bdl)+\partial[-D_0(\hw)-\bd(\hw)\cdot\bdl]+\partial \iota_{\td_n}(\bdl),
\end{equation*}
where “$+$” denotes the Minkowski sum. Moreover,
\begin{equation*}
    \partial[-D_0(\hw)-\bd(\hw)\cdot\bdl]=\{-\bd(\hw)\}.
\end{equation*}
Hence, the IC constraint \eqref{eqn:40} is equivalent to
\begin{equation}
\label{eqn:41}
    \bd(\hw)\in \partial c(\bdls)+\partial \iota_{\td_n}(\bdls).
\end{equation}
Finally, given $\bdls\in\td_n$,
\begin{align}
     \partial\iota_{\td_n}(\bdls)&=\{\bb\in\mr^n\mid \iota_{\td_n}(\bdl')\ge \iota_{\td_n}(\bdls)+\bb\cdot(\bdl'-\bdls)\ \forall \bdl'\in \mr^n\}\\
 &=\{\bb\in\mr^n\mid \iota_{\td_n}(\bdl')\ge \iota_{\td_n}(\bdls)+\bb\cdot(\bdl'-\bdls)\ \forall \bdl'\in\td_n \} \label{eqn:43}\\
 &=\{\bb\in\mr^n\mid \bb\cdot(\bdl'-\bdls)\le 0 \ \forall \bdl'\in\td_n \}. \label{eqn:44}
\end{align}
Here \eqref{eqn:43} holds since $\iota_{\td_n}(\bdl')=+\infty$ for $\bdl'\notin\td_n$, and \eqref{eqn:44} follows since $\iota_{\td_n}(\bdl')=0$ for $\bdl'\in\td_n$.

\subsubsection*{Interior Action}

Fix $\bdls\in\operatorname{int}\td_n$. By \eqref{eqn:44}, we have $\partial\iota_{\td_n}(\bdls)=\{\bo\}$. Hence, by \eqref{eqn:41}, the IC constraint is
\begin{equation}
\label{eqn:45}
    \bd(\hw)\in \partial c(\bdls).
\end{equation}
We can replace \eqref{eqn:45} by one inclusion constraint,
\begin{equation*}
    \bb=(b_1,\ldots,b_n)\in \partial c(\bdls),
    \tag{$IC_{in}$}\label{eqn:IC_{in}}
\end{equation*}
and $n$ constraints \eqref{eqn:IC_i(b_i)} for each $i\in\{1,\ldots,n\}$,
\begin{equation*}
    \int_{\cX} \hw(x)\,[f(x\mid\theta_i)-f(x\mid\theta_0)]\,dx=b_i.
    \tag{$IC_i(b_i)$}\label{eqn:IC_i(b_i)}
\end{equation*}
We can then prove $(\textup{i}) \Rightarrow (\textup{ii})$ of \cref{M-thm:2} by the expectation-matching argument in \cref{M-ap:c2}. If $F\succeq_{\textup{LB}}G$, then for each $\bb\in\partial c(\bdls)$, whenever there exists $\hw$ under $G$ satisfying all \eqref{eqn:IC_i(b_i)}, there exists $\hw$ under $F$ satisfying all \eqref{eqn:IC_i(b_i)}.

The implications $(\textup{ii}) \Rightarrow (\textup{i})$ and $(\textup{iii}) \Rightarrow (\textup{i})$ still hold. For the implication $(\textup{i}) \Rightarrow (\textup{iii})$, note that we can absorb the inclusion constraint \eqref{eqn:IC_{in}} into the principal's objective, which now is
\begin{equation*}
    \inf_{\hw\in\mathcal W_F^{\bu},\ \bb\in\mr^n}\ \int_{\cX} \gamma_{\bu}\!\big(\hw(x)\big)\,dF(x\mid\bdls)\;+\;\iota_{\partial c(\bdls)}(\bb)
\end{equation*}
subject to the IR constraint \eqref{M-eqn:7} and the IC constraints \eqref{eqn:IC_i(b_i)} for each $i$, where $\iota_{\partial c(\bdls)}(\bb)=0$ if $\bb\in\partial c(\bdls)$ and $\iota_{\partial c(\bdls)}(\bb)=+\infty$ otherwise. For each fixed $\bb\in\partial c(\bdls)$, the compactness argument in Case~1 of \cref{M-ap:c3} gives strong duality for the resulting problem with affine constraints. Apply the conjugate-dual comparison to each fixed-$\bb$ problem and then take the infimum over $\bb\in\partial c(\bdls)$ to establish $(\textup{i}) \Rightarrow (\textup{iii})$.

\subsubsection*{Boundary Action}

For simplicity, let $\bdls=\bo$ (the pure action $\theta_0$); the argument is analogous for any $\bdls\in\partial\td_n$. By \eqref{eqn:44}, we have
\begin{equation*}
    \partial\iota_{\td_n}(\bdls)=\{\bb'\in\mr^n \mid b'_i\le 0\ \forall i\}.
\end{equation*}
Hence, if $\bd(\hw)\in \partial c(\bdls)+\partial \iota_{\td_n}(\bdls)$, there must exists $\bb\in \partial c(\bdls)$ such that $D_i(\hw)\leq b_i$ for each $i$. Therefore, we can rewrite the IC constraint \eqref{eqn:41} as one inclusion constraint \eqref{eqn:IC_{in}} together with $n$ constraints \eqref{eqn:IC^0_i(b_i)} for each $i\in\{1,\ldots,n\}$,
\begin{equation*}
    \int_{\cX} \hw(x)\,[f(x\mid\theta_i)-f(x\mid\theta_0)]\,dx\le b_i.
    \tag{$IC^0_i(b_i)$}\label{eqn:IC^0_i(b_i)}
\end{equation*}
We can then establish $(\textup{i}) \Rightarrow (\textup{ii})$ of \cref{M-thm:2} by the expectation-matching argument, and $(\textup{i}) \Rightarrow (\textup{iii})$ by the conjugate-dual approach, exactly as in the interior case.

\section{Zero Densities and Unbounded Likelihood Ratios}
\label{ap:j}

In \cref{M-ap:c}, we assume that $f(x\mid\theta)>0$ and $g(y\mid\theta)>0$ for all $x\in \cX$, $y\in \cY$, and $\theta\in\Theta$, ensuring that likelihood ratios are well defined. We now drop this assumption, allowing densities to vanish and likelihood ratios to be unbounded. We show that \cref{M-thm:2} holds without the positivity assumption on signal densities. 

We start with a technical lemma.

\begin{lemma}
\label{lem:5}
Suppose $\rho:\mr\to\mr$ is convex and let $\rho'_+$ be its right derivative. If $\lim_{t\to-\infty}\rho'_+(t)=\alpha\in\mr$, $\lim_{t\to+\infty}\rho'_+(t)=\beta\in\mr$, and $\lim_{t\to-\infty}[\rho(t)-\alpha t]\in\mr$, then there exists a finite nonnegative Borel measure $\eta$ with $\eta(\mr)=\beta-\alpha$ and $\eta((-\infty,t])=\rho'_+(t)-\alpha$ for each $t\in\mr$, such that
\begin{equation}
\label{eqn:46}
\rho(t)=C+\alpha t+\int_{\mr}(t-s)_{+}\,d\eta(s)\quad\forall t\in\mr,
\end{equation}
where $C$ is a constant.
\end{lemma}

\begin{proof}
Let $\eta$ be the finite nonnegative Lebesgue--Stieltjes measure determined by $\eta((-\infty,t])=\rho'_+(t)-\alpha$. For each $t\ge 0$,
\begin{equation*}
\begin{split}
\rho(t)-\rho(0)
&=\int_{0}^{t}\rho'_+(r)\,dr
=\int_{0}^{t}\big[\alpha+\eta((-\infty,r])\big]\,dr\\
&=\alpha t+\int_{0}^{t}\eta((-\infty,r])\,dr
=\alpha t+\int_{0}^{t}\int_{\mr}\mathbf{1}_{\{s\mid s\le r\}}\,d\eta(s)\,dr\\
&=\alpha t+\int_{\mr}\int_{0}^{t}\mathbf{1}_{\{s\mid s\le r\}}\,dr\,d\eta(s)
=\alpha t+\int_{\mr}\big[(t-s)_{+}-(0-s)_{+}\big]\,d\eta(s).
\end{split}
\end{equation*}
The same identity holds for $t<0$. Letting $t\to-\infty$, the additional hypothesis and monotone convergence imply $\int_{\mr}(-s)_{+}\,d\eta(s)<\infty$. Set $C=\rho(0)-\int_{\mr}(0-s)_{+}\,d\eta(s)$. Then \eqref{eqn:46} follows.
\end{proof}

\newcommand{\oK}{\overline{K}}
\newcommand{\ok}{\overline{k}}

We now revisit Case 1 of \cref{M-ap:c1}, where we employ the positivity assumption on signal densities. Using \eqref{M-eqn:19}, under $G$, the Lagrangian is
\begin{equation}
\label{eqn:47}
    L_G(\hat w,\lambda,\boldsymbol\mu)
  = \lambda\,c(\bo) - \sum_{i=1}^n \mu_i c_i(\bo)+\oK_G(\hw,\lambda,\bmu),
\end{equation}
where
\begin{equation*}
    \oK_G(\hw,\lambda,\bmu)=\int_{\cY}\ok_G(y,\hw,\lambda,\bmu)\,dy,
\end{equation*}
\begin{equation*}
    \ok_G(y,\hw,\lambda,\bmu)=\gamma_{\bu}\!\big(\hat w(y)\big)g(y\mid\theta_0)
     - \hat w(y)\Big\{\lambda g(y\mid\theta_0)+\textstyle\sum_{i=1}^n \mu_i\big[g(y\mid\theta_0)-g(y\mid\theta_i)\big]\Big\}.
\end{equation*}

We now derive $\inf_{\hw\in\mathcal W_G^{\bu}}\oK_G(\hw,\lambda,\bmu)$ for each $\lambda \in \mr$ and $\bmu\in\mr^n$. Note that the \eqref{M-eqn:IR} and IC constraints \eqref{M-eqn:IC'^0_i} in \cref{M-ap:c3} are inequalities, so the Lagrangian multipliers are nonnegative in that setting. However, in Case 2 of \cref{M-ap:c3}, we go through variants of the principal’s problem in which some constraints are imposed as equalities, which allow the corresponding Lagrangian multipliers to be negative. Furthermore, for implementing the interior action $\bdls\in\operatorname{int}\td_n$, the IC constraints \eqref{M-eqn:IC_i} in \cref{M-ap:c1} are equalities and the associated multipliers can be negative. Hence, here we also examine the case where Lagrangian multipliers may be negative.

Let $\cY^0=\{y\in \cY \mid g(y\mid \theta_0)=0\}$ and $\cY^+=\cY\setminus \cY^0$. On $\cY^+$, we use the likelihood ratios $\ell_{G,i}(y)=g(y\mid\theta_i)/g(y\mid\theta_0)$ and obtain
\begin{equation*}
\int_{\cY^+}\ok_G (y,\hw,\lambda,\bmu)\,dy=
     \int_{\cY^+} \Big\{ \gamma_{\bu}\!\big(\hat w(y)\big)
     - \hat w(y)\Big[\lambda+\textstyle\sum_{i=1}^n \mu_i\big(1-\ell_{G,i}(y)\big)\Big] \Big\}\,dG(y\mid\theta_0).
\end{equation*}
Using \eqref{M-eqn:20},
\begin{equation}
\label{eqn:48}
    \inf_{\hw\in\mathcal W_G^{\bu}}\int_{\cY^+}\ok_G(y,\hw,\lambda,\bmu)\,dy
    = -\int_{\cY^+} \rho\!\Big(\lambda+\sum_{i=1}^n \mu_i\big(1-\ell_{G,i}(y)\big)\Big)\,dG(y\mid\theta_0),
\end{equation}
where
\begin{equation*}
    \rho(t)=\sup_{w\in\mathbb{R}}[t w-\gamma_{\bu}(w)]
  \;=\;\sup_{w\in[\underline u,\overline u]}[t w-\gamma(w)].
\end{equation*}
Note that $\rho(\cdot)$ is convex. Since $\gamma(\cdot)$ is continuous on $[\uu,\ou]$, we have $\lim_{t\to-\infty}\rho'_+(t)=\uu$ and $\lim_{t\to+\infty}\rho'_+(t)=\ou$. Moreover, $\rho(t)-\uu t$ is nondecreasing and bounded below by $-\gamma(\uu)$, so its limit as $t\to-\infty$ is finite. Therefore, by \cref{lem:5},
\begin{equation}
\label{eqn:49}
    \rho(t)=C+\uu t+\int_{\mr}(t-s)_{+}\,d\eta(s)\quad\forall t\in\mr,
\end{equation}
where $\eta(\mr)=\ou-\uu$ and $\eta((-\infty,t])=\rho'_+(t)-\uu$.

Plug \eqref{eqn:49} into \eqref{eqn:48}. After calculation, we obtain
\begin{equation}
\label{eqn:50}
\begin{split}
\int_{\cY^+}\rho\!\Big(\lambda+\sum_{i=1}^n \mu_i\big(1-\ell_{G,i}(y)\big)\Big)\,dG(y\mid\theta_0)
= C+\uu\,\lambda+\uu\sum_{i=1}^{n}\mu_i\,G(\cY^0\mid\theta_i)\\
+\int_{\cY^+}\int_{\mr}\Big(\big(\lambda+\textstyle\sum_{i=1}^{n}\mu_i-s\big)g(y\mid\theta_0)-\textstyle\sum_{i=1}^{n}\mu_i\,g(y\mid\theta_i)\Big)_{+}\,d\eta(s)\,dy.
\end{split}
\end{equation}
where $G(\cY^0\mid\theta_i)=\int_{\cY^0}dG(y\mid\theta_i)$ for each $i$.

On $\cY^0$, since $g(y\mid\theta_0)=0$, we have
\begin{equation*}
    \ok_G(y,\hw,\lambda,\bmu)=\hat w(y)\,\sum_{i=1}^n \mu_i\, g(y\mid\theta_i).
\end{equation*}
Hence the pointwise minimization over $w\in[\uu,\ou]$ yields
\begin{equation*}
    \inf_{\hw\in\mathcal W_G^{\bu}}\int_{\cY^0}\ok_G(y,\hw,\lambda,\bmu)\,dy
    = \int_{\cY^0}\min_{w\in[\uu,\ou]}\Big\{w\,\textstyle\sum_{i=1}^n \mu_i g(y\mid\theta_i)\Big\}\,dy.
\end{equation*}
Writing $S(y):=\sum_{i=1}^n \mu_i g(y\mid\theta_i)$ and $s_-:=\min\{s,0\}$, we have
\begin{equation*}
    \min_{w\in[\uu,\ou]}\{w\,S(y)\}=\uu S(y)+(\ou-\uu) S(y)_-.
\end{equation*}
Hence,
\begin{equation}
\label{eqn:51}
    \inf_{\hw\in\mathcal W_G^{\bu}}\int_{\cY^0}\ok_G(y,\hw,\lambda,\bmu)\,dy=\int_{\cY^0}\left[\uu S(y)+(\ou-\uu) S(y)_-\right]dy.
\end{equation}
Furthermore,
\begin{equation*}
    \uu\sum_{i=1}^{n}\mu_iG(\cY^0\mid\theta_i)=
    \int_{\cY^0}\uu S(y)\,dy,
\end{equation*}
and
\begin{equation*}
\begin{split}
    &\int_{\cY^0}\int_{\mr}\left(\left(\lambda+\sum_{i=1}^{n}\mu_i-s\right) g(y\mid \theta_0 )-\sum_{i=1}^{n}\mu_i g(y\mid \theta_i)\right)_{+}d\eta(s)\,dy\\
    &=\int_{\cY^0}\int_{\mr}\left(-\sum_{i=1}^{n}\mu_i g(y\mid \theta_i)\right)_{+}d\eta(s)\,dy
    =\int_{\cY^0}\int_{\mr}\big[-S(y)_-\big]\,d\eta(s)\,dy
    =-(\ou-\uu)\int_{\cY^0}S(y)_-\,dy.
\end{split}
\end{equation*}
Therefore, by \eqref{eqn:51},
\begin{equation}
\label{eqn:52}
\begin{split}
    &\inf_{\hw\in\mathcal W_G^{\bu}}\int_{\cY^0}\ok_G(y,\hw,\lambda,\bmu)\,dy\\
    =&\uu\sum_{i=1}^{n}\mu_iG(\cY^0\mid\theta_i)-\int_{\cY^0}\int_{\mr}\left(\left(\lambda+\sum_{i=1}^{n}\mu_i-s\right) g(y\mid \theta_0 )-\sum_{i=1}^{n}\mu_i g(y\mid \theta_i)\right)_{+}d\eta(s)\,dy.
\end{split}
\end{equation}

By \eqref{eqn:48}, \eqref{eqn:50} and \eqref{eqn:52},
\begin{equation}
    \label{eqn:53}
    \begin{split}
    &\inf_{\hw\in\mathcal W_G^{\bu}} \oK_G(\hw,\lambda,\bmu)=\inf_{\hw\in\mathcal W_G^{\bu}} \int_{\cY}\ok_G (y,\hw,\lambda,\bmu)\,dy\\
    =&\inf_{\hw\in\mathcal W_G^{\bu}}\int_{\cY^+}\ok_G (y,\hw,\lambda,\bmu)\,dy
     +\inf_{\hw\in\mathcal W_G^{\bu}}\int_{\cY^0}\ok_G (y,\hw,\lambda,\bmu)\,dy\\
    =&-C-\uu\lambda-\int_{\cY}\int_{\mr}\Big(\big(\lambda+\sum_{i=1}^{n}\mu_i-s\big)g(y\mid\theta_0)-\sum_{i=1}^{n}\mu_i\,g(y\mid\theta_i)\Big)_{+}\,d\eta(s)\,dy.
    \end{split}
\end{equation}

Now suppose $F\succeq_{\textup{LB}} G$. By \cref{M-lem:1}, for each $\lambda\in\mr$, $\bmu\in\mr^n$, and $s\in\mr$,
\begin{equation*}
\begin{split}
    \int_{\cX}\Big(\big(\lambda+\textstyle\sum_{i=1}^{n}\mu_i-s\big)f(x\mid\theta_0)-\textstyle\sum_{i=1}^{n}\mu_i\,f(x\mid\theta_i)\Big)_{+}\,dx\\
    \geq \int_{\cY}\Big(\big(\lambda+\textstyle\sum_{i=1}^{n}\mu_i-s\big)g(y\mid\theta_0)-\textstyle\sum_{i=1}^{n}\mu_i\,g(y\mid\theta_i)\Big)_{+}\,dy.
\end{split}
\end{equation*}
Hence, by \eqref{eqn:53},
\begin{equation*}
    \inf_{\hw\in\mathcal W_F^{\bu}} \oK_F(\hw,\lambda,\bmu)\;\leq\; \inf_{\hw\in\mathcal W_G^{\bu}} \oK_G(\hw,\lambda,\bmu)\quad \forall\,\lambda\in\mr,\ \bmu\in\mr^n.
\end{equation*}
Then, by \eqref{eqn:47},
\begin{equation*}
    \inf_{\hw\in\mathcal W_F^{\bu}} L_F(\hat w,\lambda,\bmu)\;\leq\;\inf_{\hw\in\mathcal W_G^{\bu}} L_G(\hat w,\lambda,\bmu)\quad \forall\,\lambda\in\mr,\ \bmu\in\mr^n.
\end{equation*}
Therefore,
\begin{equation*}
    \sup_{\lambda\geq 0,\ \bmu\geq 0}\ \inf_{\hat w\in\mathcal W_F^{\bu}}\,L_F(\hat w,\lambda,\bmu)\ \leq\ \sup_{\lambda\geq 0,\ \bmu\geq 0}\ \inf_{\hat w\in\mathcal W_G^{\bu}}\,L_G(\hat w,\lambda,\bmu).
\end{equation*}
By the strong duality established in \cref{M-ap:c3}, we conclude that $I_F(\bo,\mathcal{M}) \leq I_G(\bo,\mathcal{M})$.

\section{Proof of Theorem 3}
\label{ap:d}

For simplicity, we assume that $f(x|\theta)>0$ and $g(y|\theta)>0$ for all $x\in \cX$, $y\in \cY$, and $\theta\in\Theta$, ensuring that likelihood ratios are well defined. The proof for the general case follows directly from the arguments presented in \cref{ap:j}.

\subsection{An Auxiliary Result}
\label{ap:d1}

Given a screening environment $\mathcal{E}=\{\cA,\bq,\mathcal{P},\pi,\psi,v_1,v_2,u_1,\bmm\}\in\mathfrak{E}$, an allocation rule $\hat a:\mathcal{P}\to\Delta(\cA)$ is \emph{implementable} under $F$ if there exists a payment rule $\hat t:\mathcal{P}\times \cA\times \oX\to\mathbb{R}$ such that $(\hat a,\hat t)$ satisfy IR \eqref{M-eqn:11} and IC \eqref{M-eqn:12}. For fixed $\hat a$, let $\hat{A}(\bp)$ denote the random variable distributed according to $\hat{a}(\bp)$ for each $\bp\in\cp$, and define
\begin{equation*}
 D_F(\hat a,\mathcal{E})=\inf_{\hat t\in\mathcal{T}_F^{\bmm}}\mathbb{E}\!\left\{\mathbb{E}\!\left[\mathbb{E}\big[v_2(\hat t(\bP,\hat{A}(\bP),x))\mid\tilde{\theta},\hat{A}(\bP)\big]\right]\middle | \bP\right\},
\end{equation*}
subject to \eqref{M-eqn:11}–\eqref{M-eqn:12}, with $D_F(\hat a,\mathcal{E})=+\infty$ if $\hat a$ is not implementable.

\begin{proposition}\label{prop:4}
$F\succeq_{\textup{LB}}G\iff D_F(\hat a,\mathcal{E})\le D_G(\hat a,\mathcal{E})\quad\forall\,\mathcal{E}\in\mathfrak{E},\ \hat a\in(\Delta(\cA))^{\mathcal{P}}$.
\end{proposition}

\begin{proof}
\textbf{Step 1: “$\Rightarrow$”.} Consider the simplest case: the agent observes the state $\theta$ and the ex post signal is always informative ($\psi(a)\equiv1$). Under $G$, the principal commits to $\hat a:\Theta\to\Delta(\cA)$ and $\hat t:\Theta\times \cY\to\mathbb{R}$. Fold the constraint $\hat t\in[\underline m,\overline m]$ into the objective via
\begin{equation*}
  v_2^{\bmm}(t):=\begin{cases} v_2(t), & t\in[\underline m,\overline m],\\[2pt] +\infty,& \text{otherwise}.\end{cases}
\end{equation*}
For a deterministic $\hat a$,\footnote{The randomized case is identical after replacing $u_1(\hat a(\theta_j),\theta_i)$ by $\mathbb{E}[u_1(\hat{A}(\theta_j),\theta_i)]$, where $\hat{A}(\theta_j)$ is a random variable distributed according to $\ha(\theta_j)$.} the problem under $G$ is
\begin{align*}
  &D_G(\hat a,\mathcal{E})=\inf_{\hat t\in\mathcal{T}_G^{\bmm}}\ \sum_{i=0}^{n}q_i\!\int_{\cY} v_2^{\bmm}\big(\hat t(\theta_i,y)\big)\,dG(y\mid\theta_i)\\
  \text{s.t. }&\ u_1(\hat a(\theta_i),\theta_i)+\int_{\cY}\hat t(\theta_i,y)\,dG(y\mid\theta_i)\ge 0,\quad i=0,\ldots,n,\tag{$IR_i$}\label{eqn:IR_i}\\
  &\ u_1(\hat a(\theta_i),\theta_i)+\!\int_{\cY}\!\hat t(\theta_i,y)dG(y\mid\theta_i) \ge u_1(\hat a(\theta_j),\theta_i)+\!\int_{\cY}\!\hat t(\theta_j,y)dG(y\mid\theta_i),\ \ \forall\, i\ne j. \tag{$IC_{i,j}$}\label{eqn:IC_{i,j}}
\end{align*}

Let $\lambda_i\ge0$ denote multipliers for \eqref{eqn:IR_i} and $\mu_{i,j}\ge0$ for \eqref{eqn:IC_{i,j}}. Writing $\ell^{(i)}_{G,j}(y):=g(y\mid\theta_j)/g(y\mid\theta_i)$ and collecting terms in $\hat t(\theta_i,\cdot)$ yields
\begin{align*}
  &L_G(\hat t,\blb,\bmu)=\sum_{i=0}^n\left\{\tilde L_{G,i}\big(\hat t(\theta_i,\cdot),\lambda_i,\bmu\big)-\lambda_i u_1(\hat a(\theta_i),\theta_i)+\sum_{j\ne i}\mu_{i,j}\big[u_1(\hat a(\theta_j),\theta_i)-u_1(\hat a(\theta_i),\theta_i)\big]\right\},\\
  &\tilde L_{G,i}\big(\hat t(\theta_i,\cdot),\lambda_i,\bmu\big)=\int_{\cY}\!\Big\{q_i v_2^{\bmm}(\htt(\theta_i,y))-\hat t(\theta_i,y)\Big[\lambda_i+\!\sum_{j\ne i}\mu_{i,j}-\!\sum_{j\ne i}\mu_{j,i}\ell^{(i)}_{G,j}(y)\Big]\Big\}\,dG(y\mid\theta_i).
\end{align*}
Assume strict feasibility under $G$. Then strong duality holds, so
\begin{equation*}
  D_G(\hat a,\mathcal{E})= \inf_{\hat t\in\mathcal{T}_G^{\bmm}}\ \sup_{\blb\ge0,\bmu\ge0}\ L_G(\hat t,\blb,\bmu)=\sup_{\blb\ge0,\bmu\ge0}\ \inf_{\hat t\in\mathcal{T}_G^{\bmm}}\ L_G(\hat t,\blb,\bmu).
\end{equation*}
Taking the pointwise infimum in $\hat t(\theta_i,\cdot)$ introduces the convex conjugate of $q_i v_2^{\bmm}$,
\begin{equation*}
  \rho_i(k):=\sup_{t\in[\underline m,\overline m]}\{kt-q_i v_2(t)\},
\end{equation*}
and gives
\begin{align*}
  D_G(\hat a,\mathcal{E})=\sup_{\blb\ge0,\bmu\ge0}\ \sum_{i=0}^n\Big\{K_{G,i}(\lambda_i,\bmu)-\lambda_i u_1(\hat a(\theta_i),\theta_i)+\sum_{j\ne i}\mu_{i,j}\big[u_1(\hat a(\theta_j),\theta_i)-u_1(\hat a(\theta_i),\theta_i)\big]\Big\},
\end{align*}
where
\begin{equation*}
  K_{G,i}(\lambda_i,\bmu)=-\int_{\cY} \rho_i\!\Big(\lambda_i+\sum_{j\ne i}\mu_{i,j}-\sum_{j\ne i}\mu_{j,i}\,\ell^{(i)}_{G,j}(y)\Big)\,dG(y\mid\theta_i).
\end{equation*}

Suppose $F\succeq_{\textup{LB}}G$. By \cref{M-prop:3} and the expectation-matching argument in \cref{M-ap:c2}, for each $\htt\in\mathcal{T}^{\bmm}_G$ there exists $\htt'\in\mathcal{T}^{\bmm}_F$ such that $\mathbb{E}[\htt'(\theta_i,X)\mid \theta_j]=\mathbb{E}[\htt(\theta_i,Y)\mid \theta_j]$, for every $\theta_i,\theta_j\in\Theta$. Hence, strict feasibility under $G$ carries over to $F$. Moreover, as discussed in Section 3.1, LB dominance is invariant to the choice of baseline state: for each $i$, the likelihood–ratio vectors $\bl^{(i)}_F:=(\ell^{(i)}_{F,j})_{j\ne i}$ and $\bl^{(i)}_G:=(\ell^{(i)}_{G,j})_{j\ne i}$ satisfy $\bl^{(i)}_F\succeq_{\textup{lcx}}\bl^{(i)}_G$. Since $\rho_i$ is convex, it follows that $K_{F,i}(\lambda_i,\bmu)\le K_{G,i}(\lambda_i,\bmu)$ for all $(\lambda_i,\bmu)$, which implies $D_F(\hat a,\mathcal{E})\le D_G(\hat a,\mathcal{E})$.

If strict feasibility fails under $G$, we proceed as in Case~2 of \cref{M-ap:c3}: bind a suitable subset of constraints to obtain a strictly feasible variant, compare the corresponding duals as above, and finally take the minimum over subsets.

Now let the agent draw the belief $\bP$ taking values in $\mathcal P=\{\bp_0,\ldots,\bp_k\}$ with $\bp_i=(p_{i,0},\ldots,p_{i,n})\in\Delta(\Theta)$, keeping $\psi(a)\equiv 1$. Under $G$, the principal commits to $\hat a:\mathcal P\to\Delta(\cA)$ and $\hat t:\mathcal P\times \cY\to\mathbb R$. Define the induced experiment on $\mathcal P$ by $G_{\mathcal P}(\cdot\mid \bp_i):=\sum_{j=0}^n p_{i,j}\,G(\cdot\mid \theta_j)$. Re-run the dual with types in place of states—replace $\hat t(\theta_i,\cdot)$ by $\hat t(\bp_i,\cdot)$; the Lagrangian and the conjugate term are unchanged. By \eqref{M-eqn:3}, LB dominance is preserved, that is, $F\succeq_{\textup{LB}}G \Rightarrow F_{\mathcal P}\succeq_{\textup{LB}}G_{\mathcal P}$. Hence, the previous comparison implies $D_F(\hat a,\mathcal E)\le D_G(\hat a,\mathcal E)$.

With a general $\psi$, under $G$ the principal commits to $\hat a:\mathcal P\to\Delta(\cA)$ and $\hat t:\mathcal P\times \cA\times\overline \cY\to\mathbb R$, where $\overline \cY:=\cY\cup\{\perp\}$. For each $a\in \cA$, define the induced experiment on $\mathcal P$ with an uninformative outcome by $ G_{\mathcal P,a}(\cdot\mid \bp_i):=(1-\psi(a))\,\delta_{\perp}(\cdot)+\psi(a)\,G_{\mathcal P}(\cdot\mid \bp_i)$. Re-run the dual with $\hat t(\bp,a,\cdot)$; the Lagrangian and conjugate terms keep the same form with likelihood ratios of $G_{\mathcal P,a}$. By the Lorenz–zonoid characterization (Section 3.2), adding the state-independent mass at $\perp$  preserves inclusion, so $F_{\mathcal P}\succeq_{\textup{LB}}G_{\mathcal P}\Rightarrow F_{\mathcal P,a}\succeq_{\textup{LB}}G_{\mathcal P,a}$ for each $a\in \cA$, and therefore $D_F(\hat a,\mathcal E)\le D_G(\hat a,\mathcal E)$.

\textbf{Step 2: “$\Leftarrow$”.} Consider first $\Theta=\{\theta_0,\theta_1\}$.  Assume for the sake of contradiction that $F \nsucceq_{\textup{LB}} G$. By \cref{M-prop:3}, $\mathcal{Z}(F)\not\supset \mathcal{Z}(G)$. Since $\mathcal{Z}(F)$ and $\mathcal{Z}(G)$ are centrally symmetric about $(\tfrac12,\tfrac12)$, pick $\hat{\bz}\in \mathcal{Z}(G)\setminus\mathcal{Z}(F)$ and set $\hat{\bz}'=(1,1)-\hat{\bz}\in\mathcal{Z}(G)$. As $\mathcal{Z}(F)$ is compact and convex, the separating hyperplane theorem yields $\bb=(b_0,b_1)$ with
\begin{equation*}
    \bb\cdot\hat{\bz} \;>\; \sup_{\bz\in \mathcal{Z}(F)}\bb\cdot\bz
    \quad\text{and}\quad
    \bb\cdot\hat{\bz}' \;<\; \inf_{\bz\in \mathcal{Z}(F)}\bb\cdot\bz.
\end{equation*}
Hence, for all $\bz,\bz'\in \mathcal{Z}(F)$,
\begin{equation}
\label{eqn:23}
    \bb\cdot\hat{\bz}-\bb\cdot\hat{\bz}' \;>\; \bb\cdot\bz-\bb\cdot\bz'.
\end{equation}
Because $\mathcal{Z}(F),\mathcal{Z}(G)\subset[0,1]^2$ and contain $(1,1)$, the separator must have mixed signs; without loss, take $b_0>0>b_1$.

Let the agent observe $\theta$ and take $\psi\equiv 1$, $\bmm=(0,1)$, and $\cA=\{a_0,a_1\}$. Choose $u(a,\theta)$ so that
\begin{equation}
\label{eqn:24}
    u(a_0,\theta_0)\ge -\hat z_0,\qquad u(a_1,\theta_1)\ge -\hat z'_1,
\end{equation}
\begin{equation}
\label{eqn:25}
    u(a_1,\theta_0)-u(a_0,\theta_0)=\hat z_0-\hat z'_0,\qquad
    u(a_0,\theta_1)-u(a_1,\theta_1)=\hat z'_1-\hat z_1.
\end{equation}
Consider $\hat a$ with $\hat a(\theta_0)=a_0$ and $\hat a(\theta_1)=a_1$. Since $\hat{\bz},\hat{\bz}'\in\mathcal{Z}(G)$, there exists $\hat t:\Theta\times \cY\to[0,1]$ with
\begin{align*}
\int_{\cY} \hat t(\theta_0,y)\,dG(y\mid\theta_0)=\hat z_0,\ \ \int_{\cY} \hat t(\theta_0,y)\,dG(y\mid\theta_1)=\hat z_1, \\
    \int_{\cY} \hat t(\theta_1,y)\,dG(y\mid\theta_0)=\hat z'_0,\ \ \int_{\cY} \hat t(\theta_1,y)\,dG(y\mid\theta_1)=\hat z'_1.
\end{align*} 
Then \eqref{eqn:24} ensures \eqref{eqn:IR_i}, and \eqref{eqn:25} ensures \eqref{eqn:IC_{i,j}} for $(i,j)=(0,1)$ and $(1,0)$, so $\hat a$ is implementable under $G$. We claim there is no $\hat t':\Theta\times \cX\to[0,1]$ implementing $\hat a$ under $F$. Otherwise, define $\tilde{\bz}=(\tilde z_0,\tilde z_1)$ and $\tilde{\bz}'=(\tilde z'_0,\tilde z'_1)$ by
\begin{align}
    \int_{\cX} \hat t'(\theta_0,x)\,dF(x\mid\theta_0)=\tilde z_0,\ \ \int_{\cX} \hat t'(\theta_0,x)\,dF(x\mid\theta_1)=\tilde z_1,\label{eqn:26}\\
    \int_{\cX} \hat t'(\theta_1,x)\,dF(x\mid\theta_0)=\tilde z'_0,\ \ \int_{\cX} \hat t'(\theta_1,x)\,dF(x\mid\theta_1)=\tilde z'_1.\label{eqn:27}
\end{align}
IC under $F$ gives
\begin{equation*}
    \tilde z_0-\tilde z'_0 \;\ge\; u(a_1,\theta_0)-u(a_0,\theta_0)=\hat z_0-\hat z'_0,\qquad
    \tilde z'_1-\tilde z_1 \;\ge\; u(a_0,\theta_1)-u(a_1,\theta_1)=\hat z'_1-\hat z_1.
\end{equation*}
With $b_0>0>b_1$, this implies
\begin{equation*}
    \bb\cdot\tilde{\bz}-\bb\cdot\tilde{\bz}' \;=\; b_0(\tilde z_0-\tilde z'_0)+b_1(\tilde z_1-\tilde z'_1)
    \;\ge\; b_0(\hat z_0-\hat z'_0)+b_1(\hat z_1-\hat z'_1)
    \;=\; \bb\cdot\hat{\bz}-\bb\cdot\hat{\bz}',
\end{equation*}
contradicting \eqref{eqn:23}. Thus, $\hat a$ is implementable under $G$ but not under $F$, so $D_F(\hat a,\mathcal E)=+\infty$ and therefore $D_G(\hat a,\mathcal E)<D_F(\hat a,\mathcal E)$, a contradiction.

\begin{remark}
\label{rem:2}
There does not exist a sequence of allocation rules $\{\hat a_k\}_{k\ge1}$ implementable under $F$ such that $\lim_{k\to\infty}\mathbb{P}(\hat{A}_k(\theta_0)=a_0)=1$ and $\lim_{k\to\infty}\mathbb{P}(\hat{A}_k(\theta_0)=a_1)=1$, where $\hat{A}_k(\theta_0)$ and $\hat{A}_k(\theta_1)$ are random varibales distributed accordingg to $\ha_k(\theta_0)$ and $\ha_k(\theta_1)$, respectively. Otherwise, for each $k$ let $\hat t_k'$ implement $\hat a_k$ under $F$, and define $\tilde{\bz}_k=(\tilde z_{k,0},\tilde z_{k,1})$ and $\tilde{\bz}_k'=(\tilde z'_{k,0},\tilde z'_{k,1})$ by \eqref{eqn:26} and \eqref{eqn:27} analogously. Then $\lim_{k\to\infty}\big(\bb\cdot\tilde{\bz}_k-\bb\cdot\tilde{\bz}_k'\big)=\bb\cdot\hat{\bz}-\bb\cdot\hat{\bz}'$. By compactness of $\cz(F)$, passing to a subsequence if necessary, there exist $\tilde{\bz}_\infty,\tilde{\bz}'_\infty\in\cz(F)$ with $\bb\cdot\tilde{\bz}_\infty-\bb\cdot\tilde{\bz}'_\infty=\bb\cdot\hat{\bz}-\bb\cdot\hat{\bz}'$, contradicting \eqref{eqn:23}.
\end{remark}

Now let $\Theta=\{\theta_0,\ldots,\theta_n\}$. Assume for the sake of contradiction that $F\nsucceq_{\textup{LB}}G$. By \eqref{M-eqn:3}, there exists $\bb=(b_0,\ldots,b_n)\in\mathbb R^{n+1}$ such that
\begin{equation}
\label{eqn:28}
    \int_{\cX}\!\Big(\textstyle\sum_{i=0}^{n} b_i f(x\mid\theta_i)\Big)_{+}\, dx
    \;<\;
    \int_{\cY}\!\Big(\textstyle\sum_{i=0}^{n} b_i g(y\mid\theta_i)\Big)_{+}\, dy.
\end{equation}
Since $\int_{\cX} f(x\mid\theta_i)\,dx=\int_{\cY} g(y\mid\theta_i)\,dy=1$ for each $i$, there must be indices $j,k$ with $b_j>0$ and $b_k<0$. Relabel so that for some cutoff $\kappa\in\{0,\ldots,n-1\}$ we have $b_i\le 0$ for $i\le \kappa$ and $b_i>0$ for $i>\kappa$. Let $B_0:=\sum_{i\le \kappa} b_i$ and $B_1:=\sum_{i>\kappa} b_i$. Define weights $\alpha_i:=b_i/B_0$ for $i\le\kappa$ and $\beta_i:=b_i/B_1$ for $i>\kappa$. For $H\in\{F,G\}$, construct the binary experiment on $\Omega=\{\omega_0,\omega_1\}$ by $H_{\Omega}(\cdot\mid \omega_0):=\sum_{i\le \kappa}\alpha_i\,H(\cdot\mid \theta_i)$ and $H_{\Omega}(\cdot\mid \omega_1):=\sum_{i> \kappa}\beta_i\,H(\cdot\mid \theta_i)$, and let $f_{\Omega}$ and $g_{\Omega}$ denote the densities of $F_{\Omega}$ and $G_{\Omega}$. From \eqref{eqn:28},
\begin{equation*}
\int\!\Big(B_0 f_{\Omega}(x\mid\omega_0)+B_1 f_{\Omega}(x\mid\omega_1)\Big)_{+}\,dF_{\Omega}(x\mid\omega_0)
\;<\;
\int\!\Big(B_0 g_{\Omega}(x\mid\omega_0)+B_1 g_{\Omega}(x\mid\omega_1)\Big)_{+}\,dG_{\Omega}(y\mid\omega_0),
\end{equation*}
so $F_{\Omega}\nsucceq_{\textup{LB}}G_{\Omega}$ by \eqref{M-eqn:3}.

Take $\mathcal P=\{\bp_0,\bp_1\}$ with equal prior, where $\bp_0=(\alpha_0,\ldots,\alpha_{\kappa},0,\ldots,0)$ and $\bp_1=(0,\ldots,0,\beta_{\kappa+1},\ldots,\beta_n)$. Let $\psi\equiv1$, $\bmm=(0,1)$, $\cA=\{a_0,a_1\}$, and choose $u_1(a,\theta)$ constant within each group $\{\theta_i:i\le\kappa\}$ and $\{\theta_i:i>\kappa\}$. Consider $\hat a:\mathcal P\to \cA$ with $\hat a(\bp_0)=a_0$ and $\hat a(\bp_1)=a_1$. Then IR/IC depend only on the groupwise expectations, so the problem reduces to the binary experiments $(F_{\Omega},G_{\Omega})$. By the binary-state argument, since $F_{\Omega}\nsucceq_{\textup{LB}}G_{\Omega}$, we can choose the (group-constant) values of $u_1(a,\theta)$ so that there exists a payment rule under $G$ implementing $\hat a$, but none under $F$. Therefore, $D_G(\hat a,\mathcal E) < D_F(\hat a,\mathcal E)$.

\end{proof}

\subsection{Proof of Theorem 3}

The “$\Rightarrow$” direction follows immediately from \cref{prop:4}. For “$\Leftarrow$”, suppose $\Theta=\{\theta_0,\theta_1\}$. Assume for the sake of contradiction that $F\nsucceq_{\textup{LB}}G$. Let $\mathcal E$ be the screening environment constructed in Step~2 of the proof of \cref{prop:4}. The allocation rule $\ha$ with $\ha(\theta_0)=a_0$ and $\ha(\theta_1)=a_1$ is implementable under $G$ but not under $F$. Let $v_2(t)\equiv 0$ and set $v_1(a_0,\theta)=\mathbf{1}\{\theta=\theta_0\}$ and $v_1(a_1,\theta)=\mathbf{1}\{\theta=\theta_1\}$. Under $G$, the allocation rule $\ha$ yields an expected payoff of $1$ for the principal, whereas under $F$, every implementable allocation rule delivers a strictly lower payoff. By \cref{rem:2}, no sequence of implementable allocation rules under $F$ can approximate $\ha$; hence $W_G(\mathcal E)>W_F(\mathcal E)$, a contradiction. The case for a general $\Theta=\{\theta_0,\ldots,\theta_n\}$ is analogous, using the grouping construction in Step~2 of the proof of \cref{prop:4}.

\section{Continuous State Space}
\label{ap:k}

Let $\Theta=[\underline{\theta},\overline{\theta}]\subset \mathbb R$
be the state space, endowed with its Borel $\sigma$-algebra
$\mathcal B(\Theta)$. Let $\mathcal P(\Theta)$ be the set of probability measures over $\Theta$. Let $q(\cdot)\in \mathcal P(\Theta)$ be the prior belief,

As in the main text, we compare two experiments $F$ and $G$, with signal
spaces $\mathcal X=[\underline{x},\overline{x}]$ and
$\mathcal Y=[\underline{y},\overline{y}]$. For each $\theta\in\Theta$,
the conditional distribution functions $F(\cdot\mid\theta)$ and
$G(\cdot\mid\theta)$ are absolutely continuous, admitting jointly
measurable densities $f(\cdot\mid\theta)$ and $g(\cdot\mid\theta)$. For a better exposition, we make the following regularity assumption.
    \begin{assumption}
\label{asp:1}
The conditional densities are continuous in the state in $L^1$. That is,
for every sequence $\theta_k\to\theta$ in $\Theta$,
\[
\int_{\mathcal X}
\left|f(x\mid\theta_k)-f(x\mid\theta)\right|dx
\to 0
\]
and
\[
\int_{\mathcal Y}
\left|g(y\mid\theta_k)-g(y\mid\theta)\right|dy
\to 0 .
\]
\end{assumption}

Given $q$, let $F_q$ and $G_q$ denote the unconditional distributions of
$X$ and $Y$, respectively, and let $f_q$ and $g_q$ denote their densities.
For each $x\in\mathcal X$ with $f_q(x)>0$, the posterior belief induced
by $F$ is the probability measure
$\pi_F^q(\cdot\mid x)\in\mathcal P(\Theta)$ defined by
\[
\pi_F^q(B\mid x)
=
\frac{\int_B f(x\mid\theta)\,q(d\theta)}
{f_q(x)},
\qquad B\in\mathcal B(\Theta).
\]
Similarly, for each $y\in\mathcal Y$ with $g_q(y)>0$,
\[
\pi_G^q(B\mid y)
=
\frac{\int_B g(y\mid\theta)\,q(d\theta)}
{g_q(y)},
\qquad B\in\mathcal B(\Theta).
\]

Let $X\sim F_q$ and $Y\sim G_q$. The experiments and the prior induce the
random posterior beliefs
\[
\Pi_F^q:=\pi_F^q(\cdot\mid X),
\qquad
\Pi_G^q:=\pi_G^q(\cdot\mid Y),
\]
which are random elements of $\mathcal P(\Theta)$. Let $\mathcal B_b(\Theta)$ denote the set of bounded Borel-measurable
functions $\varphi:\Theta\to\mathbb R$. We now extend the definition of
the linear-Blackwell order in \cref{M-def:3} to a continuous state space.

\begin{definition}
\label{def:9}
Experiment $F$ dominates experiment $G$ in the linear-Blackwell order,
denoted $F\succeq_{LB}G$, if for every prior
$q\in\mathcal P(\Theta)$ and every $\varphi\in\mathcal B_b(\Theta)$,
\[
\int_\Theta \varphi(\theta)\,\Pi_F^q(d\theta)
\succeq_{cx}
\int_\Theta \varphi(\theta)\,\Pi_G^q(d\theta).
\]
This requires that for every convex function $C:\mathbb R\to\mathbb R$,
\[
\int_{\mathcal X}
C\left(
\int_\Theta \varphi(\theta)\,\pi_F^q(d\theta\mid x)
\right)f_q(x)\,dx
\geq
\int_{\mathcal Y}
C\left(
\int_\Theta \varphi(\theta)\,\pi_G^q(d\theta\mid y)
\right)g_q(y)\,dy .
\]
\end{definition}

Thus, for every state valuation $\varphi$, dominance in the LB order implies that the posterior expectation
\[
\mathbb E_q[\varphi(\tilde\theta)\mid X]
=
\int_\Theta \varphi(\theta)\,\pi_F^q(d\theta\mid X)
\]
induced by $F$ is a mean-preserving spread of the corresponding posterior
expectation $\mathbb E_q[\varphi(\tilde\theta)\mid Y]$ induced by $G$. This extends the posterior-expectation characterization
in \cref{M-prop:2}.

Using the hinge characterization in \eqref{M-eqn:2}, it follows that
$F\succeq_{LB}G$ if and only if, for every
$q\in\mathcal P(\Theta)$ and every $\varphi\in\mathcal B_b(\Theta)$,
\begin{equation}
\label{eqn:55}
\int_{\mathcal X}
\left(
\int_\Theta \varphi(\theta) f(x\mid\theta)\,q(d\theta)
\right)_+
dx
\geq
\int_{\mathcal Y}
\left(
\int_\Theta \varphi(\theta) g(y\mid\theta)\,q(d\theta)
\right)_+
dy .
\end{equation}

Under \cref{asp:1}, the maps
\[
q\mapsto
\int_{\mathcal X}
\left(
\int_\Theta f(x\mid\theta)\,q(d\theta)
\right)_+
dx
\quad\text{and}\quad
q\mapsto
\int_{\mathcal Y}
\left(
\int_\Theta g(y\mid\theta)\,q(d\theta)
\right)_+
dy
\]
are continuous under weak convergence of finite signed Borel measures.
Since signed measures of the form $ \varphi(\theta)d\theta$, with
$ \varphi\in\mathcal B_b(\Theta)$, are weakly dense in the set of finite
signed Borel measures on $\Theta$, \eqref{eqn:55} is equivalent to the
following condition.

\begin{lemma}
\label{lem:6}
We have $F\succeq_{LB}G$ if and only if, for every
$ \varphi\in\mathcal B_b(\Theta)$,
\begin{equation}
\label{eqn:56}
\int_{\mathcal X}
\left(
\int_\Theta  \varphi(\theta) f(x\mid\theta)\,d\theta
\right)_+
dx
\geq
\int_{\mathcal Y}
\left(
\int_\Theta  \varphi(\theta) g(y\mid\theta)\,d\theta
\right)_+
dy .
\end{equation}
\end{lemma}

\cref{lem:6} extends \cref{M-lem:1} to the continuous-state setting.

We now extend the likelihood-ratio characterization in \cref{M-prop:1}.
Fix an arbitrary base state $\theta^\ast\in\Theta$ and assume
\[
f(x\mid\theta^\ast)>0 \quad \text{for all } x\in\mathcal X,
\qquad
g(y\mid\theta^\ast)>0 \quad \text{for all } y\in\mathcal Y.
\]
Define the likelihood-ratio functions relative to $\theta^\ast$ by
\[
\ell_F^{\theta^\ast}(x,\theta)
:=
\frac{f(x\mid\theta)}{f(x\mid\theta^\ast)},
\qquad
\ell_G^{\theta^\ast}(y,\theta)
:=
\frac{g(y\mid\theta)}{g(y\mid\theta^\ast)}.
\]
Let $X^\ast\sim F(\cdot\mid\theta^\ast)$ and
$Y^\ast\sim G(\cdot\mid\theta^\ast)$. Using \eqref{M-eqn:2} and \cref{lem:6}, we obtain
\begin{equation}
    \label{eqn:57}
    F\succeq_{LB}G
\quad\Longleftrightarrow\quad
\int_\Theta
\ell_F^{\theta^\ast}(X^\ast,\theta)\varphi(\theta)\,d\theta
\succeq_{cx}
\int_\Theta
\ell_G^{\theta^\ast}(Y^\ast,\theta)\varphi(\theta)\,d\theta
\quad
\forall \varphi\in\mathcal B_b(\Theta).
\end{equation}

We can also extend the geometric characterization in \cref{M-prop:3}. Let
\[
\mathcal H_F
:=
\{h:\mathcal X\to[0,1]\mid h \text{ is Borel-measurable}\}.
\]
For each $h\in\mathcal H_F$, define the statewise expectation profile
\[
z_F(h)(\theta)
:=
\int_{\mathcal X} h(x)f(x\mid\theta)\,dx,
\qquad \theta\in\Theta.
\]
Under \cref{asp:1}, $z_F(h)$ is continuous in $\theta$, so
$z_F(h)\in C(\Theta)$. The Lorenz zonoid of experiment $F$ is
\[
Z(F)
:=
\{z_F(h)\mid h\in\mathcal H_F\}.
\]
Moreover, $Z(F)$ is a compact and convex subset of $C(\Theta)$. Define
$\mathcal H_G$, $z_G(h)$, and $Z(G)$ analogously.

\begin{proposition}
\label{prop:11}
\[
F\succeq_{LB}G
\quad\Longleftrightarrow\quad
Z(F)\supseteq Z(G).
\]
\end{proposition}

The proof is analogous to that of \cref{M-prop:3}.

We further relate the continuous-state order to its finite-state restrictions. For any finite subset 
$S=\{\theta^1,\ldots,\theta^m\}\subseteq\Theta$, let $F|_S$ and $G|_S$ denote the finite-state experiments obtained by restricting $F$ and $G$ to the state space $S$.

\begin{proposition}
\[
F\succeq_{LB}G
\quad\Longleftrightarrow\quad
F|_S\succeq_{LB}G|_S
\quad\text{for every finite }S\subseteq\Theta .
\]
\end{proposition}

The ``only if'' direction is immediate. For the converse, it is enough to verify the continuous-state positive-part inequality for every finite signed Borel measure on $\Theta$. Any such measure can be approximated by finitely supported signed measures, and the desired inequality for these finitely supported measures follows from the LB dominance of the corresponding finite-state restrictions. Passing to the limit gives $F\succeq_{LB}G$.

Finally, we extend results in applications. Regarding the decision-making application, \cref{M-lem:2} and \cref{M-thm:1} continue to hold when the state space is continuous. The proofs extend directly: the binary-action result follows from the Lorenz-zonoid inclusion, and the ex ante payoff comparison in QCC problems follows from the same binary-subproblem decomposition. The strategy-wise payoff comparison follows from the same separating-hyperplane argument, with finite-dimensional separating vectors replaced by positive Borel measures, which can be normalized into priors.

Regarding the principal--agent applications, \cref{M-thm:2,M-thm:3} also extend to the continuous-state setting. In the moral-hazard problem, the agent's mixed action set becomes $\mathcal P(\Theta)$, and each incentive scheme $w:X\to[\underline u,\bar u]$ induces a statewise expected-utility profile $\theta\mapsto \mathbb E_F[w(X)\mid \theta]$. Since IR and IC depend on $w$ only through this profile, the flexibility part follows from the Lorenz-zonoid inclusion $Z(F)\supseteq Z(G)$. The effectiveness part follows by applying the finite-state argument to arbitrary finite restrictions of the mixed-action space $\mathcal P(\Theta)$ and passing to the limit using bounded schemes and compactness. The screening result extends analogously from the same finite-restriction argument.

\section{Weak-Linear-Blackwell Order}
\label{ap:l}

\subsection{Characterization}

We return to the finite-state setting
$\Theta=\{\theta_0,\ldots,\theta_n\}$ and introduce a weaker
ranking of experiments based on their Lorenz zonoids.

\begin{definition}
\label{def:10}
The experiment $F$ dominates the experiment $G$ in the
\textit{weak-linear-Blackwell (wLB) order}, and we write
$F\succeq_{\textup{wLB}}G$, if for each
$\bz=(z_0,\ldots,z_n)\in\cz(G)$ and each $i\in\{0,\ldots,n\}$,
there exists $\bz^{i}\in\cz(F)$ such that
\begin{equation*}
    z_i^{i}\geq z_i
    \quad\text{and}\quad
    z_j^{i}\leq z_j
    \quad\forall\,j\in\{0,\ldots,n\}\setminus\{i\}.
\end{equation*}
\end{definition}

To motivate this definition, consider the moral hazard problem in
Section~\ref{M-sec:5} of the main text when the agent is restricted
to pure actions. Suppose that an incentive scheme
$\hw\in\mathcal{W}^{\bu}_{G}$ implements $\theta_i$ under $G$.
If $F\succeq_{\textup{wLB}}G$, then there exists
$\hw'\in\mathcal{W}^{\bu}_{F}$ that yields a weakly higher expected
utility from payment in state $\theta_i$ and weakly lower expected
utilities from payment in all other states. Since action costs
are unchanged, the agent's expected payoff from the target action
weakly increases, while his expected payoff from each deviation
weakly decreases. Thus, $\hw'$ preserves both IR and IC and
implements $\theta_i$ under $F$. Preserving implementability for the pure action $\theta_i$
therefore does not require the statewise expectation matching
implied by the LB order.

We next establish a geometric characterization of the wLB order.
For a nonempty set $S\subseteq\mathbb{R}^{n+1}$, define its
\emph{coordinatewise-max closure} by
\begin{equation*}
    \operatorname{mcl}(S):=
    \left\{\bz^1\vee\cdots\vee\bz^m
    \mid m\geq 1,\ \bz^1,\ldots,\bz^m\in S\right\},
\end{equation*}
where the $i$th coordinate of $\bz^1\vee\cdots\vee\bz^m$ is
$\max_{1\leq k\leq m}z_i^k$. Thus, $\operatorname{mcl}(S)$ is
the smallest set containing $S$ that is closed under coordinatewise
maxima. Since there are $n+1$ coordinates, every element of
$\operatorname{mcl}(S)$ can be expressed as the coordinatewise
maximum of at most $n+1$ elements of $S$.

\begin{proposition}
\label{prop:13}
We have
\begin{equation*}
    F\succeq_{\textup{wLB}}G
    \iff
    \operatorname{mcl}(\cz(F))
    \supseteq\operatorname{mcl}(\cz(G)).
\end{equation*}
Equivalently, $F\succeq_{\textup{wLB}}G$ if and only if
$\operatorname{mcl}(\cz(F))
    \supseteq\cz(G)$.
\end{proposition}

\begin{proof}
Suppose first that $F\succeq_{\textup{wLB}}G$, and fix
$\bz\in\cz(G)$. For each $i\in\{0,\ldots,n\}$,
\cref{def:10} yields $\bz^{\prime i}\in\cz(F)$ such that
$z_i^{\prime i}\geq z_i$ and $z_j^{\prime i}\leq z_j$
for every $j\neq i$. If $z_i>0$, let
$\tilde{\bz}^{\,i}:=(z_i/z_i^{\prime i})\bz^{\prime i}$;
if $z_i=0$, let $\tilde{\bz}^{\,i}:=\bo$.
Since $\cz(F)$ is convex and contains $\bo$, we have
$\tilde{\bz}^{\,i}\in\cz(F)$. Moreover, all coordinates are
nonnegative, so $\tilde z_i^{\,i}=z_i$ and
$\tilde z_j^{\,i}\leq z_j$ for every $j\neq i$. Hence
\begin{equation*}
    \bz=\tilde{\bz}^{\,0}\vee\cdots\vee\tilde{\bz}^{\,n}
    \in\operatorname{mcl}(\cz(F)).
\end{equation*}

Conversely, suppose that
$\cz(G)\subseteq\operatorname{mcl}(\cz(F))$.
For any $\bz\in\cz(G)$, write
$\bz=\bz^1\vee\cdots\vee\bz^m$, where
$\bz^1,\ldots,\bz^m\in\cz(F)$. For each $i$, some $k$
satisfies $z_i^k=z_i$, while $z_j^k\leq z_j$ for every
$j\neq i$. Thus, $\bz^k$ satisfies the inequalities in
\cref{def:10}, and $F\succeq_{\textup{wLB}}G$.

Finally, since $\operatorname{mcl}(\cz(F))$ is closed under
coordinatewise maxima, it contains $\cz(G)$ if and only if
it contains $\operatorname{mcl}(\cz(G))$.
\end{proof}

By \cref{M-prop:3,prop:13}, the LB order compares the Lorenz
zonoids themselves, whereas the wLB order compares their
coordinatewise-max closures. In particular,
$F\succeq_{\textup{LB}}G$ implies
$F\succeq_{\textup{wLB}}G$.

We can also characterize the wLB order by comparing the support
functions of the Lorenz zonoids in a restricted set of directions.

\begin{proposition}
\label{prop:14}
We have $F\succeq_{\textup{wLB}}G$ if and only if, for each
$i\in\{0,\ldots,n\}$ and every
$\bb=(b_0,\ldots,b_n)\in\mr^{n+1}$ satisfying $b_i\geq 0$
and $b_j\leq 0$ for all $j\neq i$,
\begin{equation}
\label{eqn:58}
    \max_{\bz\in\cz(F)}\bb\cdot\bz
    \geq
    \max_{\bz\in\cz(G)}\bb\cdot\bz.
\end{equation}
Equivalently, the same characterization holds with the sign
restrictions $b_i\leq 0$ and $b_j\geq 0$ for all $j\neq i$.
\end{proposition}

\begin{proof}
Suppose first that $F\succeq_{\textup{wLB}}G$. Fix $i$ and
a vector $\bb$ satisfying $b_i\geq 0$ and $b_j\leq 0$
for all $j\neq i$. For each $\bz\in\cz(G)$,
\cref{def:10} yields $\bz^{\prime i}\in\cz(F)$ such that
$z_i^{\prime i}\geq z_i$ and $z_j^{\prime i}\leq z_j$
for all $j\neq i$. Hence
$\bb\cdot\bz^{\prime i}\geq\bb\cdot\bz$.
Maximizing over $\bz\in\cz(G)$ gives \eqref{eqn:58}.

Conversely, suppose that \eqref{eqn:58} holds in every such
direction, but $F\nsucceq_{\textup{wLB}}G$. Then there exist
$\bz\in\cz(G)$ and $i\in\{0,\ldots,n\}$ such that $\cz(F)$
is disjoint from the closed convex set
\begin{equation*}
    K_i(\bz):=
    \left\{\bz'\in\mr^{n+1}\mid
    z_i'\geq z_i,\ z_j'\leq z_j\ \forall\,j\neq i\right\}.
\end{equation*}
Since $\cz(F)$ is compact and convex, the separating hyperplane
theorem yields $\bb\in\mr^{n+1}$ such that
\begin{equation*}
    \max_{\bz'\in\cz(F)}\bb\cdot\bz'
    <\inf_{\bz'\in K_i(\bz)}\bb\cdot\bz'.
\end{equation*}
The infimum is finite only if $b_i\geq 0$ and $b_j\leq 0$
for every $j\neq i$, in which case it equals $\bb\cdot\bz$.
Therefore,
\begin{equation*}
    \max_{\bz'\in\cz(F)}\bb\cdot\bz'
    <\bb\cdot\bz
    \leq\max_{\bz'\in\cz(G)}\bb\cdot\bz',
\end{equation*}
contradicting \eqref{eqn:58}.

Finally, central symmetry of $\cz(F)$ about
$(\frac12,\ldots,\frac12)$ implies
\begin{equation*}
    \max_{\bz\in\cz(F)}\bb\cdot\bz
    =\sum_{j=0}^{n}b_j
    +\max_{\bz\in\cz(F)}(-\bb)\cdot\bz.
\end{equation*}
The same identity holds for $G$. Thus, \eqref{eqn:58} holds
for $\bb$ if and only if it holds for $-\bb$, which establishes
the equivalent characterization with the opposite sign restrictions.
\end{proof}

As shown in the proof of \cref{M-prop:3},
\begin{equation*}
    \max_{\bz\in\cz(F)}\bb\cdot\bz
    =\int_{\cX}
    \left(\sum_{j=0}^{n}b_j f(x\mid\theta_j)\right)_{+}\,dx,
\end{equation*}
and analogously for $G$. Hence, \cref{prop:14} states that
the positive-part inequality in \eqref{M-eqn:3} characterizes
the wLB order when restricted to these directions, whereas
the LB order requires it for every $\bb\in\mr^{n+1}$.

We now relate the wLB order to a restriction of the linear convex
order.

\begin{definition}
\label{def:11}
The random vector $\bmu$ dominates $\bnu$ in the
\textit{positive linear convex order}, and we write
$\bmu\succeq_{\textup{plcx}}\bnu$, if
$\bb\cdot\bmu\succeq_{\textup{cx}}\bb\cdot\bnu$
for each $\bb\in(0,\infty)^n$.
\end{definition}

For a reference state $\theta_r$ with positive signal densities,
define
\begin{equation*}
    l_{F,j}^{(r)}(x):=
    \frac{f(x\mid\theta_j)}{f(x\mid\theta_r)},
    \qquad
    \bl_F^{(r)}(x):=
    \left(l_{F,j}^{(r)}(x)\right)_{j\neq r}.
\end{equation*}
Let $X$ be distributed according to $F(\cdot\mid\theta_r)$.
As in the main text, we suppress $X$ from the notation for the
induced random vector and write it as $\bl_F^{(r)}$.
Define $\bl_G^{(r)}$ analogously. In particular,
$\bl_F^{(0)}=\bl_F$ and $\bl_G^{(0)}=\bl_G$.

\begin{proposition}
\label{prop:15}
Assume that $f(x\mid\theta_i)>0$ and $g(y\mid\theta_i)>0$
for all $x\in\cX$, $y\in\cY$, and $i\in\{0,\ldots,n\}$.
Then
\begin{equation*}
    F\succeq_{\textup{wLB}}G
    \iff
    \bl_F^{(r)}\succeq_{\textup{plcx}}\bl_G^{(r)}
    \quad\forall\,r\in\{0,\ldots,n\}.
\end{equation*}
\end{proposition}

\begin{proof}
Fix $r\in\{0,\ldots,n\}$ and
$\bb=(b_j)_{j\neq r}\in(0,\infty)^n$.
Each component of $\bl_F^{(r)}$ and $\bl_G^{(r)}$
has expectation one. Hence, $\bb\cdot\bl_F^{(r)}$
and $\bb\cdot\bl_G^{(r)}$ have the same expectation
$\sum_{j\neq r}b_j$. By \eqref{M-eqn:1}, their
convex-order comparison is equivalent to
\begin{equation*}
    \mathbb{E}\left[
        \left(\bb\cdot\bl_F^{(r)}-t\right)_{+}
    \right]
    \geq
    \mathbb{E}\left[
        \left(\bb\cdot\bl_G^{(r)}-t\right)_{+}
    \right]
    \quad\forall\,t\in\mr.
\end{equation*}
For $t\leq 0$, both sides equal $\sum_{j\neq r}b_j-t$.
For $t>0$, the inequality can be written as
\begin{equation*}
\begin{split}
    &\int_{\cX}
    \left(
        \sum_{j\neq r}b_j f(x\mid\theta_j)
        -t f(x\mid\theta_r)
    \right)_{+}\,dx\\
    &\qquad\geq
    \int_{\cY}
    \left(
        \sum_{j\neq r}b_j g(y\mid\theta_j)
        -t g(y\mid\theta_r)
    \right)_{+}\,dy.
\end{split}
\end{equation*}
These are precisely the support-function comparisons in directions
whose $r$th coordinate is strictly negative and whose other
coordinates are strictly positive. Zero coefficients are included
by continuity of the support functions. Letting $r$ range over
all states therefore yields the characterization in \cref{prop:14}.
\end{proof}

\Cref{prop:15} treats all states symmetrically: under wLB dominance,
the positive linear convex comparison holds whichever state is
chosen as the reference. The requirement that it hold for every
reference state is essential, however. Unlike the unrestricted
linear convex comparison in \cref{M-prop:1}, a positive linear
convex comparison at one fixed reference state alone need not
imply the corresponding comparisons at other reference states.

\subsection{Applications}

We now consider two restrictions of the principal--agent models
in the main text, maintaining their other assumptions.

\subsubsection*{Moral Hazard with Pure Actions}

Consider the moral hazard problem in Section~\ref{M-sec:5}
when the agent can only choose a pure action
$\theta\in\Theta=\{\theta_0,\ldots,\theta_n\}$.
The agent incurs cost $c(\theta)$, where $c:\Theta\to\mr$.
Let $\mathfrak{M}_{\textup{p}}$ denote the class of environments
$\mathcal{M}=\{\bu,c,\gamma\}$ with the same utility bounds
and assumptions on $\gamma$ as in the main text.
For each $\theta_i\in\Theta$, let $I_F(\theta_i,\mathcal{M})$
denote the principal's minimum disutility from implementing
$\theta_i$, subject to IR and IC against all other actions
in $\Theta$. We set $I_F(\theta_i,\mathcal{M})=+\infty$
if $\theta_i$ is not implementable under $F$.

\begin{proposition}
\label{prop:16}
The following statements are equivalent:
\begin{enumerate}[label=\textup{(\roman*)}]
    \item $F\succeq_{\textup{wLB}}G$.

    \item
    $I_F(\theta_i,\mathcal{M})
    \leq I_G(\theta_i,\mathcal{M})$
    for every $\theta_i\in\Theta$ and
    $\mathcal{M}\in\mathfrak{M}_{\textup{p}}$.

\end{enumerate}
Each statement also implies that every pure action implementable
under $G$ remains implementable under $F$.
\end{proposition}

The minimum-disutility comparison follows from the dual argument
in \cref{M-ap:c}: with pure actions, the likelihood ratios of
deviations relative to the target enter through nonnegative
linear combinations. The positive linear convex comparisons in
\cref{prop:15} therefore suffice. Necessity holds even for linear
disutility, since suitable action costs and a binding participation
constraint recover the statewise comparisons in \cref{def:10}.
Optimizing over target actions gives the maximal-payoff comparison,
while arbitrary gross payoffs allow any target action to be
singled out.

Preservation of pure-action implementability alone is weaker than
wLB dominance. To preserve IR and IC, it suffices that expected
utility from the target action does not decrease and that the
expected utility gap between the target action and each deviation
weakly increases. Thus, expected utility from a deviation may
increase, provided that the increase is no larger than that from
the target action. Consequently, wLB dominance is sufficient, but
not necessary, for preserving pure-action implementability.
However, wLB dominance is necessary for the minimum-disutility
comparison in \cref{prop:16}. Necessity follows even when the
principal and the agent share the same risk attitude: consider
the case in which both the principal and the agent share the same risk attitude, so that $\gamma(w)=w$
after normalization. For any contract under $G$, choose action
costs so that IR and all IC constraints bind. Keeping disutility
weakly lower under $F$ then requires expected utility from the
target action to remain unchanged, while IC requires expected
utility from each deviation to weakly decrease. Varying the
original expected-utility profile and the target action recovers
the comparisons in \cref{def:10}.

\subsubsection*{Screening When the Agent Observes the State}

Let $\mathfrak{E}_{\Theta}$ denote the class of screening
environments in Section~\ref{M-sec:6} in which the agent observes
the realized state. We identify his type with $\theta\in\Theta$,
so an allocation rule is a mapping $\ha:\Theta\to\Delta(\cA)$.
Recall that $D_F(\ha,\mathcal{E})$, defined in \cref{ap:d1},
is the principal's minimum disutility from implementing $\ha$,
and $W_F(\mathcal{E})$ is her maximal expected payoff.

\begin{proposition}
\label{prop:17}
The following statements are equivalent:
\begin{enumerate}[label=\textup{(\roman*)}]
    \item $F\succeq_{\textup{wLB}}G$.

    \item
    For every $\mathcal{E}\in\mathfrak{E}_{\Theta}$ and every
    allocation rule $\ha:\Theta\to\Delta(\cA)$, implementability
    under $G$ implies implementability under $F$.

    \item
    $D_F(\ha,\mathcal{E})\leq D_G(\ha,\mathcal{E})$
    for every $\mathcal{E}\in\mathfrak{E}_{\Theta}$ and every
    allocation rule $\ha:\Theta\to\Delta(\cA)$.

    \item
    $W_F(\mathcal{E})\geq W_G(\mathcal{E})$
    for every $\mathcal{E}\in\mathfrak{E}_{\Theta}$.
\end{enumerate}
\end{proposition}

For each report $\theta_i$, the wLB order allows the principal
to increase the expected transfer conditional on $\theta_i$
while decreasing its expectation conditional on every other state.
Applying this comparison separately to each report and selected
alternative preserves IR and IC. The minimum-disutility comparison
follows from the dual argument in \cref{ap:d1}, where the
likelihood ratios again enter through nonnegative linear
combinations. Optimizing over allocation rules yields the
maximal-payoff comparison. Conversely, if wLB dominance fails,
one can choose allocation utilities and transfer bounds for which
an allocation rule is implementable under $G$ but not under $F$,
and choose the principal's allocation payoffs so that this failure
reduces her maximal payoff. The conclusions continue to allow
allocation-dependent signal arrival, since adding the common
uninformative outcome associated with $\psi(a)$ preserves
wLB dominance.

\end{appendix}

\bibliographystyle{ecta-fullname} 
\bibliography{supp}  

